\documentclass[a4paper]{article}
\pdfoutput=1
\usepackage{jcappub}

\usepackage{graphicx,color,amsmath}
\usepackage{graphicx}
\usepackage[caption=false]{subfig}
\usepackage{dcolumn}
\usepackage{multirow}
\usepackage{tabularx}
\usepackage{bm}
\usepackage{amsmath}
\usepackage{leftindex}
\usepackage{amssymb}
\usepackage{comment}
\usepackage{setspace}
\usepackage{physics}
\usepackage[dvipsnames]{xcolor}
\usepackage{enumerate}
\usepackage{booktabs}
\graphicspath{{figs/}}

\definecolor{mypurple}{RGB}{164,64,214}

\newcommand{\ga}{g_{{\rm a} \gamma\gamma}}

\newcommand{\ompl}{m_{\gamma}}
\newcommand{\ma}{m_{\rm a}}
\newcommand{\nhat}{\hat{n}}
\newcommand{\res}{{\rm res}}
\newcommand{\dom}{{\rm dom}}

\newcommand{\asc}{\ensuremath{{\rm a}}}

\newcommand{\ie}{\textit{i.e.}\, }
\newcommand{\eg}{e.g.\ }

\newcommand{\perimeter}{Perimeter Institute for Theoretical Physics, 31 Caroline St N, Waterloo, ON N2L 2Y5, Canada}

\newcommand{\UW}{Department of Physics and Astronomy, University of Waterloo, Waterloo, ON N2L 3G1, Canada}

\newcommand{\york}{Department of Physics and Astronomy, York University, Toronto, ON M3J 1P3, Canada}

\defcitealias{Pirvu:2023lch}{\text{Paper} I}

\begin{document}

\title{Axion-Induced Patchy Screening of the Cosmic Microwave Background}

\author[a]{Cristina Mondino}
\author[a,b]{, Dalila~P\^irvu}
\author[a]{, Junwu~Huang}
\author[a,c]{, and Matthew~C.~Johnson}

\affiliation[a]{\perimeter}
\affiliation[b]{\UW}
\affiliation[c]{\york}

\emailAdd{cmondino@perimeterinstitute.ca}
\emailAdd{dpirvu@perimeterinstitute.ca}
\emailAdd{jhuang@perimeterinstitute.ca}
\emailAdd{mjohnson@perimeterinstitute.ca}

\date{\today}

\abstract{Cosmic Microwave Background (CMB) photons can undergo resonant conversion into axions in the presence of magnetized plasma distributed inside non-linear large-scale structure (LSS). This process leads to axion-induced patchy screening: secondary temperature and polarization ani\-sotropies with a characteristic non-blackbody frequency dependence that are strongly correlated with the distribution of LSS along our past light cone. We compute the axion-induced patchy screening contribution to two- and three- point correlation functions that include CMB anisotropies and tracers of LSS within the halo model. We use these results to forecast the sensitivity of existing and future surveys to photon-axion couplings for axion masses between $2\times 10^{-13}$ eV and $3\times 10^{-12}$ eV, using a combination of empirical estimates from Planck data of the contribution from instrumental noise and foregrounds as well as modeled contributions on angular scales only accessible with future datasets. We demonstrate that an analysis using Planck and the unWISE galaxy catalogue would be complementary to the most sensitive existing astrophysical axion searches, probing couplings as small as $3\times 10^{-12} \, {\rm GeV}^{-1}$, while observations from a future survey such as CMB-S4 could extend this reach by almost an additional order of magnitude.}

\maketitle

\tableofcontents

\section{Introduction and Summary}

The QCD axion and axion-like particles, generally referred to as axions, are among the most well-motivated additions to the Standard Model (SM)~\cite{axion1,axion2,axion3,Peccei:1977ur,Svrcek:2006yi,Arvanitaki:2009fg}. Axions provide a solution to the strong CP problem~\cite{axion1,axion2,axion3,Peccei:1977ur}, and can be good dark matter candidates~\cite{Preskill:1982cy, Abbott:1982af, Dine:1982ah}. Axions can couple to the SM through the strong or electromagnetic force. Here, we will be concerned with the coupling between the axion and the photon, described by the Lagrangian:
\begin{equation}\label{eq:lagr}
    \mathcal{L}_{\asc \gamma} = - \frac{1}{4}\ga \, \asc \, F^{\mu\nu}\tilde{F}_{\mu\nu} = \ga \, \asc \, \mathbf{E} \cdot \mathbf{B},
\end{equation}
where $F^{\mu\nu}$ is the electromagnetic field-strength tensor and $\tilde{F}^{\mu\nu} \equiv 1/2 \, \epsilon_{\mu\nu\alpha\beta}F^{\alpha\beta}$ its dual. The above coupling allows the photon to oscillate into an axion in the presence of an external magnetic field transverse to the photon three-momentum.

Searches for photon-axion conversion have been conducted over a wide range of the electromagnetic spectrum and axion parameter space both in terrestrial laboratory experiments and through astrophysical observations~\footnote{For a comprehensive summary of ongoing effort to look for axions, see~\cite{ParticleDataGroup:2022pth,ciaran_o_hare_2020_3932430}.}. These efforts have probed axions from effectively zero mass to masses up to a TeV, and down to a coupling of $\ga \sim 10^{-11}  \, {\rm GeV}^{-1}$. In this paper, we focus on very low-mass axions, $m_{\asc} \sim \mathcal{O}({\rm peV})$, where the best existing limit on the axion-photon couplings $\ga$ comes from the CERN Axion Solar Telescope (CAST) experiment~\cite{CAST:2004gzq} and various astrophysical observations. CAST uses a large magnetic field to induce the conversion of axions produced in the Sun to photons, placing a limit of $\ga < 6.6 \times 10^{-11}  \, {\rm GeV}^{-1}$~\cite{CAST:2017uph} at low mass. The strongest astrophysical constraints arise from scenarios where sources of axions (from stars~\cite{Dessert:2020lil, Ning:2024eky} or supernovae~\cite{Hoof:2022xbe}) are converted to photons in the galactic magnetic field or scenarios where photons from a background source are converted to axions in extragalactic magnetic fields (\eg~\cite{Reynolds:2019uqt}). The strongest existing astrophysical constraints are from the latter category, yielding $\ga < 6-8 \times 10^{-13}  \, {\rm GeV}^{-1}$ for $m_{\asc} \lesssim 10\ {\rm peV}$, from the impact on AGN spectra of photons converting to axions in the magnetized intracluster medium~\cite{Reynolds:2019uqt}. Constraints that are independent of the axion-photon coupling in this mass range also comes from black hole superradiance~\cite{Arvanitaki:2014wva}, which can be affected by axion self-interactions~\cite{Baryakhtar:2020gao}.

The cosmic microwave background (CMB) is an exquisitely calibrated source: it has an almost perfect blackbody frequency spectrum, anisotropies are small and follow simple, Gaussian statistics, and it is only weakly polarized. Measurements of the CMB are therefore extremely sensitive to {\it secondary} anisotropies and spectral distortions produced by the interactions of CMB photons with large-scale structure (LSS) as they propagate through cosmic history to our telescopes. Secondary anisotropies are a primary target of future CMB surveys such as Simons Observatory~\cite{Ade:2018sbj} (SO), CMB-S4~\cite{abazajian_cmb-s4_2016}, and CMB-HD~\cite{Sehgal:2019ewc} which lie on the high-resolution, low-noise frontier; their true potential will be unleashed through cross-correlation with upcoming galaxy surveys performed by Vera Rubin Observatory Legacy Survey of Space and Time (LSST)~\cite{0912.0201}, Dark Energy Spectroscopic Instrument~\cite{DESI:2019jxc} (DESI), Euclid~\cite{Laureijs2011}, and Spectro-Photometer for the History of the Universe, Epoch of Reionization, and Ices Explorer~\cite{SPHEREx:2014bgr} (SPHEREx). Standard Model sources of secondary anisotropies include Sunyaev Zel'dovich effects (scattering from charges) and weak lensing (scattering from masses). Measurements of these secondaries have broad application, from determining the sum of neutrino masses to narrowing down the properties of inflationary cosmology~\cite{Ade:2018sbj,abazajian_cmb-s4_2016,Sehgal:2019ewc}. Any beyond the SM (BSM) physics scenarios that involves new interactions between CMB photons and LSS will lead to new sources of secondary anisotropies. The high sensitivity and resolution of existing and upcoming surveys motivates identifying the range of BSM models that lead to new CMB secondaries and designing optimal search techniques for their signatures. 

In this paper we investigate scenarios where CMB photons are converted to axions in magnetic fields associated with LSS. As CMB photons transit a magnetic field $\mathbf{B}_{\perp}$, perpendicular to the photon's direction of propagation, the polarization state along $\mathbf{B}_{\perp}$, $A_{\parallel}$, mixes with an ultra-relativistic axion ($\omega \gg m_{\asc}$) according to the equation of motion~\cite{Raffelt:1987im, Deffayet:2001pc, Mirizzi:2005ng}
\begin{equation}
    \left[\omega - i\partial_z + \frac{1}{2}\begin{pmatrix} -\ompl^2/\omega & \ga |\mathbf{B}_{\perp}| \\ \ga |\mathbf{B}_{\perp}| & -\ma^2/\omega\\ \end{pmatrix} \right]  \begin{pmatrix} A_{\parallel} \\ \asc \\ \end{pmatrix}  = 0,
\end{equation}
where $m_{\asc}$ is the axion mass and $\ompl^2 = e^2 n_e/m_e$ denotes the photon plasma mass in an ionized medium with electron density $n_e$. The probability for a CMB photon to resonantly convert to an axion is computed using the Landau-Zener expression~\cite{PhysRevLett.57.1275,Mirizzi:2009nq, Tashiro:2013yea}:
\begin{equation}\label{eq:prob}
    P_{A_{\parallel}\rightarrow \asc }^{\res} \simeq \frac{\pi\omega \ga^2 |\mathbf{B}_{\perp}|^2}{m_{\asc}^2}\left|\frac{\dd \ln \ompl^2}{\dd t}\right|_{t_{\res}}^{-1},
\end{equation}
which is a good approximation for the scenarios considered throughout this paper~\footnote{We provide a detailed derivation of this formula and its range of validity in the context of this work in App.~\ref{app:conversion}. A complimentary scenario where the axion mass is too light (effective massless) to satisfy the resonant condition anywhere in the universe is discussed in App.~\ref{app:massless_axion}.}. CMB photons propagating along different lines of sight encounter varying magnetic fields in media with varying density. From Eq.~\eqref{eq:prob}, the removal of CMB photons due to conversion into axions therefore leads to an anisotropic spectral distortion of the CMB intensity and polarization. 

Early work on the imprint of photon-axion conversion in the CMB used the absence of significant spectral distortions of the CMB monopole observed by COBE/FIRAS~\cite{1996ApJ...473..576F} to rule out proposals for axion-induced supernova dimming~\cite{Csaki:2001yk}~\footnote{It was also pointed out that, due to the photon plasma mass, the conversion probability acquires a frequency dependence, making it hard to account for achromatic dimming of SNe~\cite{Deffayet:2001pc}.}. Ref.~\cite{DAmico:2015snf} proposed to look for {\it anisotropic} spectral distortions of the CMB due to inhomogeneous plasma densities and magnetic fields, but assumed unrealistically shallow plasma density gradients to maintain the resonance condition over long distances, leading to an overestimate and incorrect frequency dependence of the conversion probability. Ref.~\cite{Schlederer:2015jwa} studied the spectral distortion in intensity from individual clusters, and obtained upper limits from Planck CMB temperature anisotropies of $\ga \lesssim \mathcal{O}(10^{-11}  \, {\rm GeV}^{-1})$, subject to assumptions about magnetic field profile in clusters. 

Subsequently, Ref.~\cite{Mukherjee:2018oeb} performed a detailed study of the CMB polarization and intensity signature from resonant and non-resonant conversion in the Milky Way's magnetic fields (with resonant conversion happening in coherent magnetic domains and non-resonant conversion in turbulent domains). Achieving a strong constraint from this signal requires high spectral resolution. The near-term space-based Lite-Bird~\cite{2012SPIE.8442E..19H} mission was forecasted to yield limits down to $\ga \sim 10^{-12}  \, {\rm GeV}^{-1}$~\cite{Mukherjee:2018oeb}, with stronger limits requiring futuristic missions such as PIXIE~\cite{A_Kogut_2011}. An analysis of the non-resonant signal using Planck temperature anisotropies provided a far weaker constraint of $\ga \lesssim 10^{-9}  \, {\rm GeV}^{-1}$~\cite{Mukherjee:2018oeb}. Ref.~\cite{Mukherjee:2019dsu} revisited the extragalactic signal first examined in~\cite{Schlederer:2015jwa}, modeling the detectability of the polarization signal. Subject to assumptions about magnetic fields, gas profiles, and the number of detectable clusters, they demonstrate that sensitivity of beyond $\ga \sim 10^{-13}  \, {\rm GeV}^{-1}$ could be achieved with next-generation CMB experiments such as CMB-S4. These works indicate that the CMB signature of photon to axion conversion could be competitive with the laboratory and astrophysical constraints described above.

This paper proposes a new framework to look for the spectral secondary CMB temperature and polarization anisotropies, sourced by the resonant conversion of CMB photons into axion radiation within the magnetic field of structure in the late universe. The effect is analogous to the case of CMB photons converting into dark photons studied in Ref.~\cite{Pirvu:2023lch}, hereafter~\citetalias{Pirvu:2023lch}, and manifests as an {\it anisotropic} absorption optical depth with a characteristic {\it linear frequency dependence}, $\tau^{\asc}(\omega, \hat{n}) \propto \omega$ that is strongly correlated with LSS.
The radial profile of the electron density inside halos provides a natural scanner of the photon plasma mass, which allows for resonant conversion between CMB photons and light axions in the halo magnetic fields for more than a decade in axion masses around $\ma \simeq 10^{-12}\, {\rm eV}$. Following~\citetalias{Pirvu:2023lch}, we use the halo model to compute two-point correlation functions of the resulting secondary CMB temperature anisotropies, and their correlations with tracers of LSS\footnote{Note that, in this work, we assume the axion makes up a negligible fraction of the dark matter abundance, and
the standard $\Lambda$CDM halo model is used here. An axion in this mass range or lighter can be a perfect dark matter candidate. If it were to be the dark matter in the universe, then interesting effects of axion clustering and structure suppression can also lead to modifications to the halo model~\cite{Hlozek:2014lca}, however these effects are relevant only for ultra-light axions, well below the masses considered in this work.}. Photon-axion conversion generates an anisotropic polarization signal from the unpolarized CMB monopole. We derive the corresponding CMB two-point function and the CMB polarization-LSS three-point function. To compute the signal strength, halo magnetic fields are modelled according to state of the art hydrodynamical cosmological simulations. Depending on the axion mass, photon-axion conversion predominantly occurs at different halo radii, inducing a characteristic scale dependence in the correlations. Our results provide a simple framework for computing correlation functions, which can be easily adapted to future analyses with differing assumptions. As in~\citetalias{Pirvu:2023lch}, the known frequency dependence of the signal is crucial to disentangle the axion-induced CMB secondary anisotropies from the primary anisotropies. We use the signal two- and three-point functions to project the sensitivity of CMB and LSS surveys to the axion-photon coupling.

We forecast that current data from Planck and unWISE galaxies are complementary to the best existing constraints from AGN spectra described above. The component-separated temperature-galaxy correlator $\ev{T^{\asc} g}$ is the most sensitive, and can in principle achieve $\ga \lesssim 3\times 10^{-12}  \, {\rm GeV}^{-1}$. Other correlators have a slightly weaker sensitivity, which can be helpful in confirming any possible detection. This strongly motivates an analysis using existing measurements, which we pursue in a separate publication~\cite{Goldstein:2024mfp}. Even though the region in the axion parameter space that can be currently probed seems disfavoured by other astrophysical searches, we stress that, given the assumptions required to derive those bounds, it is nevertheless interesting to have complementary probes which rely on a different set of assumptions and completely different observations. Moreover, we find that a future search using results from the CMB-S4 experiment could be sensitive to up to an order of magnitude smaller couplings compared to Planck, providing the most sensitive probe of axions in this mass range. Data from ACT~\cite{Louis_2017} and Simons Observatory~\cite{Ade:2018sbj} will continuously extend the reach in parameter space as we approach the S4 era.

The paper is organized as follows. We describe photon-axion conversion inside an individual halo in Sec.~\ref{sec:halo} and the resulting sky-averaged optical depth in Sec.~\ref{sec:skyopticaldepth}. In Sec.~\ref{sec:cmbsign} we derive the CMB temperature and polarization anisotropies, computing the temperature and polarization auto-correlation functions in Sec.~\ref{sec:cmb_auto}, the temperature-galaxy cross-correlation function in Sec.~\ref{sec:cmb_cross}, and the polarization-galaxy bispectrum in Sec.~\ref{sec:cmb_bispectrum}. In Sec.~\ref{sec:ilc} we investigate how CMB maps at different frequencies can be used to separate the photon-axion conversion signal from the primary CMB as well as galactic and extragalactic foregrounds. We forecast the sensitivity of existing and future CMB and galaxy surveys to photon-axion conversion in Sec.~\ref{sec:results} and comment on the limit of (effectively) massless axions in Sec.~\ref{sec:massless_axion}. Finally, we conclude in Sec.~\ref{sec:conclusion}. We include a set of appendices containing various technical discussions and derivations. In App.~\ref{app:conversion} we derive the photon-axion conversion probability and discuss the domain of validity in the context of our analysis. In App.~\ref{app:pol_cell} and~\ref{app:pol_bispectra} we derive the two- and three-point correlators involving polarization. In App.~\ref{app:gal_powerspectra} we detail the model used for the galaxy distribution and their power spectra. In App.~\ref{app:foregroundandnoise} we describe the foreground and instrumental noise models used in our forecasts. In App.~\ref{app:massless_axion} we sketch the utility of our formalism to the study of photon-axion conversion for (effectively) massless axions. In App.~\ref{app:estimate} we provide a qualitative order-of-magnitude estimate of the effect and the expected sensitivity. Finally, in App.~\ref{app:likelihood} we derive the likelihood for the axion signal. Natural units are used throughout, with $\hbar = c = k_B = 1$.

\section{Photon-axion conversion inside large-scale structure}\label{sec:conversion}

To model photon-to-axion conversion, we first need a model for the distribution of LSS. Here, we work within the halo model (see~\eg Refs.~\cite{2002PhR...372....1C,asgari2023halo} for a review), where dark and baryonic matter is assumed bound in virialized halos. The mass and redshift of halos determines the properties of the baryonic matter and galaxies that inhabit them. Correlation functions are then computed from the distribution of matter between and within halos. The halo model is extremely flexible, allowing for a unified framework to incorporate a wide variety of observables. Below, we first compute the photon-axion conversion probability in individual halos and then compute the sky-averaged (monopole) signal over all halos.

\subsection{Individual halo conversion}\label{sec:halo}

In this section, we derive the rate for resonant photon-axion conversion inside an individual halo, which is the main ingredient needed to compute the axion-induced CMB spectral distortions and anisotropies we study in the following sections. An important feature of the conversion into axions is that only the photon polarization parallel to the magnetic field mixes with the axion~\cite{Raffelt:1987im}; therefore there are two types of signal that can be looked for: a reduction (screening) in the intensity of the CMB and an induced polarization. 

In analogy with~\citetalias{Pirvu:2023lch}, we compute the probability that photons traveling along the direction $\hat{n}$ convert to axions inside a halo with mass $m_i$, at comoving distance $\chi_i(z_i)$, and redshift $z_i$ according to the Landau-Zener expression Eq.~\eqref{eq:prob}. We assume halos to be spherically symmetric and centered at $\hat{n}_i$. While individual halos could be far from spherical, we will only be concerned with ensemble averages below, where spherical symmetry is a good approximation. Due to the gradient in the halo's gas density profile, at some distance $r_{\res}(\chi_i, m_i)$ from the halo center, the resonance condition will be satisfied, with $\ompl(r_{\res}) = m_{\asc}$. Within these assumptions, the conversion probability is azimuthally symmetric with respect to the halo center, and can be written as
\begin{align}\label{eq:haloprob}
    P^{i}_{\gamma \rightarrow \asc}(\chi_i, m_i, \hat{n}_i - \hat{n}) & = P(\chi_i, m_i) \, N_{\res}(\chi_i, m_i) \, u(\hat{n}_i - \hat{n} | \chi_i, m_i) \, \gamma(\hat{n} | \chi_i),
\end{align}
where
\begin{gather}
    P(\chi_i, m_i) = \pi \omega(1+z_i) \ga^2 |\mathbf{B}(r_{\res},z_i,m_i)|^2 \left|\frac{\dd \ompl^2(r,z_i,m_i)}{\dd r}\right|_{r_{\res}}^{-1}, \\
    N_{\res}(\chi_i, m_i) =     
    \begin{cases}
      2, \quad r_{\res} < r_{\mathrm{vir}}, \\
      1, \quad r_{\res} = r_{\mathrm{vir}}, \\
    \end{cases}\, \\
    u(\hat{n}_i - \hat{n}| \chi_i, m_i) = \left[1-\frac{(\chi_i \theta/ r_{\res})^2}{(1+z_i)^2}\right]^{-1/2}.
\end{gather}
In the above expressions, $\theta \simeq |\hat{n}_i - \hat{n}| \le r_{\res}(1+z_i)/\chi_i \ll 1$ is the small angle between the halo center and the photon trajectory. $N_{\res}$ counts the number of resonance crossings for $r_{\res}$ within the virial radius $r_{\rm vir}$; it is set to one for $r_{\rm res} = r_{\rm vir}$ to smooth the sharp transition,~\ie the conversion only happens half of the time at the edge. $|\mathbf{B}(r_{\res},z_i,m_i)|$ denotes the magnitude of the magnetic field inside the halo at the resonance radius. In general, the magnetic field within the halo has a finite coherence length -- much smaller than $r_{\mathrm{vir}}$ or $r_{\mathrm{res}}$ -- and will take a random orientation in different domains. To account for the random angle of the magnetic field along each photon propagation direction, we multiply the conversion probability by $\gamma(\hat{n})$. This function takes a different form -- and has different statistical properties -- depending on whether we are computing the contribution of the conversion to the intensity or polarization signals, since a different combination of the magnetic field components enters in each case. We will therefore write $\gamma(\hat{n})$ explicitly in Sec.~\ref{sec:cmbsign}, when computing the axion-induced signal to CMB intensity and polarization anisotropies.

The radial profile of the photon plasma mass within a spherically symmetric halo can be modelled using the baryonic gas density profile $\rho_{\rm gas}(r,z_i,m_i)$ based on hydrodynamical cosmological simulations from Ref.~\cite{Battaglia:2016xbi}, the widely-used Battaglia et al. AGN Feedback profile (see~\eg Ref.~\cite{Smith:2018bpn} for an example in a different context)\footnote{As shown in the Appendix of ~\citetalias{Pirvu:2023lch}, the assumption of the specific AGN Feedback profile only mildly affects the final sensitivity to the signal, making our forecasts robust.}. Assuming that protons account for all the baryonic mass and that there is an equal number of electrons and protons, $\ompl^2(r,z_i,m_i) = e^2 \rho_{\rm gas}(r,z_i,m_i)/(m_e m_p)$, where $e$ is the electric charge, $m_e$ and $m_p$ the electron and proton masses, and the expression for $\rho_{\rm gas}$ is given in Sec.~2.2 of~\citetalias{Pirvu:2023lch}.

To model the magnetic field profile within halos we use recent results from the high resolution cosmological magneto-hydrodynamical zoom simulations from the Auriga project. The structure of the magnetic field in the circumgalactic medium and its time evolution has been analysed for Milky Way-like galaxies~\cite{2020MNRAS.498.3125P} and a broad range of halo masses~\cite{2023arXiv230913104P}. We use the interpolated magnetic field radial profiles provided by the authors of Ref.~\cite{2023arXiv230913104P} for $z < 1.9$ in 7 halo mass bins between $10^{10}\, {\rm M}_{\odot}$ and $10^{13}\, {\rm M}_{\odot}$; for heavier halos, we conservatively use a flat extrapolation,~\ie we assume the same magnetic field profile for all halo masses above the highest mass bin available. The magnetic field we use only includes the smooth halo component and no contributions from sub-structure (such as satellite galaxies) within the halos, which represent additions to the smooth density profile. For Milky Way-like halos, the magnetic field at $z=0$ reaches a value of about $0.1 \, \mu{\rm G}$ at the virial radius, with larger (smaller) values for heavier (lighter) halos. While in an individual halo the magnetic field is far from being spherically symmetric, we are only interested in statistically averaged quantities, in which case the averaged $|\mathbf{B}(r)|$ profiles should give a good approximation.

Similar to~\citetalias{Pirvu:2023lch}, we compute only the conversion in the smooth circumgalactic medium, in regions where the density is well characterized by the Battaglia density profile~\cite{Battaglia:2016xbi}.  Making use of, for example, the central  region of disk galaxies will extend the sensitivity to higher axion masses. Similarly, the sensitivity can be extended to lighter axions by utilizing the regions outside the virial radius of a halo, where the baryon density slowly decreases to the average density of the universe. Both extensions come with more modeling uncertainties associated with the density profile of matter and amplitude of the magnetic field in these regions. As a result, we defer these studies to a future publication.

We emphasise that the resonance conversion formula in Eq.~\eqref{eq:haloprob} is valid even for a finite coherence length of the magnetic field, as long as the latter changes slowly compared to the geometric mean of $2\pi \omega/m_{\asc}^2$ and $\left|\dd \ln \ompl^2/\dd r\right|_{r_{\res}}^{-1}$ \cite{Marsh:2021ajy}. For the smallest axion masses considered here, this is equivalent to a minimum coherence length of about a hundred parsec, which is much smaller than the smallest spatial resolution of the cosmological simulations, of $\mathcal{O}({\rm kpc})$. The magnetization of the circumgalactic medium is driven by galactic outflows transporting magnetised gas from the disk into the halo and later amplified by a turbulent dynamo acting in the halo. Both of these processes operate at length scales much larger than a parsec and strongly suggests that the magnetic field to be coherent over long enough length scales for Eq.~\eqref{eq:haloprob} to be valid. The results of the simulations additionally show that the magnetic energy power spectra are dominated by scales $\gtrsim 10$ kpc~\cite{2020MNRAS.498.3125P}, which support the assumption made here that the magnetic field is dominated by relatively large-scale fluctuations, while rapidly oscillating components can be neglected. Importantly, in this regime, the conversion probability scales linearly with frequency and it is insensitive to the exact value of the magnetic field coherence length. For a more detailed discussion on the derivation of the conversion probability and its range of validity see App.~\ref{app:conversion}.

\subsection{Conversion monopole and optical depth}\label{sec:skyopticaldepth}

The total photon-axion conversion probability is given by the sum of the individual halo contributions from Eq.~\eqref{eq:haloprob} along the line of sight. Since the magnetic field orientation changes randomly in each halo (and also between the two resonance crossings in the same halo), the axion-induced polarization will have a random positive or negative sign at each crossing of the conversion surface, yielding a zero mean but non-zero variance. The photon-axion conversion always removes photons, therefore reducing the intensity and {\it screening} the CMB monopole~\footnote{This is analogous to Thomson scattering of CMB photons from free electrons in the post-reionization universe screening the primary CMB anisotropies. Two crucial differences here are that while Thomson screening preserves the blackbody spectrum, axion-induced screening does not and while Thomson screening couples only to temperature anisotropies, axion-induced screening couples to the temperature {\em monopole}.}. On average, only one component of the magnetic field contributes to the intensity axion-induced screening, such that the $\gamma$ factor in Eq.~\eqref{eq:haloprob} averages to $1/3$. We define the axion-induced screening optical depth in the direction $\nhat$ as
\begin{align}\label{eq:tau_n}
    \tau^{\asc}(\nhat) & \equiv \sum_i P^{i}_{\gamma \rightarrow \asc}(\chi_i, m_i, \hat{n}_i - \hat{n}) = \sum_i P(\chi_i, m_i) N_{\res}(\chi_i, m_i) u(\hat{n}_i - \hat{n} | \chi_i, m_i) \gamma(\nhat | \chi_i)  \nonumber \\ & = \int_{z_{\rm min}}^{z_{\rm max}} \dd z\, \frac{\dd \tau^\asc(\nhat, \chi)}{\dd z},
\end{align}
where we have defined the differential optical depth as a function of redshift
\begin{align}\label{eq:dtaudz}
    \frac{\dd \tau^\asc}{\dd z}
    & \equiv \frac{\chi^2}{H} \int \dd^2 \nhat^{\prime}\, \dd m \sum_i \frac{\delta(\chi - \chi_i)}{\chi^2}\delta^2(\nhat^{\prime}-\nhat_i)\delta(m-m_i)\ \frac{1}{3}P(\chi, m) N_{\res}(\chi, m) u(\hat{n}^{\prime} - \hat{n} | \chi, m), 
\end{align}
where $H$ is the Hubble parameter at redshift $z$, we performed a change of variables from comoving distance $\chi(z)$ to redshift, and we replaced explicitly the average value of the $\gamma$ factor. Notice that, similarly to the dark photon case of~\citetalias{Pirvu:2023lch}, the axion-induced screening has a simple scaling with frequency and photon-axion coupling, in this case $\tau^{\asc} \propto \omega \ga^2$. We will leverage this frequency dependence to appropriately combine CMB measurements across multiple frequency channels and maximize the signal-to-noise ratio, as described in Sec.~\ref{sec:cmbsign}. The ensemble average of the optical depth is  
\begin{equation}\label{eq:tau_sky_avg}
    \begin{aligned}
        \ev{\tau^{\asc}(\nhat)} & = \int_{z_{\rm min}}^{z_{\rm max}} \dd z \, \ev{\frac{\dd \tau^\asc(\nhat, z)}{\dd z}} \\
        & = \int_{z_{\rm min}}^{z_{\rm max}} \dd z \frac{\chi^2}{H} \int \dd m  \, n(\chi, m) \frac{1}{3} P(\chi, m) N_{\res}(\chi, m) \int \dd^2 \nhat^{\prime} \,  u(\hat{n}^{\prime} - \hat{n} | \chi, m),
    \end{aligned}
\end{equation}
where $n(\chi, m)$ denotes the isotropic average halo number density per volume per halo mass,~\ie the halo mass function. Since the halo number density is isotropic, the average optical depth in Eq.~\eqref{eq:tau_sky_avg} does not depend on the direction and we can evaluate it at the north pole, $\hat{n}=\hat{z}$, to get the expected screening monopole 
\begin{align}\label{eq:monopole}
    \bar{\tau}^{\asc} \equiv \ev{\tau^{\asc}(0)} & = \int_{z_{\rm min}}^{z_{\rm max}} \dd z \frac{\chi^2}{H} \int \dd m \, n(\chi, m)  \frac{1}{3}P(\chi, m) N_{\res}(\chi, m) \int \dd^2 \hat{n} \,   u(\hat{n} | \chi, m) \nonumber \\
    & = \int_{z_{\rm min}}^{z_{\rm max}} \dd z \frac{\chi^2}{H} \int \dd m \, n(\chi, m) \tau_{00}(z, m).
\end{align}
For later convenience, we have introduced the notation
\begin{align}
    \label{eq:def_tau00}
    \tau_{00}(z, m) & \equiv \frac{\sqrt{4\pi}}{3} N_{\res}(\chi, m) P(z, m) u_{00}(z, m),  \\
    \label{eq:def_u00}
    u_{00}(z, m) & \equiv \frac{1}{\sqrt{4\pi}}\int \dd^2 \hat{n}\, u(\theta|z,m) = \sqrt{\pi}\frac{(1+z)^2r_{\res}^2}{\chi(z)^2}, 
\end{align}
for the optical depth monopole at each redshift and halo mass, and the monopole of the angular part of the conversion probability $u$, respectively. In the next sections, these will be generalized to higher multipoles. 

We choose the limits of integration to range between a minimum redshift of $z_{\rm{min}} = 0.005$ and up to a maximum redshift for which the halo magnetic field profile is available $z_{\rm max} = 1.9$, which is well below the redshift at which reionization is complete. We checked for all our observables that the lowest redshift bins give a subdominant contribution; however note that, in general, the effect from individual particularly nearby objects can be significant~\cite{Schlederer:2015jwa}, in particular, if we can resolve and model the central regions of these objects, for the upper end of the axion masses. To perform numerical computations we assume the mass-function of~\cite{tinker2008} that fixes the bias function~\cite{Tinker_2010}, and the concentration-mass relation from~\cite{Bhattacharya_2013}, which fixes the free parameters in the halo density profile. We also work under the assumption that the halo boundary is at the virial radius where the overdensity is $\approx 178$ greater than the background density, so that the halo mass is defined in a sphere of radius $r_{\rm vir}$ in units of $M_{\odot}$. We assume $100$ halo mass bins logarithmically spaced in the $10^{11} - 10^{17} {\rm \, M_{\odot}}$ interval. Our numerical halo-model computations use a modified version of the code \textit{hmvec}~\footnote{\url{https://github.com/simonsobs/hmvec}}. Finally, throughout this work we assume a flat $\Lambda$CDM cosmology, with parameters fixed by the best-fit Planck $2018$ data~\cite{Planckcollab_cosmoparams}: $\Omega_{\rm cdm} = 0.11933$, $\Omega_{\rm b} = 0.02242$, $H_0 = 67.66\ {\rm km/s/Mpc}$, $\ln(10^{10} A_{s}) = 3.047$, $n_{s} = 0.9665$ and $\tau_{\rm reio} = 0.0561$.

\begin{figure}[!]
    \centering
    \includegraphics[width=0.5\textwidth]{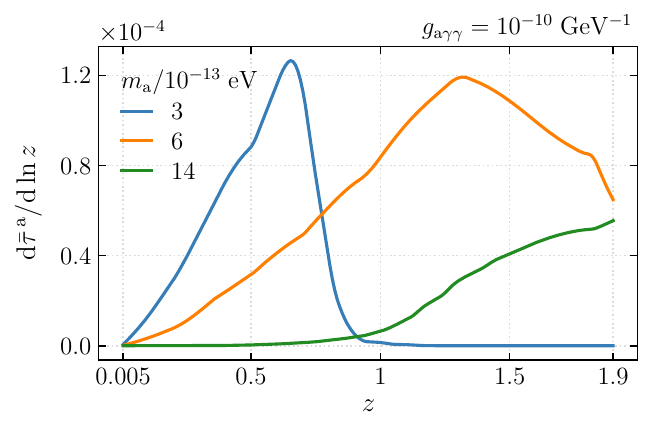}
    \caption{The axion-induced differential optical depth (see Eq.~\eqref{eq:monopole}) as a function of redshift for three choices of axion mass $m_\asc$. The coupling constant $\ga$ is fixed as labeled and we use a reference frequency $\omega/(2\pi)=145$ GHz. The latter is adopted for all the figures throughout this work. At the low-end of the axion masses accessible in our analysis, the conversion takes place in the outermost regions of low-redshift halos, where electron densities reach the lowest amplitude. For heavier axions, the density required to undergo resonant conversion increases, and a broader range of redshift becomes relevant. The non-smooth features in these curves are due to the magnetic field profiles evaluated in discrete coarse halo mass bins as provided by Ref.~\cite{2023arXiv230913104P}. We work under the assumption that there are no significant magnetic fields in the circumgalactic medium for virialized halos beyond $z>1.9$. This represents a boundary on the axion-induced screening that results in a natural cutoff at $\ma \simeq 3\times 10^{-12}$ eV for the range of axion masses accessible with this method.}\label{fig:dtdz}
\end{figure}

Fig.~\ref{fig:dtdz} shows the differential optical depth $\dd \bar{\tau}^{\asc}/\dd \ln z$, the integrand of Eq.~\eqref{eq:monopole}, as a function of redshift for three choices of axion masses. The conversion mostly happens at low redshifts for the smallest masses and at high redshift for the heavier masses, as expected following the redshift evolution of the photon plasma mass. The range of redshifts where we have a reliable model for the circumgalactic magnetic field limits the range of heavy axion masses accessible in our analysis. Therefore, our result for the integrated optical depth is conservative, as it would in principle receive additional contributions at higher redshifts.

\begin{figure}[h!]
    \centering
    \includegraphics[width=0.6\textwidth]{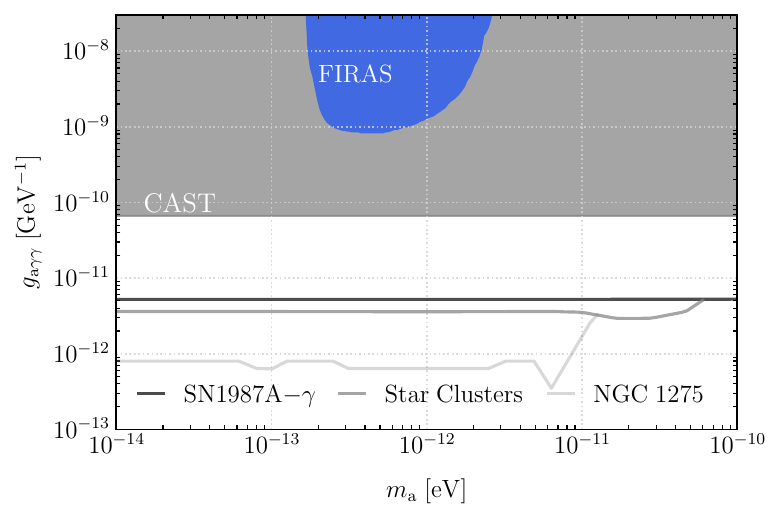}
    \caption{The region of interest in the parameter space of axion mass $m_{\rm{a}}$ and coupling to photons $\ga$. In grey we show the existing bounds reproduced from the repository~\cite{ciaran_o_hare_2020_3932430}, which include: the CAST helioscope~\cite{CAST:2017uph} (shaded gray), the non-observation of $\gamma$-rays from axions produced in the SN1987A that convert to photons in the galactic magnetic field~\cite{Hoof:2022xbe} (solid dark gray), the non-observation of X-rays from axions produced in stars that convert to photons in the galactic magnetic field~\cite{Dessert:2020lil} (solid gray), and the absence of spectral distortions in the X-ray spectra of cluster-hosted quasars due to photon-axion conversion~\cite{Reynolds:2019uqt} (solid light gray). Note that Ref.~\cite{Ning:2024eky} recently placed a limit comparable to the NCG 1275 line from the non-observation of X-rays from stellar axions produced in M82 and M87 which is not shown here. The blue shaded region represents the parameters excluded at $99\%$ confidence level from requiring that the CMB spectral distortions from photon-axion conversion inside halos is compatible with the COBE/FIRAS measurements~\cite{1996ApJ...473..576F}, using the optical depth described in Sec.~\ref{sec:skyopticaldepth} (see Eq.~\eqref{eq:monopole}) and a $\chi^2$-squared test identical to what was implemented in~\citetalias{Pirvu:2023lch}.}\label{fig:bounds_and_FIRAS}
\end{figure}

The total optical depth over the sky from Eq.~\eqref{eq:monopole} represents a spectral distortion of the blackbody CMB spectrum from photon-axion conversion inside structure in the late universe, and can be used to derive a bound on the axion-photon coupling from COBE/FIRAS~\cite{1996ApJ...473..576F}. The resulting constraint obtained from a $\chi^2$ test identical to what was implemented in ~\citetalias{Pirvu:2023lch} is shown in Fig.~\ref{fig:bounds_and_FIRAS}. Spectral distortions can only probe couplings as small as $10^{-9}\, \mathrm{GeV}^{-1}$, which are well within the excluded region from the axion helioscope CAST~\cite{CAST:2017uph}. Note that COBE/FIRAS can only exclude an optical depth of $\bar{\tau}^\asc \gtrsim 10^{-2}$ (see Fig.~\ref{fig:dtdz}); the reason for such a weak bound is that the effect from an optical depth with linear frequency scaling can be partially compensated by increasing the best fit black body temperature. 
In fact, in the high frequency tail, where the axion-induced screening effect is strongest, $\exp[-\omega/(\bar{T}+\Delta \bar{T})] \simeq \exp(-\omega/\bar{T}) (1 + \omega\ \Delta \bar{T}/\bar{T}^2)$; for example, we find that a change of the best fit CMB temperature $\bar{T}$ by a small fractional amount of $\Delta \bar{T}/\bar{T} \lesssim 10^{-4}$ is enough to wash out the axion screening spectral distortions for $\ga = 10^{-10}$ GeV. The bound is therefore coming from higher order terms and from the low frequency tail, where the effect is weaker. This constraint could be strengthened slightly if additional contributions to the spectral distortion were taken into account, such as contributions from higher redshift, the conversion in the intercluster medium (which contains a larger volume compared to structure, but also weaker magnetic fields), and the contribution to the conversion inside the Milky Way. 

However, as shown in the rest of this work, the sensitivity to the photon-axion coupling can be improved by a few orders of magnitude by considering the {\it anisotropies} induced by the photon-axion conversion inside structure. In the next section, we introduce the anisotropic axion-screened CMB temperature and polarization fields and compute the most promising observables to look for an axion signal: the two-point auto-correlation functions, the two-point temperature and LSS cross-correlation function, and the polarization and LSS three-point function.

\section{Axion-induced patchy screening}\label{sec:cmbsign}

Conventionally, the CMB temperature and polarization anisotropies are given with respect to the mean brightness temperature,~\ie the CMB blackbody temperature $\bar{T} = 2.726 \, \mathrm{K}$. The photon-axion conversion effect from Eq.~\eqref{eq:haloprob} removes CMB photons, or equivalently reduces the intensity of the blackbody spectrum, in a frequency dependent way. The corresponding axion-induced fluctuations to the temperature and polarization Stokes parameters are:
\begin{align} 
    T^{\asc}(\hat n) &= -\frac{1-e^{-x}}{x} \bar{T} \int_{z_{\rm min}}^{z_{\rm max}} \dd z  \, \frac{\dd  \tau^{\asc}(\hat n, \chi)}{\dd z} \gamma^{I}(\hat{n}, \chi)\label{eq:tfluc}, \\
    (Q\pm i U)^{\asc}(\hat n) &= -\frac{1-e^{-x}}{x} \bar{T} \int_{z_{\rm min}}^{z_{\rm max}} \dd z  \, \frac{\dd  \tau^{\asc}(\hat n, \chi)}{\dd z} \gamma^{\pm}(\hat{n}, \chi),\label{eq:pfluc}
\end{align}
where $x\equiv \omega / \bar{T}$, the multiplicative factor arises when converting from intensity to temperature units~\cite{DAmico:2015snf}, and the differential contribution to axion-induced screening is defined in Eq.~\eqref{eq:dtaudz}. In the equations above we have considered only the dominant contributions proportional to the CMB temperature monopole $\bar{T}$, neglecting the screening of CMB anisotropies; because the axion-induced screening couples to the monopole, we can hope to differentiate it from the primordial CMB anisotropies, which are about 5 orders of magnitude smaller,  despite the suppression from the small photon-axion coupling.

The fields $\gamma^{I,\pm}(\hat{n},\chi)$ in Eq.~\eqref{eq:tfluc} and~\eqref{eq:pfluc} encode the information about the magnetic field components perpendicular to the line of sight that contribute to the conversion. Explicitly,
\begin{equation}\label{eq:gammas} 
    \begin{aligned}
        \gamma^{I}(\hat{n}, \chi) & = \frac{3}{2}\frac{B_\theta(\hat{n}, \chi)^2 + B_\phi(\hat{n}, \chi)^2}{{|\mathbf{B}(\hat{n}, \chi)}|^2}, \\
        \gamma^{\pm}(\hat{n}, \chi) & = \frac{3}{2} \frac{(B_\theta(\hat{n}, \chi) \mp i B_\phi(\hat{n}, \chi))^2}{ {|\mathbf{B}(\hat{n}, \chi)}|^2},
    \end{aligned}
\end{equation}
where $B_{\theta,\phi}$ are the polar and azimuthal components along each line of sight $\hat{n}$ and ${|\mathbf{B}|}$ is total magnitude (evaluated at the location of the resonance at comoving distance $\chi$). The overall factor of $3$ is just due to the choice of normalization of $\dd\tau^{\asc}/\dd z$ to include a factor of $1/3$ (see Eq.~\eqref{eq:dtaudz}). For simplicity, we model the magnetic field as a random Gaussian field roughly constant over a domain of characteristic physical size $r_{\dom} \in [1, 10]$ kpc, which is independent of halo mass and redshift. If the magnetic fields have a random orientation in each domain, averaging over many domains gives
\begin{align}\label{eq:gammamon}
    \ev{\gamma^{I}(\hat{n}, \chi)} = 1, \quad
    \ev{\gamma^{\pm}(\hat{n}, \chi)} = 0.
\end{align}
The corresponding two-point functions are  
\begin{gather}
    \ev{\gamma^{I}(\hat{n}_1, \chi)^{\ast} \gamma^{I}(\hat{n}_2, \chi)}  \simeq \ev{\left|\gamma^{I}(\hat{n}_1)\right|^2 }  = 1,\\
    \ev{\gamma^{\pm}(\hat{n}_1, \chi)^{\ast} \gamma^{\pm}(\hat{n}_2, \chi)}  = \ev{ \left| \gamma^{\pm}(\hat{n}_1, \chi) \right|^2 } \times e^{- |\hat{n}_1-\hat{n}_2|^2/ (2\theta_\dom^2)} = \frac{9}{N_{\res}}\frac{2}{15} e^{- |\hat{n}_1-\hat{n}_2|^2/(2\theta_\dom^2)}, \label{eq:gamma_corr_pm}
\end{gather}
where $\theta_\dom(z) = r_{\dom}(1+z) / \chi(z)$. Any cross-correlation between $I, +$ or $-$ vanishes\footnote{For the temperature two-point function $\ev{\gamma^{I}(\hat{n}_1)^{\ast} \gamma^{I}(\hat{n}_2)}$ we neglect a second, sub-leading, contribution of $4/5 \, e^{- |\hat{n}_1-\hat{n}_2|^2 \theta_\dom^2 / 2}$ that is non-zero only for small angular separations $|\nhat_1-\nhat_2|$ inside the same domain, which are mostly unresolved. Depending on the domain size and the resolution of the CMB experiment, this term could contribute detectable small-scale power.}. The factor of $N_{\res}$ in the denominator of Eq.~\eqref{eq:gamma_corr_pm} is to account for the fact that, for  polarization, there is no cross-correlation between the resonant crossings going in and out of the halo (so that the polarization correlation should scale as $N_{\res}$ and not $N_{\res}^2$)\footnote{Strictly speaking, $N_{\res}$ is not the same in each halo, so it cannot be factorized outside of the sum over halos; however it is just equal to 2 in most cases and we adopt this factorization for simplicity, so as to use the same $\dd\tau^{\asc}/\dd z$ in Eqs.~\eqref{eq:tfluc}-\eqref{eq:pfluc} and keep the same notation in the computation of the temperature and polarization power spectra. Operationally, this means that we use the $\tau^{\asc}$ power spectra with the appropriate power of $N_{\res}$ in each case.}.

\begin{figure}[!]
    \centering
    \includegraphics[width=0.9\textwidth]{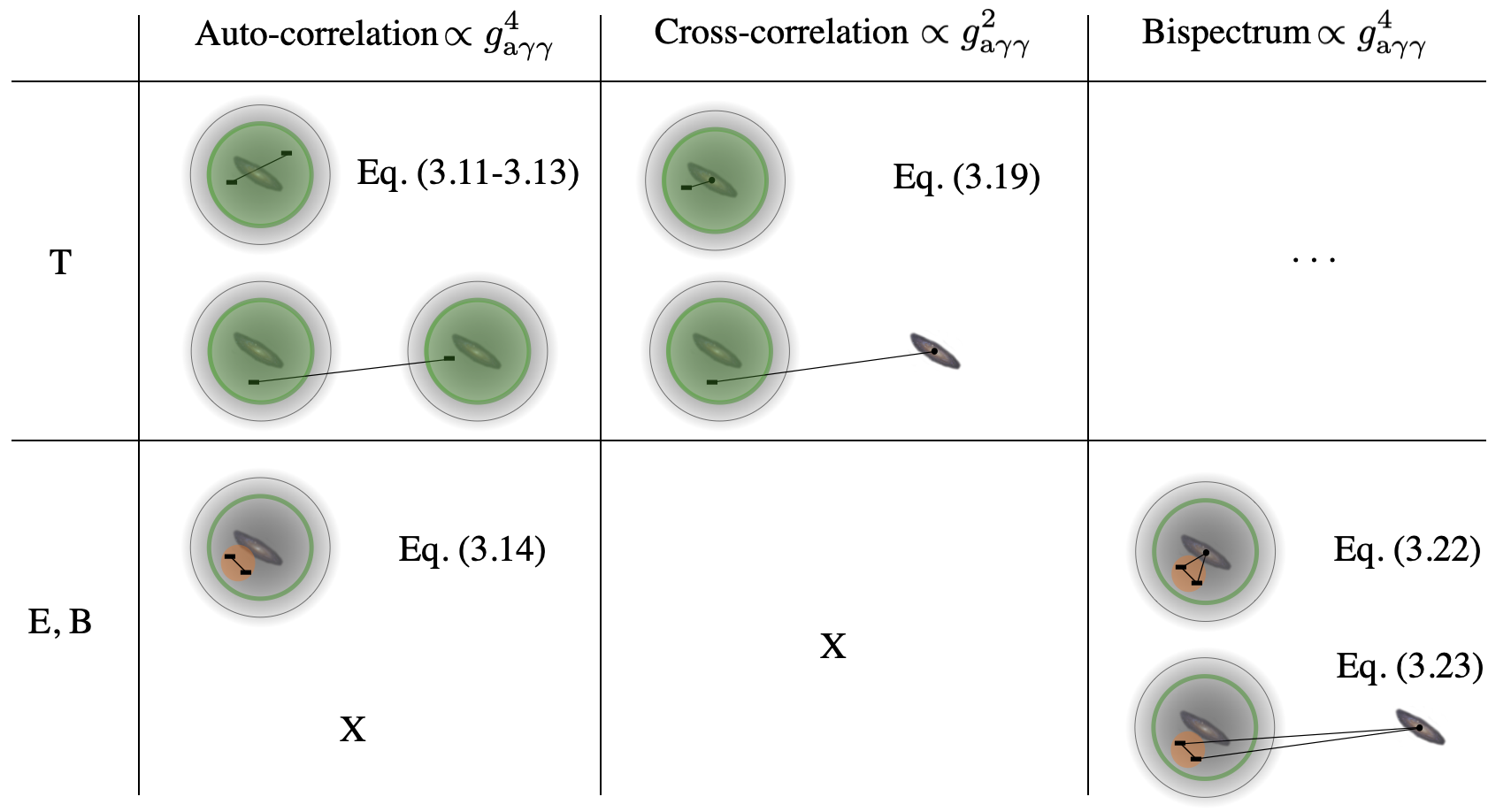}
    \caption{Schematic representation of the axion-induced screening n-point functions considered in this work. The green circle represents the distance from the halo's center at which $\ompl=\ma$; all the photon trajectories crossing the halo within the circle undergo resonant photon-axion conversion. For the temperature (top row) the leading-order observables in $\ga$ are: $\ev{T^{\asc}T^{\asc}}$ auto-correlation (left, see  Sec.~\ref{sec:cmb_auto}) and $\ev{T^{\asc} g}$ cross-correlation (center, see  Sec.~\ref{sec:cmb_cross}), both with non-vanishing 1-halo and 2-halo terms. We checked that the bispectrum $\ev{T^{\asc} T^{\asc} g}$ does not improve the sensitivity compared to the two-point functions considered. For polarization (bottom row) the leading-order observables are: $\ev{B^{\asc} B^{\asc}}$ auto-correlation (left, see Sec.~\ref{sec:cmb_auto}), with the 1-halo term only, and $\ev{B^{\asc} B^{\asc} g}$ bispectrum (right, see Sec.~\ref{sec:cmb_bispectrum}), with 1-halo and 2-halo terms (both dominated by squeezed triangles). The orange shaded region in the bottom row represents a magnetic field domain, where the polarization signal is correlated (see Eq.~\eqref{eq:gamma_corr_pm}). Note that the figures are not to scale and are given for illustrative purposes only.}\label{fig:schema}
\end{figure}

In the following subsections we present the expressions for the signal contribution to the relevant two- and three- point functions in harmonic space. These will be used to forecast the sensitivity of current and future CMB and LSS surveys to axions that couple to photons in Sec.~\ref{sec:sensitivity}. A schematic representation of the terms that enter the axion-signal n-point statistics computed here is given in Fig.~\ref{fig:schema} for illustrative purposes.

\subsection{CMB temperature and polarization auto-correlation functions}\label{sec:cmb_auto}

In the previous section we explored the effect of resonant axion-photon conversion on the temperature and polarization of the CMB at the field level. Here, we derive the two-point angular correlation functions of these fields. In analogy with~\citetalias{Pirvu:2023lch} we use a halo model approach~\cite{1974ApJ...187..425P,2000MNRAS.318..203S,2002PhR...372....1C,asgari2023halo}. Below we report power spectra in harmonic space following the usual notation for statistically isotropic correlators, with
\begin{align}\label{eq:def_cell}
\quad \ev{X_{\ell m}^* X^{\prime}_{\ell^{\prime} m^{\prime}}} = C_\ell^{XX^{\prime}}\delta_{\ell \ell^{\prime}}\delta_{m m^{\prime}},
\end{align}
where $X^{({\prime})}$ denotes a general field on the sky, $X^{({\prime})}_{\ell m}$ its corresponding spherical harmonic coefficients and $\delta_{\ell \ell^\prime}$ is a Kronecker delta. Real space correlators can be written in terms of their power spectra as
\begin{align}\label{eq:def_celnorm}
    \xi^{XX^\prime}(\nhat_1 -\nhat_2) = \sum_{\ell=0}^{\ell_{\rm max}}\sum_{m=-\ell}^{\ell} C_\ell^{XX^\prime} Y^{*}_{\ell m}(\nhat_1)Y_{\ell m}(\nhat_2) = \sum_{\ell=0}^{\ell_{\rm max}}\frac{2\ell+1}{4\pi}C_\ell^{XX^\prime} \mathcal{P}_\ell(\nhat_1 \cdot \nhat_2),
\end{align}
where $Y_{\ell m}$ are the spherical harmonic functions and $\mathcal{P}_{\ell}$ denotes the Legendre polynomial of degree $\ell$ and $\ell_{\rm max}$ is set by the angular resolution of the survey. If $X$ is a spin-2 function, analogous expressions hold for expansions in spin-2 spherical harmonics.

The first quantities to model are anisotropies in axion-induced screening $\tau^{\asc}$, specifically the  angular power spectrum, $C_{\ell}^{\tau\tau}$. This quantity carries two crucial features of the signal: the small-scale dependence on the halo profile and the large-scale clustering of structure. In the halo model, these are captured by the 1-halo and 2-halo terms contributing to the power spectrum, respectively. The full expression was derived in detail in~\citetalias{Pirvu:2023lch}. Here we simply report the result, expressing the harmonic-space screening optical depth as
\begin{align}
    \label{eq:tau_multipoles}
    \tau^{\asc}_{\ell 0}(z,m) & \equiv \sqrt{\frac{4\pi}{2\ell+1}} \frac{1}{3} N_{\res}(z, m) P(z, m) u_{\ell 0}(z,m), \\
    u_{\ell 0}(z, m) & \equiv \sqrt{\frac{2\ell+1}{4\pi}}\int \dd^2 \hat{n}\, u(\theta|z,m) \mathcal{P}_{\ell}(\cos\theta),
\end{align}
which generalizes Eqs.~\eqref{eq:def_tau00} and~\eqref{eq:def_u00} to higher multipoles. The power spectrum is given by
\begin{equation}
\begin{gathered}\label{eq:tautaumasterspectra}
    C_\ell^{\tau\tau}  = C_\ell^{\tau\tau, \, 1-{\rm halo}} + C_\ell^{\tau\tau, \, 2-{\rm halo}}, \\
	C_\ell^{\tau\tau, \, 1-{\rm halo}} = \int_{z_{\rm min}}^{z_{\rm max}} \dd z \, \frac{\chi(z)^2}{H(z)} \int \dd m \, n(z,m) \left[\tau^{\asc}_{\ell 0}(z, m)\right]^2, \\
    C_\ell^{\tau\tau, \, 2-{\rm halo}} = \int_{z_{\rm min}}^{z_{\rm max}} \dd z \, \frac{\chi(z)^2}{H(z)} \left[ \int \dd m \, n(z, m) b(z, m) \tau^{\asc}_{\ell 0}(z,m) \right]^2 P^{\rm lin}\left(\frac{\ell+\frac{1}{2}}{\chi(z)}, z\right).
\end{gathered}
\end{equation}
where $P^{\rm lin}$ is the linear matter power spectrum evaluated at comoving wavenumbers $k = \left( \ell+ \frac{1}{2}\right) / \chi $ and redshift $z$, $b(z,m)$ is the linear halo bias and $n(z,m)$ is the halo mass function. The expression given here is the result obtained after taking the Limber approximation to simplify the halo-halo power spectrum~\cite{1953ApJ...117..134L, PhysRevD.88.063526}.\footnote{We have verified numerically that, over the range of scales considered here, the Limber approximation is equivalent to the full expression given in App.~B2 of~\citetalias{Pirvu:2023lch},
\begin{equation}
\begin{gathered}\label{eq:tautau_2halo_full}
    C_\ell^{\tau\tau, \, 2-{\rm halo}} = \left[ \prod_{i=1,2} \int_{z_{\rm min}}^{z_{\rm max}} \dd z_i \, \frac{\chi(z_i)^2}{H(z_i)}  \int \dd m_i \, n(z_i, m_i) b(z_i, m_i) \tau^{\asc}_{\ell 0}(z,m_i) \right] \mathcal{C}_\ell^{\rm lin}\left(z_1, z_2\right), \\
    \mathcal{C}_\ell^{\rm lin}\left(z_1, z_2\right) = \frac{2}{\pi} \int \dd k k^2 j_\ell(k \chi_1) j_\ell(k \chi_2) \sqrt{P^{\rm lin}(k, \chi_1)P^{\rm lin}(k, \chi_2)}.
\end{gathered}
\end{equation}
Further details on how to reduce the result above to Eq.~\eqref{eq:tautaumasterspectra} are also given in App.~\ref{app:pol_bispectra} (see Eq.~\eqref{eq:Cell_linear_hh}-\eqref{eq:Cell_linear_hh_2}).
}

The axion-induced screening CMB power spectra are based on $C_{\ell}^{\tau\tau}$ and the properties of the $\gamma^{I,\pm}$ coefficients. Due to the latter, the only non-vanishing terms are the temperature and polarization auto-correlations, while any cross-correlation vanishes. 

The temperature auto-correlation function takes a simple form, given that the coefficients $\gamma^{I}$ are defined with unit variance and do not add any angular dependence. Therefore, the temperature power spectrum is simply proportional to the $\tau^\asc$ screening auto-power spectrum and reads
\begin{equation}\label{eq:cellTscTsc}
    C_{\ell}^{T^{\asc}T^{\asc}} = \left(\frac{1-e^{-x}}{x} \bar{T} \right)^2 C_{\ell}^{\tau\tau},
\end{equation}
including both 1-halo and 2-halo contributions.\footnote{The monopole of the axion-induced temperature is given by $\ev{T^{\asc}} = -(1-e^{-x})/x \, \bar{T} \, \bar{\tau}^{\asc}$.}

The polarization auto-correlation has a slightly more involved expression. Due to the finite magnetic field coherence length, the polarization signal is correlated only on small angular scales, at or below the angular size of the projected magnetic field domain -- see Eq.~\eqref{eq:gamma_corr_pm}. In recent simulations, the magnetic field energy power spectra are dominated by scales between 1 and $100$ kpc over the halo masses considered~\cite{2023arXiv230913104P}. Therefore, except for the nearest and more massive halos, the projected angular size of a magnetic field domain, $\theta_{\dom}$, is small and will not be resolved with existing and future surveys ($\ell \sim 10^4$ corresponds to a physical scale of about 100 kpc at a distance of 1 Gpc).

The scaling of the axion-induced polarization with magnetic field domain size can be understood as follows: the contribution to the signal integrated over the angular area of one halo of size $\theta_{\rm vir}$ adds up incoherently from the unresolved $N_\dom$ domains within the halo and scales as $B^\asc \propto \theta_\dom^2 \sqrt{N_\dom}$; on the effectively two dimensional conversion surface there are $N_\dom \simeq (\theta_{\rm vir}/\theta_\dom)^2$ domains; therefore, the polarization two-point function will scale as $(\theta_\dom \theta_{\rm vir})^2$. Despite the suppression from the small angle $\theta_{\dom}$, there are several handles on the polarization signal that can be leveraged for detection: it induces $B$-modes (the signal contributes equally to $E$ (curl-free) and $B$ (gradient-free) modes, but as we see below, the noise is lower for $B$-modes), it has a characteristic frequency dependence, and it is correlated with the location of galaxies. For these reasons, we find it can be competitive and complementary to the temperature observables, although it is sensitive to the unknown value of the magnetic field coherence length.

Here we report the power spectra for the $E$ and $B$ modes. These are derived in App.~\ref{app:pol_cell} from the two-point functions of the Stokes parameters $Q$ and $U$, following the standard expansion in spin-2 spherical harmonics for the polarization tensor. The final result is\footnote{Note that within this assumption of randomly oriented magnetic field in each domain, the correlation function $\ev{E^\asc B^\asc}$ vanishes.}
\begin{equation}\label{eq:cellEscEsc}
    C_\ell^{E^{\asc} E^{\asc}} = C_\ell^{B^{\asc} B^{\asc}} = \left(\frac{1-e^{-x}}{x} \bar{T}\right)^2  \int \dd z \, \frac{\chi(z)^2}{H(z)} \int \dd m \, n(z,m) \sum\limits_{\substack{L L^{\prime}}} \left(W_{\ell L^{\prime} L}^{220} \right)^2 \left[\tau^{\asc}_{L0}(z, m)\right]^2 C_{L^{\prime}}^{\rm pol}(z,m),
\end{equation}
where the factor $W_{\ell L^{\prime} L}^{220}$, defined in Eq.~\eqref{eq:w_def}, arises due to the appropriate weighting by Wigner 3j-symbols when combining the product of spherical harmonics, and $C_{\ell}^{\rm pol}$ denotes the power spectrum of the appropriate combination of $\gamma$ functions from Eq.~\eqref{eq:gammas} with two-point function from Eq.~\eqref{eq:gamma_corr_pm}, which captures the correlation length of magnetic domains. These are
\begin{equation}\label{eq:cell_pol_f_dom}
    C_{\ell}^{\rm pol}(z,m) \equiv \frac{1}{4} \left[ C_{\ell}^{+}(z,m) + C_{\ell}^{-}(z,m) \right] = \frac{9}{N_{\res}(z,m)} \frac{1}{15} \, 2\pi\theta_\dom^2(z) \exp\left[ -\ell(\ell+1) \theta_\dom^2(z) / 2 \right]
\end{equation}
where $\theta_\dom(z) = r_{\dom}(1+z) / \chi(z)$ and we fix the physical size of the magnetic field domain $r_{\dom} \in [1, 10]$ kpc. Comparing Eq.~\eqref{eq:tautaumasterspectra} with~\eqref{eq:cellEscEsc}, notice that there is no contribution from the 2-halo term for the polarization screening. Since the magnetic field direction varies randomly in different domains, the polarization signal has a non-zero correlation only for points within the same domain.

\begin{figure}[!]
    \centering
    \includegraphics[width=0.6\textwidth]{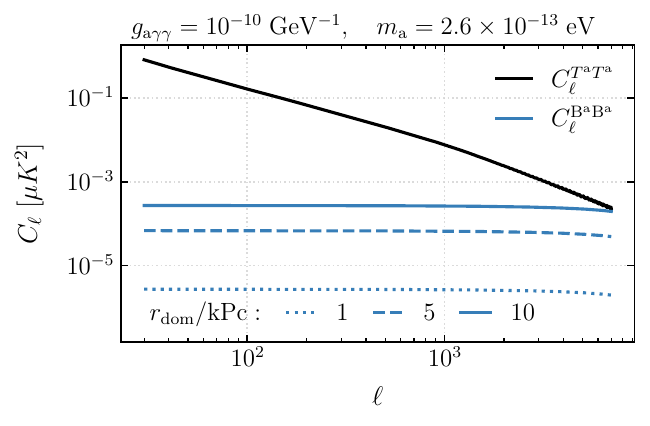}
    \caption{Axion-screening induced CMB power spectra for fixed axion parameters as labeled and reference frequency $\omega/(2\pi) = 145$ GHz. The temperature power spectrum $C_{\ell}^{T^{\asc}T^{\asc}}$ is given in Eq.~\eqref{eq:cellTscTsc} and shown by the black solid line, while the polarization power spectrum $C_{\ell}^{B^{\asc}B^{\asc}}$ is given in~\eqref{eq:cellEscEsc} and shown by the blue dotted, dashed, and solid lines, for magnetic domain size of $r_{\rm dom} = 1, 5$, and $10$ kpc, respectively. The amplitude of the power spectrum scales as $r_{\dom}^2$, as expected. The ratio between the temperature and polarization power spectra at fixed $\ell$ changes depending on the axion mass. In general, for the range of magnetic domain sizes chosen, the temperature power spectrum is stronger on all scales. This is in part due to the fact that the polarization auto-correlator does not receive inter-halo contributions proportional to the linear matter power spectrum. On small scales of $\ell>\mathcal{O}(10^3)$, where the 1-halo term dominates, the polarization signal can be competitive with the temperature, especially at low axion mass where there is no redshift-dependent suppression from $C_{\ell}^{\rm pol}$ (see Eq.~\eqref{eq:cell_pol_f_dom}).}\label{fig:cell_auto}
\end{figure}

The resulting power spectra are shown in Fig.~\ref{fig:cell_auto}, where the temperature power spectrum from Eq.~\eqref{eq:cellTscTsc} is compared to the polarization power spectra from~\eqref{eq:cellEscEsc}, for one choice of axion mass and coupling. From the flat shape of the $C_{\ell}^{E^{\asc} E^{\asc}}/ C_\ell^{B^{\asc} B^{\asc}}$, we see that the polarization signal is predominantly coming from unclustered positive and negative sources. At large-scales, where the 2-halo term dominates, the signal in $\ev{T^{\asc} T^{\asc}}$ is significantly larger than $\ev{E^{\asc}E^{\asc}}, \ev{B^{\asc}B^{\asc}}$. Notice that the relative size between the temperature and polarization power spectra will change slightly at different values of the axion mass: lighter axions correspond to conversions at smaller redshifts, when the angular size of the magnetic field domain is less suppressed for nearby halos; therefore the polarization signal is stronger (relative to $\ev{T^{\asc}T^{\asc}}$) for smaller masses and becomes more suppressed at larger masses. This will be apparent when comparing the sensitivity to photon-axion couplings from $\ev{T^{\asc}T^{\asc}}$ and $\ev{B^{\asc}B^{\asc}}$, whose relative importance will depend on the axion mass (see Sec.~\ref{sec:results}). Finally, the fact that the signal in polarization is weaker than the one in temperature at almost all scales, does not mean that the polarization does not contribute to the overall signal sensitivity. In fact, $\ev{B^{\asc}B^{\asc}}$ gives a much cleaner channel to look for non-SM effects, due to the small amplitude of lensing $B$-modes, making it possible to distinguish a smaller signal compared to the $\ev{T^{\asc}T^{\asc}}$ channel.

\subsection{CMB temperature-LSS cross-correlation}\label{sec:cmb_cross}

Since axion-induced screening occurs inside LSS, the cross-correlation of the CMB with a tracer of LSS will be more sensitive than the CMB auto-correlations discussed in the previous section. On one hand, the cross-correlation contains only one power of the small coupling squared $\ga^2$ - compared to the double insertion in the auto-correlation functions - which translates into a more favourable scaling of the sensitivity with the highest accessible multipole $\ell_{\rm max}$, as can be seen from the signal-to-noise ratio described below in Sec.~\ref{sec:results} (see also App.~\ref{app:estimate} for a qualitative order-of-magnitude comparison of the sensitivity of different observables). Additionally, having a template based on the distribution of LSS helps in the detection of a weak signal. Finally, cross-correlation minimizes the impact of uncorrelated foregrounds and systematics that contribute strongly to the individual auto-spectra, increasing the sensitivity of the measurement. The simple cross-correlation between the axion-induced polarization signal and LSS vanishes due to the random orientation of the magnetic field in different domains (there are, however, non-vanishing higher-point functions that will be described in Sec.~\ref{sec:cmb_bispectrum}). Therefore, here we focus on the two-point correlation function between the axion-induced temperature signal and LSS,
\begin{equation}\label{eq:Tg_main}
    \ev{T^{\asc}(\hat{n}_1)g(\hat{n}_2)},
\end{equation}
where $T^{\asc}(\nhat)$ is defined in Eq.~\eqref{eq:tfluc} and $g(\hat{n})$ represents the projected galaxy overdensity field. Since the axion-induced screening signal in the CMB is projected along the line-of-sight and receives contribution from a wide range of redshifts, we do not require precise redshift measurements. The ideal tracer has a high number-density, to leverage both the clustering signal on large angular scales and the small-scale structure within halos. In the following, we adopt the unWISE galaxy sample~\cite{Schlafly_2019,Krolewski_2020} as our fiducial tracer. We focus on the blue sample, which contains $\sim 50$ million objects over roughly $60\%$ of the sky with a well-characterized redshift distribution. The galaxy field template $g(\nhat)$ is defined as an overdensity in the counts of a galaxy survey, weighted by the fractional number of galaxies in the sample per redshift bin and integrated over the line-of-sight~\cite{10.1093/mnras/stz3351, 10.1093/mnras/sts006}:
\begin{equation}\label{eq:galaxy_field}
    g(\nhat) = \int \dd z\, \frac{\dd N_g}{\dd z}\frac{n_g(z, \nhat) -\bar{n}_g(z)}{\bar{n}_g(z)},
\end{equation} 
where $\bar{n}_g$ is the mean number density of galaxies per redshift bin and $\dd N_g/\dd z$ is the galaxy redshift distribution normalized so that $\int \dd z \frac{\dd N_g}{\dd z} = 1$. 
To model how the observed galaxies populate the underlying dark matter halo distribution, we use the Halo Occupation Distribution (HOD)~\cite{2000MNRAS.318.1144P} as described in Ref.~\cite{unwiseHOD_newer}. The full details of the HOD can be found in App.~\ref{app:gal_powerspectra}; here we report only the resulting galaxy field auto power-spectrum and the cross-correlation with the axion-induced signal.

Within this framework, the unWISE galaxy sample is modeled by a population of galaxies at the centre of their dark matter halo, `centrals', and a population of `satellite' galaxies distributed according to the dark matter density profile in each halo. Following Ref.~\cite{unwiseHOD}, the result is 
\begin{equation}\label{eq:celgg_full_unwise}
\begin{gathered}
    C_{\ell}^{g g} = C_{\ell}^{g g, \, 1-{\rm halo}} + C_{\ell}^{g g, \, 2-{\rm halo}} + A_{\rm SN}, \\
	C_{\ell}^{g g, \, 1-{\rm halo}} = \int \dd z \frac{\chi(z)^2}{H(z)}  \int  \dd m \, n(z,m)  \ev{ \left|u_{\ell}^g(z,m)\right|^2 }, \\
	C_{\ell}^{g g, \, 2-{\rm halo}} = \int \dd z \frac{\chi(z)^2}{H(z)} \left[ \int \dd m \, n(z,m) b(z,m) u_{\ell}^g(z, m) \right]^2 P^{\rm lin}\left(\frac{\ell+\frac{1}{2}}{\chi(z)}, z\right),
\end{gathered}
\end{equation}
where $u_{\ell}^g(z, m)$ describes the mean distribution of galaxies and their distribution inside halos, $\ev{\left|u_{\ell}^g(z,m)\right|^2}$ is the second moment of the distribution, and $A_{\rm SN}$ denotes the shot noise contribution and is an empirically determined parameter of the model. These functions depend on the details of the HOD, and are defined in App.~\ref{app:gal_powerspectra}.

The cross-power between the axion-induced screening temperature anisotropies and unWISE galaxies is a straightforward generalization of the galaxy-galaxy power spectrum. Noting that $T^{\asc}_{\ell}(z,m) \propto \tau^{\asc}_{\ell} (z,m)$, the cross-power is 
\begin{equation}\label{eq:galaxy_temp_cell}
\begin{gathered}
    C_\ell^{T^{\asc}g} = \frac{1-e^{-x}}{x} \bar{T}  \left[ C_\ell^{g \tau, \, 1-{\rm halo}} + C_\ell^{g \tau, \, 2-{\rm halo}} \right], \\
    C_\ell^{g \tau, \, 1-{\rm halo}} = \int \dd z \frac{\chi(z)^2}{H(z)} \int \dd m \, n(z,m) \tau^{\asc}_{\ell}(z,m) u_{\ell}^g (z,m), \\
    C_\ell^{g \tau, \, 2-{\rm halo}} = \int \dd z \frac{\chi(z)^2}{H(z)} \left[ \prod_{i=1,2} \int \dd m_i \, n(z, m_i) b(z, m_i) \right] \tau^{\asc}_{\ell} (z,m_1) u_{\ell}^g (z,m_2)  P^{\rm lin}\left(\frac{\ell+\frac{1}{2}}{\chi(z)}, z\right).
\end{gathered}
\end{equation}
The cross-power depends on the redshift overlap in the axion-induced optical depth and the distribution of unWISE galaxies described by $u_{\ell}^g (z,m)$. As discussed in App.~\ref{app:gal_powerspectra}, $u_{\ell}^g (z,m)$ is proportional to the redshift distribution of galaxies in the unWISE blue sample, which spans the range $0.2 \lesssim z \lesssim 0.8$. Comparing with Fig.~\ref{fig:dtdz}, the overlap is greatest at low axion mass $m_{\asc}$. The strength of the auto- and cross-correlations depends on the statistical power of the halo model: when the distribution of structure is known to high precision, the signal-to-noise of the estimator is enhanced. In principle, the best sensitivity is achieved when the location of all halos and distribution of galaxies therein is known.

\begin{figure}[!]
    \centering
    \includegraphics[width=1\textwidth]{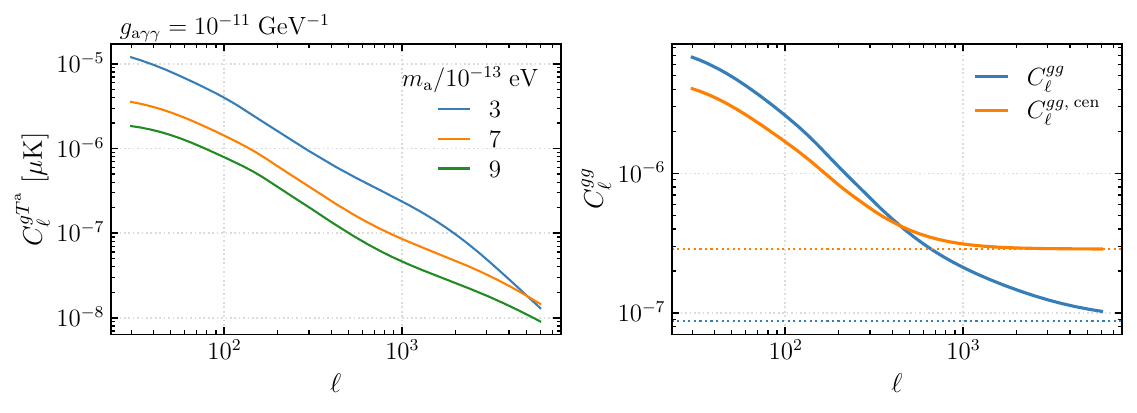}
    \caption{({\it left}) Cross-correlation of the axion-induced CMB temperature screening with a tracer of LSS for fixed photon-axion coupling as labeled and reference frequency $\omega/(2\pi) = 145$ GHz. The signal angular cross-power spectrum is defined in Eq.~\eqref{eq:galaxy_temp_cell} and shown by the blue, orange, and green solid lines for $\ma = 3, 7,$ and $9 \times 10^{-13}$ eV, respectively. On large scales, the main difference in the three spectra is the amplitude, while the spectral shape is dictated by the proportionality with the linear matter power spectrum $P_\ell^{\rm lin}$. On small scales, the spectra hold some information about the conversion radius $r_{\rm res}$ characteristic for each axion mass. ({\it right}) Galaxy-galaxy and centrals-only power-spectra $C_\ell^{gg}$ (blue solid line) and $C_\ell^{gg, {\rm \, cen}}$ (orange solid line), defined in Eq.~\eqref{eq:celgg_full_unwise} and Eq.~\eqref{eq:celgg_centralsonly_unwise} respectively. On large-scales, the galaxy auto-spectra trace the linear matter power spectrum, while at small scales they asymptote to the shot noise terms $A_{\rm SN}$ and $A^{\rm cen}_{\rm SN}$ (also shown as dotted lines). The HOD parameters used to model the properties of the galaxies are the best-fit values from the unWISE blue catalogue obtained in Ref.~\cite{unwiseHOD_newer} (see  App.~\ref{app:gal_powerspectra}).
    }\label{fig:cell_Tg}
\end{figure}

An example of the axion-induced cross-correlation power spectra $C_\ell^{T^\asc g}$ at fixed coupling constant $\ga$ is shown in the left panel of Fig.~\ref{fig:cell_Tg} for three different values of the axion mass $m_{\asc}$. For $\ell < 1000$, the curves follow the linear matter power spectrum $P_\ell^{\rm lin}$ (through the 2-halo term) and have similar shapes, but different amplitudes. On smaller angular scales, where the 1-halo term dominates, the cross-spectra hold information about the resonance scale characteristic of each mass. The right panel of Fig.~\ref{fig:cell_Tg} shows the galaxy auto-spectra corresponding to the full unWISE sample $C_{\ell}^{gg}$ defined in Eq.~\eqref{eq:celgg_full_unwise}, as well as for central galaxies only $C_{\ell}^{g g, {\rm cen}}$ defined in Eq.~\eqref{eq:celgg_centralsonly_unwise}.
As explained in the next section, when deriving expressions for the axion-induced bispectrum, we use a centrals-only galaxy template to simplify the calculation. This simplification should be a conservative choice, which only mildly affects the forecasted sensitivity to the axion coupling strength. Notice that on large-scales both curves trace the matter power spectrum, but with a slightly larger amplitude for the full galaxy power spectrum; the reason for this discrepancy is that both the number of satellite galaxies and the bias function are larger for heavier halos, while the number of centrals is constant above a certain mass threshold; therefore, removing satellites removes a preferential weighting of large-bias halos, in turn reducing the amplitude of the 2-halo term (cf. definitions Eq.~\eqref{eq:celgg_full_unwise} and Eq.~\eqref{eq:celgg_centralsonly_unwise}). At small scales, on the other hand, $C_{\ell}^{g g, {\rm cen}}$ becomes larger due to a larger shot noise term in a sample with fewer galaxies, $A^{\rm cen}_{\rm SN} > A_{\rm SN}$.

\subsection{CMB polarization and LSS bispectrum}\label{sec:cmb_bispectrum}
 
Similarly to the CMB-LSS cross-correlation observable constructed in the previous section for the axion-induced temperature map, it is possible to leverage the fact that the polarization signal originates in structure. In this case, however, the leading-order non-vanishing observable is the three-point function containing two polarization fields and one galaxy field. The simple cross-correlation vanishes because it has only one insertion of the polarization field. Therefore we are interesting in obtaining the signal contribution to 
\begin{eqnarray}\label{eq:QQg_main}
    \ev{(Q \pm i U)^{\asc}(\hat{n}_1)(Q \pm i U)^{\asc}(\hat{n}_2) g(\hat{n}_3)},
\end{eqnarray}
where $(Q \pm i U)^{\asc}(\nhat)$ is defined in Eq.~\eqref{eq:pfluc} and $g(\hat{n})$ represents the projected galaxy overdensity field as defined in Eq.~\eqref{eq:galaxy_field}. To simplify the calculation of the bispectrum, we only include galaxies at the center of each halo, neglecting satellite galaxies. The HOD modeling with centrals-only is outlined in App.~\ref{app:gal_powerspectra} and the corresponding auto-power spectrum is given in Eq.~\eqref{eq:celgg_centralsonly_unwise} and shown in the right panel of Fig.~\ref{fig:cell_Tg}.

Since the polarization signal is non-zero only for small angular separations that fall within the same magnetic field domain (within the same halo), the three point function above will receive a non-vanishing contribution only for $|\nhat_1-\nhat_2| \lesssim \theta_\dom(\chi)$. There are then two contributions to Eq.~\eqref{eq:QQg_main}: a 1-halo term, where $\nhat_3$ is taken to be at the center of a halo while $\nhat_1$ and $\nhat_2$ are two points in the same halo, and a 2-halo term, where $\nhat_3$ is taken to be at the center of a different halo than $\nhat_1$ and $\nhat_2$. Since the magnetic domain size is typically much smaller than both the characteristic resonance conversion radius and the separation between two halos, both 1-halo and 2-halo terms are dominated by squeezed  configurations of the bispectrum. Similar to the case of the auto-correlation functions, the 1-halo term will be sensitive to the small-scale shape of the halos, while the 2-halo term holds information about the modulation of small-scale power by the density field as traced by LSS. We find the two terms are comparable, but the 2-halo terms slightly dominates across the full range of axion masses considered.

The full derivation is presented in App.~\ref{app:pol_bispectra} and here we report the final expressions for the angle-averaged bispectrum, which is the quantity appearing in the signal-to-noise ratio used for the sensitivity forecast. In terms of the $B$-mode (the same expression applies to the $E$-mode), the result is
\begin{equation}\label{eq:red_bispectrum_B}
\begin{aligned}
    \mathcal{B}^{B^\asc B^\asc g}_{\ell \ell^{\prime} \ell^{\prime\prime}} &= \mathcal{B}^{B^\asc B^\asc g, \, 1-{\rm halo}}_{\ell \ell^{\prime} \ell^{\prime\prime}} + \mathcal{B}^{B^\asc B^\asc g, \, 2-{\rm halo}}_{\ell \ell^{\prime} \ell^{\prime\prime}}, \\  
\end{aligned}
\end{equation}
\begin{equation}\label{eq:red_bispectrum_B_1halo}
\begin{aligned}
    \mathcal{B}^{B^\asc B^\asc g, \, 1-{\rm halo}}_{\ell \ell^{\prime} \ell^{\prime\prime}} &= \sqrt{\frac{(2 \ell + 1)(2 \ell^{\prime} + 1)(2 \ell^{\prime\prime} + 1)}{4\pi}} 
    \begin{pmatrix}
			\ell & \ell^{\prime}  & \ell^{\prime\prime} \\
			+ 2 & - 2 & 0 
    \end{pmatrix} e_{\ell \ell^{\prime} \ell^{\prime\prime}} \times \\
	 &\left(\frac{1-e^{-x}}{x} \bar{T}\right)^2 \int \dd z \, \frac{\chi(z)^2}{H(z)} \int \dd m \, n(\chi,m) u^{g, {\rm cen}}(\chi,m) \times \\
    &\sum_{L L^{\prime}}  \left(W_{\ell^{\prime\prime} L L^{\prime}}^{000}\right)^2  \tau^\asc_{L0}(\chi, m)\tau^\asc_{L^{\prime}0}(\chi, m) \frac{\mathcal{C}^{\rm pol}_{\ell}(\chi) + \mathcal{C}^{\rm pol}_{\ell^{\prime}}(\chi)}{2}, \\
\end{aligned}
\end{equation}
\begin{equation}\label{eq:red_bispectrum_B_2halo}
\begin{aligned}
    \mathcal{B}^{B^\asc B^\asc g, \, 2-{\rm halo}}_{\ell \ell^{\prime} \ell^{\prime\prime}}  &= \sqrt{\frac{(2 \ell + 1)(2 \ell^{\prime} + 1)(2 \ell^{\prime\prime} + 1)}{4\pi}} 
    \begin{pmatrix}
			\ell & \ell^{\prime}  & \ell^{\prime\prime} \\
			+ 2 & - 2 & 0 
	\end{pmatrix} e_{\ell \ell^{\prime} \ell^{\prime\prime}} \times  \\
    &\left(\frac{1-e^{-x}}{x} \bar{T}\right)^2 \int \dd z \, \frac{\chi(z)^2}{H(z)} \left[ \prod_{i=1,2} \int \dd m_i \, n(z, m_i) b(z, m_i) \right] u^{g, \, {\rm cen}}(\chi, m_2) \times  \\
    &\sum_{L L^{\prime}} \frac{\left(W_{\ell L^{\prime} L}^{220}\right)^2 + \left(W_{\ell^{\prime} L^{\prime} L}^{220}\right)^2}{2} \left[\tau^\asc_{L0}(\chi, m_1)\right]^2 \,  \mathcal{C}^{\rm{pol}}_{L^{\prime}}(\chi) \, P^{\rm lin} \left(\frac{\ell^{\prime\prime}+\frac{1}{2}}{\chi(z)}, z \right),
\end{aligned}
\end{equation}
where $W_{\ell L^{\prime} L}^{220}$ is defined in Eq.~\eqref{eq:w_def} and $e_{\ell \ell^{\prime} \ell^{\prime\prime}}$ in Eq.~\eqref{eq:e_def} and all the other quantities have been introduced in Sec.~\ref{sec:cmb_auto}-~\ref{sec:cmb_cross}. Since we have included central galaxies only in our LSS tracer for the bispectrum calculation, our prediction of the signal is conservative, and additional contributions are expected from satellites galaxies in the sample. The above bispectrum contribution to the signal have been derived in the limit of the squeezed triangles $\ell, \ell^\prime \gg \ell^{\prime\prime}$ and are symmetric under the exchange of $\ell \leftrightarrow \ell'$, as expected.

The axion-induced n-point statistics derived above will be used in the next section to estimate the sensitivity of Planck and CMB-S4 measurements, together with the unWISE galaxy catalogue, to photon-axion couplings.

\section{Sensitivity forecasts}\label{sec:sensitivity}

In this section, we forecast the ability of existing and near-term datasets to detect the axion-induced screening signal. We first assess how measurements of the CMB at multiple frequencies can be used to isolate the spectral dependence of the axion-induced screening signal from foregrounds and the primary CMB. We then determine the sensitivity of the correlation functions described in the previous section to $\ga$ as a function of $\ma$ for existing CMB data from Planck and the unWISE blue galaxy distribution, as well as for a future CMB experiment -- CMB-S4~\cite{abazajian_cmb-s4_2016}.

\subsection{Isolating the axion-induced screening signal using CMB component separation}\label{sec:ilc}

The multi-frequency information available in a CMB experiment can be used to enhance the axion-induced screening signal relative to the primary CMB and astrophysical foregrounds. Similar to the approach taken in~\citetalias{Pirvu:2023lch}, we estimate the ability of the harmonic internal linear combination (ILC) method~\cite{ILCpaper} to isolate this signal. The ILC is a weighted sum of individual frequency maps in harmonic space, with weights chosen to minimize the variance of a signal with known frequency dependence. The ability of this method to isolate the axion-induced screening signal is limited by the available frequency channels, instrumental noise, and the spectral dependence/amplitude of foregrounds. Inevitably, there will be some residual with which the signal must compete. We estimate the residual contribution to the ILC for two CMB experiments: the completed Planck satellite mission and CMB-S4. The assumed resolution of each frequency channel, quantified by the width of a Gaussian beam, for each experiment is recorded in Table~\ref{tab:noisemodel} of App.~\ref{app:foregroundandnoise}. 

The input to the ILC is the full $N_{\rm freq} \times N_{\rm freq}$ covariance matrix between the power spectra measured in $N_{\rm freq}$ channels:
\begin{equation}\label{eq:covmatILC}
    \mathbf{C}_{\ell} = \mathbf{\Omega}^{-1} \mathbf{C}^{TT}_{\ell} + \mathbf{e}\mathbf{e}^{\dagger} \zeta(\omega_0)^{2} C_\ell^{T^{\asc} T^{\asc}}(\mathbf{\omega}=1) + \mathbf{\Omega}^{-1} \left( \mathbf{N}^{TT}_{\ell}(\mathbf{\omega})/\mathbf{G}_{\ell}(\mathbf{\omega})\right),
\end{equation}
where $\mathbf{C}^{TT}_{\ell}$ is the primary blackbody CMB angular power spectrum,$\mathbf{C}^{T^{\asc} T^{\asc}}_{\ell}$ was defined in~\eqref{eq:cellTscTsc}, $\mathbf{N}_\ell^{TT}$ is the overall noise covariance (defined to include instrumental noise and astrophysical foregrounds) and $\mathbf{G}_{\ell}$ the beam model. We also used the notation $\zeta(\omega) \equiv \left(1-e^{-x(\omega)} \right) \omega / x(\omega)$ for $x \equiv \omega / \bar{T}$. The second term in Eq.~\eqref{eq:covmatILC} represents the axion-induced screening contribution in temperature. This term can be neglected in the small-signal limit we are interested in when computing the ILC. Equivalent expressions can be written for the $E$ and $B$-mode polarization spectra. We also defined the matrix $\mathbf{\Omega^{-1}}$ with entries $\Omega^{-1}_{ij} = \zeta(\omega_0)^{2}/ \left( \zeta(\omega_i) \zeta(\omega_j) \right)$, $\mathbf{e} = \left(1,1, \dots, 1\right)$ which characterize the frequency dependence of the axion-induced screening signal. The ILC method identifies the linear combination of harmonic coefficients that minimizes the variance of a map with the target frequency dependence. This residual is given by
\begin{equation}\label{eq:ILCnoise}
    \tilde{N}_{\ell}^{{T^{\asc} T^{\asc}}} = \mathbf{w}_{\ell}^{\dagger} \cdot \left( \mathbf{\Omega}^{-1} \mathbf{C}^{TT}_{\ell} + \mathbf{\Omega}^{-1} \mathbf{N}^{TT}_{\ell}/\mathbf{G}_{\ell} \right) \cdot \mathbf{w}_{\ell},
\end{equation}
where the weights $\mathbf{w}_\ell$ satisfy
\begin{equation}\label{eq:weightsss}
    \mathbf{w}_\ell = \frac{\left( \mathbf{C}_{\ell} \right)^{-1} \mathbf{e}}{\mathbf{e}^{\dagger} \left( \mathbf{C}_{\ell} \right)^{-1} \mathbf{e}}.
\end{equation}
To evaluate the residual Eq.~\eqref{eq:ILCnoise} we need the lensed primary CMB contribution $\mathbf{C}^{TT}_{\ell}$, which is independent of experiment and computed using CAMB~\cite{Lewis_2000} with cosmological parameters described in Sec.~\ref{sec:cmb_auto}. Additionally, we need a model for the noise covariance $\mathbf{N}_\ell^{TT}$. We refer the reader to App.~\ref{app:foregroundandnoise} for a complete description of our methodology. For Planck, we estimate $\mathbf{N}_\ell^{TT}$ from publicly available CMB-subtracted maps from the Public Data Release 3~\cite{Planck:2018nkj} (PR3) at 30, 44, 70, 100, 143, 217, and 353 GHz (we do not include the strongly foreground-dominated 545 and 857 GHz channels in our analysis; we confirmed that including them does not change our results). The entries of $\mathbf{N}_\ell^{TT}$ are simply the auto and cross-power spectra of these maps after applying a galactic cut retaining $40 \%$ of the sky. An analogous analysis is performed for polarization. For CMB-S4, the low-$\ell$ spectra are expected to be the same as those measured by Planck. We fit the low-$\ell$ CMB-subtracted spectra from Planck to a power law. We add this component to a noise power spectrum incorporating atmospheric effects, with parameters appropriate for CMB-S4 (see Table~\ref{tab:noisemodel}). On small-angular scales and at high frequencies, the cosmic infrared background (CIB) is expected to be the dominant foreground. We use simulated CIB maps from the Websky suite of simulations~\cite{Stein_2020} to compute the contributions to $\mathbf{N}_\ell^{TT}$ at high-$\ell$. For polarization, we include the Planck low-$\ell$ power-law spectra and instrumental noise only. For both Planck and CMB-S4, we chose a baseline frequency $\omega/(2\pi)=145$ GHz, and assume $40 \%$ sky coverage in the forecasts below, together with a maximum resolution of $\ell_{\rm max} = 3000$ for Planck and $\ell_{\rm max} = 6000$ for S4.

\begin{figure}[!]
    \centering
    \includegraphics[width=1\textwidth]{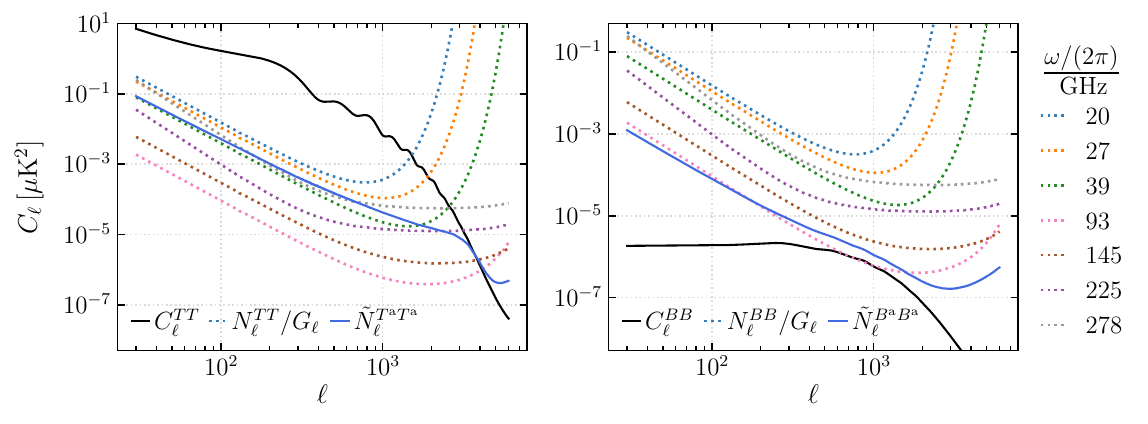}
    \caption{Comparison between lensed blackbody CMB and noise model before and after implementing the ILC for CMB-S4 temperature (left panel) and polarization (right panel). Without an ILC procedure, the axion-induced screening signal competes against the sum of the measured CMB $C_{\ell}^{XX}$ (solid black) and our estimate for the noise model in each frequency channel $N_{\ell}^{XX}/G_\ell$ (coloured dotted), where $X \in \{T,E,B \} $. The residual noise post-ILC is $\tilde{N}_{\ell}^{X^{\asc}X^{\asc}}$ (solid blue). Note that its amplitude is proportional to the $\Omega^{-1}$ matrix defined in the text. In the temperature case shown on the left, at our chosen baseline frequency $\omega/(2\pi) = 145 $ GHz, the ILC removes three orders of magnitude in the total noise amplitude (cf. the sum of the dotted lines with the black line versus the blue line). For polarization, the ILC simply minimizes the total noise so that it follows the optimal (\ie least noisy) channel at any given scale.}\label{fig:ILC_improvement}
\end{figure}

\begin{figure}[!]
    \centering
    \includegraphics[width=.9\textwidth]{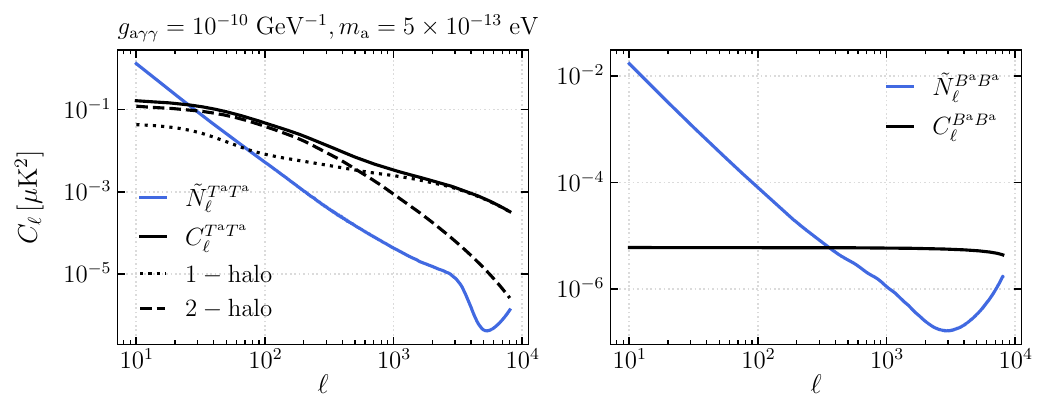}
    \caption{Axion-induced auto-correlation functions for temperature (black lines, left) and polarization (black line, right) compared to the corresponding ILC residual noise (blue lines). The signal amplitude is calculated for fixed axion parameters as labeled; both signal and noise are evaluated at the reference frequency $\omega/(2\pi) = 145 $ GHz. The noise residuals are given in Eq.~\eqref{eq:covmatILC} and correspond to CMB-S4 observations. The signal is clearly detectable in both channels for this choice of parameters and can be rescaled as $\ga^4$ to infer the magnitude at different couplings. For temperature, we show the total $C_{\ell}^{T^{\asc}T^{\asc}}$ (solid black) as well as the 1-halo (dotted black) and 2-halo (dashed black) terms individually -- see Eqs.~\eqref{eq:tautaumasterspectra} and~\eqref{eq:cellTscTsc}. The signal-to-noise ratio is driven by small angular scales (high-$\ell$ modes), where the 1-halo term dominates. For polarization $B$-modes, the signal contains only the 1-halo term -- see Eq.~\eqref{eq:cellEscEsc} -- and is above the noise level on small angular scales.}\label{fig:ILC_signal_comparison}
\end{figure}

For both experiments, the frequencies in the $\sim 100 - 200$ GHz range are weighted most strongly in the ILC, as they have the lowest overall noise, and they are favoured by the increase of the axion signal with frequency. The highest frequencies,~\ie $353$ GHz for Planck and $278$ GHz for S$4$, suffer from large foreground contamination (mainly CIB) and are penalized. Together with frequencies below $\sim 70$ GHz, these channels do not contribute significantly to the sensitivity. Fig.~\ref{fig:ILC_improvement} illustrates the total noise level for the ILC used in our CMB-S4 forecast. As mentioned, the scaling of the axion-induced screening signal with frequency causes the ILC residuals  $\tilde{N}_\ell^{T^{\asc}T^{\asc}}$ and $\tilde{N}_\ell^{B^{\asc}B^{\asc}}$ to trace the least noisy and highest-frequency channels.

In Fig.~\ref{fig:ILC_signal_comparison} we compare the ILC residual noise in temperature and polarization $B$-mode with a sample curve for each auto-correlation axion signal $C_{\ell}^{T^{\asc} T^{\asc}}$ and $C_{\ell}^{B^{\asc} B^{\asc}}$ respectively, at fixed coupling and axion mass. The ratio between the axion-induced dark screening signal and residual noise is most favourable on small angular scales, $\ell > 1000$, where most of the sensitivity of a CMB experiment will come from.

Beyond the noise modeling considered above, a real analysis would have to contend with several complexities. One must estimate the full-sky power spectrum in the presence of mode-coupling induced by masking galactic and extragalactic foregrounds, e.g. using a PCL estimator. In addition, it is possible to explicitly deproject extragalactic foregrounds, the most worrisome being CIB, from the ILC to avoid biases in the cross-correlation. These and other systematic effects have been addressed in a very similar analysis of dark photon dark screening~\cite{McCarthy:2024ozh}, and we leave a detailed analysis for the case of axion dark screening to a forthcoming paper~\cite{Goldstein:2024mfp}.

\subsection{Results}\label{sec:results}

To determine the sensitivity of the two- and three-point correlation functions described above to the coupling constant $\ga$ as a function of $\ma$, we assume that the measurements follow a Gaussian likelihood and are compatible with the hypothesis of no axion-signal,~\ie $\ga = 0$. For a given axion mass, all the observables considered in this work have a simple dependence on the coupling $\propto \ga^n$, with $n =2$ or $4$. The likelihood in this case takes the form $\ev{\log \mathcal{L}} \propto \ga^{2n}$. Following a Bayesian approach and adopting a flat prior for $\ga \geq 0 $, we derive the posterior over $\ga$ in App.~\ref{app:likelihood}. To define the sensitivity of a measurement, we compute the value of $\ga$ containing 
 $68\%$ of the posterior probability (for the Gaussian case, equivalent to a 1-$\sigma$ bound). The best sensitivity would be given by summing over all observables that contain an axion signal. However, to understand which one is the most sensitive, here we consider them separately. 

The results for CMB auto-correlation functions $C^{X^{\asc} X^{\asc}}_{\ell}$, with $X\in\{T,E,B\}$ -- see Eq.~\eqref{eq:cellTscTsc} and~\eqref{eq:cellEscEsc}, CMB temperature-LSS cross-correlation function $C^{T^{\asc} g}_\ell$ -- see Eq.~\eqref{eq:Tg_main}, and CMB polarization-LSS bispectrum $\mathcal{B}^{B^{\asc} B^{\asc} g}$ -- see Eq.~\eqref{eq:red_bispectrum_B}, are
\begin{align} 
    \label{eq:sigma_XX}
    \sigma_{\ga} & \simeq 0.7 \left(\sigma_{4, XX}\right)^{1/4}, & \sigma_{4, XX}^2 & = \left\lbrace f_{\rm sky} \sum_\ell \frac{2 \ell + 1}{2} \left[ \frac{C_{\ell}^{X^{\asc} X^{\asc}}(\ga = 1)}{\tilde{N}_{\ell}^{X^{\asc} X^{\asc}}} \right]^2\right\rbrace^{-1}, \\
    \label{eq:sigma_Tg}
    \sigma_{\ga} & \simeq 0.76 \left(\sigma_{2, Tg}\right)^{1/2}, & \sigma_{2, Tg}^2  & = \left\lbrace  f_{\rm sky} \sum_\ell  \left(2 \ell + 1\right) \frac{\left[ C^{T^{\asc} g}_\ell (\ga = 1)\right]^2 }{C_{\ell}^{gg}  \tilde{N}^{T^{\asc} T^{\asc}}_\ell} \right\rbrace^{-1}, \\
     \label{eq:sigma_BBg}
    \sigma_{\ga} &\simeq 0.7 \left(\sigma_{4, BBg}\right)^{1/4}, & \sigma_{4, BBg}^2 & = \left\lbrace  f_{\rm sky} \sum_{\ell \ell^{\prime}\ell^{\prime\prime}}   \frac{1}{2} \frac{\left[\mathcal{B}^{B^{\asc} B^{\asc} g}_{\ell \ell^\prime \ell^{\prime\prime}} (\ga = 1)\right]^2 }{\tilde{N}^{B^{\asc} B^{\asc}}_{\ell} \tilde{N}^{B^{\asc} B^{\asc}}_{\ell^\prime} C^{gg, {\rm \, cen}}_{\ell^{\prime\prime}}}\right\rbrace^{-1}.
\end{align}
where the derivation of the expressions for $\sigma_{\ga}$ is presented in App.~\ref{app:likelihood}. In the equations above, $f_{\rm sky}$ represents the fraction of the sky covered and is fixed to $0.4$ throughout. The CMB noise terms $\tilde{N}^{X^{\asc} X^{\asc}}$ are the corresponding ILC-minimized noise levels defined in Eq.~\eqref{eq:ILCnoise}, while the LSS noise terms $C^{gg}_\ell$ and $C^{gg, {\rm \, cen}}_{\ell^{\prime\prime}}$ correspond to the unWISE galaxy auto-power spectra described in App.~\ref{app:gal_powerspectra} and given in Eq.~\eqref{eq:celgg_full_unwise} (all) and Eq.~\eqref{eq:celgg_centralsonly_unwise} (centrals-only). Here we have neglected the noise contribution to $\ev{T^\asc g}$ and $\ev{B^\asc g}$, but we note that residual foregrounds in the component-separated CMB, in particular the CIB, have non-negligible correlation with unWISE galaxies~\cite{yan2023star}, which should be taken into account in a data analysis. The factor of 1/2 in the signal-to-noise ratio of the bispectrum in Eq.~\eqref{eq:sigma_BBg} comes from having two indistinguishable $B^\asc$ fields (see~\eg Ref.~\cite{Liguori_2010} for an overview).

From Eqs.~\eqref{eq:sigma_XX},~\eqref{eq:sigma_Tg} and~\eqref{eq:sigma_BBg}, it can be seen that the sensitivity to the photon-axion coupling from CMB auto-correlation functions and the bispectrum scale as signal-to-noise ratio squared to the negative power of $1/8$, while for CMB temperature-galaxy cross-correlation there is a negative power of $1/4$. For this reason, the latter observable benefits more from the sum over many $\ell$ modes and will give the best sensitivity.

\begin{figure}[!]
    \centering
    \includegraphics[width=0.75\textwidth]{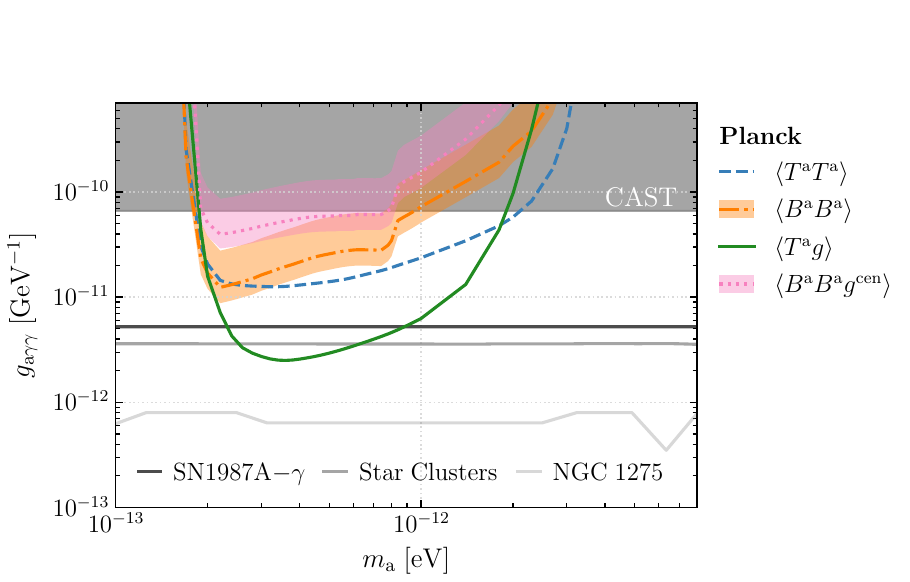}
    \caption{Forecasted sensitivity to the axion-photon coupling for Planck CMB measurements and the unWISE blue galaxy sample. The $\ev{T^{\asc} T^{\asc}}$ (blue dashed) and $\ev{B^{\asc} B^{\asc}}$ (orange dashed-dotted) sensitivities are obtained from Eq.~\eqref{eq:sigma_XX} for $X=T$ and $X=B$, respectively. The strongest sensitivity is from $\ev{T^{\asc} g}$ (solid green), which is obtained from Eq.~\eqref{eq:sigma_Tg}. We also show the three point function (pink dotted) between CMB $B$-modes and the unWISE template for central galaxies, from Eq.~\eqref{eq:sigma_BBg}. For both observables involving polarization fields, the coloured shaded band shows the effect of varying the magnetic domain size $r_{\dom}$ between $1$ and $10$ kpc, and the central line corresponds to $5$ kpc. The bump in the orange and pink contours around $m_\asc \approx 8 \times 10^{-13}$ eV is due to jumps in the magnetic field amplitude between discrete and wide bins in halo mass at a redshift $z\approx 1.3$; the effect does not appear in all the observables due to different weightings of each halo contribution in each case. Existing bounds on the photon-axion coupling (all in gray) are also shown and described in Fig.~\ref{fig:bounds_and_FIRAS}.} \label{fig:Planck_sensitivity_plot}

    \includegraphics[width=0.75\textwidth]{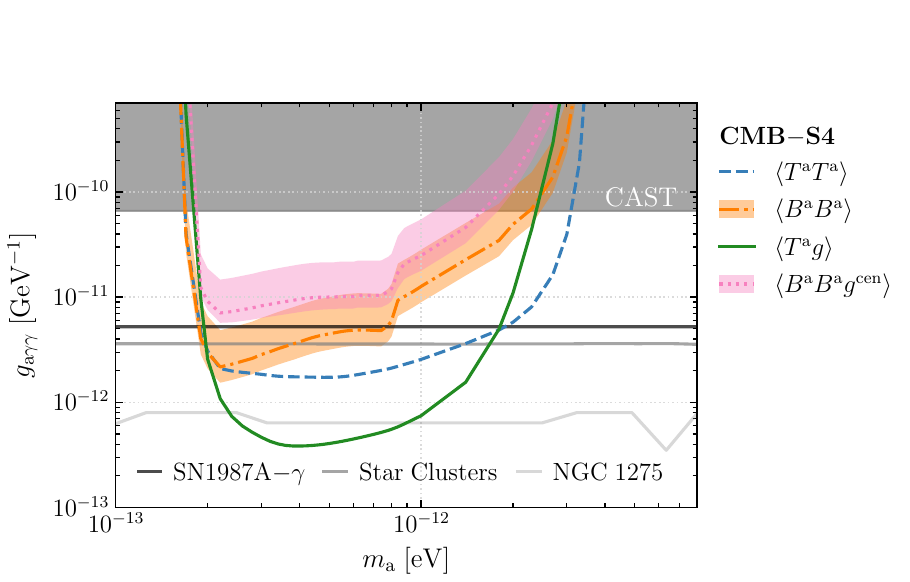}
    \caption{Same as Fig.~\ref{fig:Planck_sensitivity_plot}, but assuming CMB-S$4$ measurements and the unWISE blue galaxy sample. The lower CMB instrumental noise level and higher angular resolution translate into an improved sensitivity by a factor of $\mathcal{O}(10)$ compared to the Planck forecast, with a similar hierarchy between different observables.}\label{fig:S4_sensitivity_plot}
\end{figure}

Fig.~\ref{fig:Planck_sensitivity_plot} and Fig.~\ref{fig:S4_sensitivity_plot} show the sensitivities for the observables computed above in the $\ga$-$\ma$ parameter space for Planck and CMB-S4 respectively. We project that, with current data, our method can be complementary to existing astrophysical searches for axions, improving significantly compared to the most sensitive existing laboratory experiment, while future observations can marginally exceed the strongest astrophysical bounds currently in place.

Regarding auto-correlators, we show results for $X=T$ (blue dashed) and $X = B$ (orange dashed-dotted), since the signal for $E$ and $B$ modes is the same, but the latter has lower noise. The three auto-correlators can be easily combined by summing the individual signal-to-noise ratios squared, if one neglects $TE$ correlations in the primary CMB, which provide a negligible contribution to the final answer. The signal-to-noise ratio for both temperature and polarization is dominated by very high $\ell$-modes (see Fig.~\ref{fig:ILC_signal_comparison}). This is different from the dark photon case considered in~\citetalias{Pirvu:2023lch}, where the ILC residual noise was more suppressed at large scales and the signal-to-noise ratio in $TT$ was dominated by intermediate values of $\ell$ where the 2-halo term dominates. For polarization, the strength of the axion-induced signal is suppressed and sensitive to the small size of the coherent magnetic domains. We incorporate this uncertainty as a shaded band whose upper and lower bounds correspond to $r_\dom = 1$ and $r_\dom = 10\, {\rm kpc}$, respectively. Unless a significant component of the magnetic field is coherent on larger scales, the $TT$ channel always dominates the total sensitivity of CMB auto-correlation functions (with the exception of axion masses at the lower boundary of the range considered, where $TT$ and $BB$ have comparable sensitivities).

As expected, the CMB temperature cross-correlation with the unWISE galaxy catalogue is more sensitive than CMB auto-correlators, as can be seen from the green solid lines in Fig.~\ref{fig:Planck_sensitivity_plot} and~\ref{fig:S4_sensitivity_plot}. The performance of this observable should improve as catalogues with larger number of galaxies become available. Additionally, with a catalogue that has better redshift resolution, one could properly weight the cross-correlators in different redshift bins leveraging the redshift dependence of the photon-axion conversion at different axion masses (see Fig.~\ref{fig:dtdz}) to further increase the signal-to-noise ratio. For CMB-S4 (Fig.~\ref{fig:S4_sensitivity_plot}), the increased sensitivity is enough to go beyond existing astrophysical constraints on $\ga$ by a factor of up to $\approx 1.7$ for axion masses around $m_\asc \approx 4 \times 10^{-13}\,{\rm eV}$. 

Finally, the pink dotted lines in Fig.~\ref{fig:Planck_sensitivity_plot} and~\ref{fig:S4_sensitivity_plot} show the sensitivity of the CMB polarization-galaxy bispectrum $\ev{B^{\asc} B^{\asc} g}$. Similarly to the $\ev{B^{\asc} B^{\asc}}$ observable, the shaded region corresponds to magnetic field correlation lengths between $r_\dom = 1$ and $r_\dom = 10\, {\rm kpc}$. We find that the three-point function including a tracer of LSS does not improve the sensitivity compared to the CMB-only polarization auto-spectrum. Naively, this finding seems to contradict the expectation that the correlation with galaxies should enhance the signal-to-noise ratio, by properly weighting regions of the sky where the signal is expected. In practice, however, existing galaxy catalogues are incomplete and therefore the signal from the halos whose central galaxy is not seen by unWISE is effectively removed in $\ev{B^{\asc} B^{\asc} g}$. We checked that computing the bispectrum with a `perfect' galaxy catalogue that contains all central galaxies, improves the sensitivity beyond the auto-spectrum. Future surveys could be expected to lie between these two limiting cases. We find that the 1-halo and 2-halo terms in the bispectrum (see Eq.~\eqref{eq:red_bispectrum_B_1halo} and~\eqref{eq:red_bispectrum_B_2halo}) give a comparable contribution to the total signal, with the latter larger by a factor of a few. Note that, since we have included central galaxies only to simplify the derivation of the bispectrum, we are not leveraging the fact that axion-induced screening traces the angular profile of the satellite distribution and we are neglecting part of the signal, particularly at small scales where most of the signal-to-noise ratio should come from. If a signal is present in the data, it could therefore be larger than what is estimated here. Finally, we would like to stress the importance of having multiple correlators involving different maps of comparable sensitivity, in particular to confirm any possible future detection.

The results described in this section are consistent with, and complementary to the study of axion-induced polarization signals in clusters presented in Ref.~\cite{Mukherjee:2019dsu}. There, the authors propose to use aperture photometry on identified clusters to detect the square of the polarization signal induced by resonant photon-axion conversion. The bispectrum $\ev{B^{\asc} B^{\asc} g}$ considered here should be similar in spirit to the stacking technique considered in Ref.~\cite{Mukherjee:2019dsu}\footnote{For CMB-LSS bispectra involving the kinematic Sunyaev Zel'dovich contribution to CMB temperature, it is possible to demonstrate that constraints from the bispectrum are formally equivalent to constraints from a variety of estimators~\cite{Smith:2018bpn}. A similar set of equivalences may be identified in the context of axion-induced screening as well.}. However, it is difficult to perform a direct comparison due to differing assumptions about the coherence length of the magnetic field, a parameter which strongly affects the magnitude of the signal. Another crucial difference is that the correlators considered here target statistical signals, while aperture photometry or other filtering techniques target only the objects contributing most strongly to the signal. Arguably, the statistical signal is less sensitive to assumptions about the magnetic field and density profiles around individual halos, a primary uncertainty in any approach. Finally, the authors of Ref.~\cite{Mukherjee:2019dsu} cite the difficulty in distinguishing extragalactic CMB foregrounds from the signal as a challenge for using temperature data in correlation with LSS. Although our analysis does not account for this correlation explicitly in the $\ev{T^a g}$ correlator, we note that it is possible to exactly deproject foregrounds with known spectral energy distribution (see e.g. Ref.~\cite{McCarthy:2023hpa}), and that the axion signal in temperature always corresponds to a reduction in intensity (in contrast to e.g. radio or CIB {\em emission}). An interesting future direction would be to study in more detail the synergy of n-point correlators and various filtering and stacking techniques for both the temperature and polarization axion-induced screening signals.

\subsection{Extension to the case of effectively massless axions}\label{sec:massless_axion}

So far, we have focused on axion masses where resonant conversion happens inside galactic halos. For lighter axion masses, between $2\times 10^{-13} \,{\rm eV}$ and roughly $10^{-15} \,{\rm eV}$, the resonance condition can be met in regions outside the boundary of halos, where diffuse ionized gas traces the cosmic web. Independent of the exact location where resonant conversion occurs, axion-induced screening has the same frequency scaling and can therefore be searched for using the same maps obtained with the procedure outlined in Sec.~\ref{sec:ilc}, and cross correlating with the appropriate tracers of LSS.

For axions with even lighter masses, resonant conversion cannot happen in any astrophysical medium in the late universe (nor in the early universe, where densities were higher). However, non-resonant conversion can still occur in the presence of magnetic fields and we describe this scenario in some detail in App.~\ref{app:massless_axion}. In this case, the conversion probability still depends strongly on the photon plasma mass (scaling as $m_\gamma^{-4}$) and can therefore also give rise to axion-induced patchy screening. The axion-induced screening optical depth in the effectively massless axion case also has a characteristic dependence on the CMB photon frequency
\begin{equation}
    \tau^{\asc} \sim \omega^2,
\end{equation}
and the ILC method~\cite{ILCpaper} can be used to isolate anisotropies with this spectral dependence. In contrast to resonant conversion, for an effectively massless axion, the sign and size of the axion-induced screening signal strongly depends on the detailed properties of the inter-cluster gas density profile, and magnetic field power spectrum, neither of which are well-known. We provide an estimate for the amplitude of this signal in App.~\ref{app:massless_axion}, demonstrating that, in principle, anisotropies can improve upon existing constraints from the CMB monopole~\cite{Mirizzi:2005ng}. Future measurements sensitive to the distribution of ionized gas~\cite{CHIME:2022dwe,Madhavacheril:2019buy} or numerical simulations could be used to build more reliable models of photon-axion conversion in this regime and therefore extend the region of axion masses probed by the CMB.

\section{Concluding remarks}\label{sec:conclusion}

In this paper, we projected the sensitivity of current and future CMB experiments combined with LSS surveys to probe light axions that couple to the photon. The axion-photon coupling leads to axion-induced screening of the CMB, which is imprinted in new spectral anisotropies in temperature and polarization. We computed the resulting temperature auto-correlation $\ev{T^{\asc} T^{\asc}}$, polarization auto-correlation $\ev{B^{\asc} B^{\asc}}$, CMB temperature-galaxy cross-correlation $\ev{T^{\asc} g}$ and the polarization-galaxy bispectrum $\ev{B^{\asc} B^{\asc} g}$. The reach of the existing Planck and unWISE datasets on the axion photon coupling is significantly better than the current laboratory constraints from the axion helioscope CAST~\cite{CAST:2017uph}, and complement astrophysical constraints in the same mass range. Data from upcoming CMB-S4 experiment and future LSS surveys can further improve the sensitivity by up to about a factor of 2 in coupling compared to the best existing bounds, for $m_\asc \approx 4 \times 10^{-13}\,{\rm eV}$. 

The search we propose in this paper has different systematic uncertainties than other astrophysical probes. 
Similar to astrophysical searches that look for axion-induced irregularities on the X-ray spectra of AGN and quasars, our search also relies on the effect of photons oscillation into axions in the magnetic field of galaxies. Despite the effect being suppressed by the small frequency of CMB photons (see Eq.~\eqref{eq:haloprob}), the high quality of existing and future measurements of CMB anisotropies provides a competitive probe of axions. 

The cosmological search proposed in this work is complementary to existing analysis based on X-ray observations of the super star cluster~\cite{Dessert:2020lil} (see also the recent search for X-ray emissions from M82 and M87~\cite{Ning:2024eky}), the cluster-hosted quasar H1821+643~\cite{Reynes:2021bpe}, AGN NGC 1275~\cite{Reynolds:2019uqt}, and M87~\cite{Marsh:2017yvc}, due to several crucial differences. 
First, the most sensitive observable considered here -- $\ev{T^{\asc} g}$ -- is quite insensitive to the unknown magnetic field coherence length, as long as it is larger than $\mathcal{O}$(pc), because the relevant length scale for us is the vacuum oscillation length; at higher photon frequency, the latter becomes larger and the conversion probability becomes sensitive to the smallest length scale in the problem,~\ie the size of the magnetic field domain. 
Second, although our approach also requires a modeling of the magnetized circumgalactic medium, similarly to the models of the intracluster magnetic fields used in Refs.~\cite{Marsh:2017yvc, Reynolds:2019uqt, Reynes:2021bpe}, the axion-induced screening of CMB photons is a statistically {\it average} effect from many halos and therefore less affected by systematic uncertainties in the modeling of individual objects. Indeed, Ref.~\cite{Libanov:2019fzq} suggested that the bound from NGC 1275 relaxes considerably if the intracluster magnetic field is completely ordered and the unknown contribution from the turbulent and ordered component should be interpreted as a large uncertainty on the constraint. Similar uncertainties also exist for searches for photon to axion conversion with CMB polarization with small numbers of tracers~\cite{Mukherjee:2018oeb,Mukherjee:2018zzg,Mukherjee:2019dsu}. However, our proposal requires modeling of the magnetic field profiles of distant objects, which are much harder to measure directly compared to the MW magnetic field relevant for the limits obtained in Ref.~\cite{Dessert:2020lil}. Finally, the source property for our search, that is, the spectrum and the polarization of the blackbody CMB, is particularly well-known, compared to studies based on polarization of magnetized stars in~\cite{Lai:2006af,Dessert:2022yqq}, and potentially also SN1987A-$\gamma$~\cite{Payez:2014xsa, Hoof:2022xbe} (see also~\cite{Bar:2019ifz}).

For the reasons we listed above, a study based on CMB temperature and polarization anisotropies and their correlations with LSS will complement the existing searches in placing robust constraints on the parameter space of axion like particles. In the event of a discovery, the search proposed in this paper can be used to narrow down the exact mass and coupling of the axion that gives rise to this observed signal, by taking advantage of the $\ell$-dependence of the signal, the relative size of the different estimators, as well as tomographic information. This is certainly of particular importance for proposed laboratory searches of axion dark matter that are expanding their sensitivity to smaller axion masses~\cite{Chaudhuri:2014dla,Kahn:2016aff,DMRadio:2022pkf,Berlin:2020vrk}. The methodologies in this paper can be adapted to different axion masses (see for example, App.~\ref{app:massless_axion}) by correlating with different tracers with a wide range of plasma frequency, such as filaments and voids~\cite{Clampitt:2014lna,Galarraga-Espinosa:2023zmv,2011IJMPS...1...41V}, the details of which we leave to future studies.

The cases presented in~\citetalias{Pirvu:2023lch}, the main text of this paper, as well as App.~\ref{app:massless_axion} are three different examples of a spectral energy distribution (SED) that can be produced by dark screening of the CMB. Whereas Thomson screening produces optical depth $\tau \sim \omega^0$, resonant photon to dark photon conversion produces $\tau \sim \omega^{-1}$, and resonant (non-resonant) photon to axion conversion produces $\tau \sim \omega^{1}$ ($\tau \sim \omega^{2}$). Obtaining these different dark screening temperature and polarization maps from current and future CMB observations, and studying their correlations with the underlying large-scale structure tracers, has the potential to reveal the existence of light bosons.\\

Note added: right after this paper was posted on the arXiv, two related works also appeared~\cite{Mehta:2024wfo, Mehta:2024pdz}. There, the authors forecast the sensitivity of future CMB surveys to polarization anisotropies induced by photon-axion resonant conversion within LSS and compute the axion-induced polarization power spectrum from resolved~\cite{Mehta:2024wfo} and unresolved~\cite{Mehta:2024pdz} galaxy clusters, relying on ILC cleaning to further enhance the signal-to-noise ratio. These studies are analogous to the polarization observables $\ev{B^\asc B^\asc g}$ and $\ev{B^{\asc} B^{\asc}}$ considered in our work and the results appear to be compatible, when taking into account the different modeling of the halos' gas density and magnetic field. We stress that, as shown above, the polarization signal is more sensitive to the unknown coherence length of the magnetic field compared to the temperature signal.

\section*{Acknowledgements}

We thank Simone Ferraro, Colin Hill, and Moritz M$\ddot{\text{u}}$nchmeyer for interesting conversations; Rahul Kannan, Simon May, Rudiger Pakmor, and Freeke van de Voort for useful discussions about the circumgalactic magnetic field and Rudiger Pakmor for providing the numerical magnetic field profiles. MCJ and JH are supported by the Natural Sciences and Engineering Research Council of Canada through a Discovery Grant. Research at Perimeter Institute is supported in part by the Government of Canada through the Department of Innovation, Science and Economic Development Canada and by the Province of Ontario through the Ministry of Research, Innovation and Science.

\appendix

\section{Photon-axion conversion}\label{app:conversion}

\subsection{Homogeneous magnetic field}

An axion field $\asc$, with mass $m_{\asc}$, can interact with the SM photon as
\begin{equation}\label{eq:lagr1}
    \mathcal{L}_{\asc\gamma} = - \frac{1}{4}\ga F^{\mu\nu}\tilde{F}_{\mu\nu} \, \asc = \ga \mathbf{E} \cdot \mathbf{B} \, \asc,
\end{equation}
where $F^{\mu\nu}$ is the electromagnetic field-strength tensor, and $\tilde{F}^{\mu\nu} \equiv 1/2\epsilon_{\mu\nu\alpha\beta}F^{\alpha\beta}$ its dual. This term induces photon-axion oscillations in the presence of an external magnetic field transverse to the photon propagation direction. In particular, for a constant transverse magnetic field $\mathbf{B}$, the photon state polarized along the $\mathbf{B}$ field, $A_{\parallel}$, mixes with an ultra-relativistic axion ($\omega \gg m_\asc$) according to the equation of motion~\cite{Raffelt:1987im, Deffayet:2001pc, Mirizzi:2005ng}
\begin{equation}
    \left[\omega - i\partial_z + \begin{pmatrix} \Delta_{\rm pl}  & \Delta_{\asc\gamma} \\ \Delta_{\asc\gamma} & \Delta_{\asc} \\ \end{pmatrix} \right]  \begin{pmatrix} A_{\parallel} \\ \asc \\ \end{pmatrix}  = 0, 
\end{equation}
where $\Delta_{\rm pl} = -\ompl^2/(2\omega)$, $\Delta_{\asc} = -\ma^2/(2\omega)$, $\Delta_{\asc\gamma} = \ga |\mathbf{B}|/2$, and $\ompl^2 = e^2 n_e/m_e$ denotes the photon plasma mass in an ionized medium with electron density $n_e$. Notice that we have neglected the Cotton-Mouton birefringence of fluids in the presence of an external magnetic field and Faraday rotation that couples the two photon polarizations, as they do not affect the rest of the discussion. The mixing matrix can be diagonalized by rotating the fields by an angle 
\begin{equation}\label{eq:mix_angle}
    \theta = \frac{1}{2} \arctan \frac{2\Delta_{\asc\gamma}}{\Delta_{\rm pl} - \Delta_{\asc}},
\end{equation}
and the probability of the $A_{\parallel}$ state to convert into an axion after traveling a distance $r$ can be obtained, similarly to the case of neutrino oscillations~\cite{RevModPhys.61.937}, as
\begin{equation}\label{eq:prob_app}
    P(A_{\parallel}\rightarrow \asc) = (\Delta_{\asc\gamma} r)^2 \frac{\sin^2 (\Delta_{\rm{osc}} r/2)}{(\Delta_{\rm{osc}} r/2)^2},
\end{equation}
where $\Delta_{\rm{osc}}^2 \equiv (\Delta_{\rm pl} - \Delta_{\asc})^2 + 4\Delta_{\asc\gamma}^2$ is the oscillation wavenumber, so that one complete oscillation is obtained after a distance $l_{\rm{osc}} = 2\pi/\Delta_{\rm{osc}}$. The oscillation length varies significantly between vacuum ($m_{\asc}^2 \gg \ompl^2$) and a resonance region ($m_{\asc}^2 \simeq \ompl^2$):
\begin{align}
    l_{\rm osc} = \begin{cases}
        \frac{4\pi \omega}{m_{\asc}^2} \simeq 0.01 \, {\rm pc}\, \left(\frac{\omega}{10^{-4} {\rm eV} }\right) \left(\frac{10^{-12}\, {\rm eV}}{ m_{\asc} }\right)^2, \quad & |\delta m^2| \simeq m_{\asc}^2 \gg 2\omega\ga |\mathbf{B}|, \\
        \frac{2\pi}{\ga |\mathbf{B}|} \simeq 10 \, {\rm Mpc}\, \left(\frac{10^{-12} {\rm GeV}^{-1}}{\ga}\right) \left(\frac{0.1 \, \mu {\rm G}}{|\mathbf{B}|}\right), \quad & |\delta m^2| \ll 2\omega\ga |\mathbf{B}|, \end{cases}
\end{align}
where $\delta m^2 \equiv m_{\asc}^2 - \ompl^2$. If the photon path crosses a region where the resonance condition is met, the total conversion probability is dominated by the resonance contribution, while multiple oscillations before and after the resonance average out and can be neglected. Similarly to the well-known case of MSW neutrino resonant conversion in a medium~\cite{PhysRevD.17.2369, Mikheyev:1985zog}, the conversion probability of a photon into an axion is then given by~\cite{PhysRevLett.57.1275}
\begin{equation}\label{eq:probres}
    P(A_{\parallel}\rightarrow \asc)_{\res} = 1-p,
\end{equation}
where $p$ is the level crossing probability. This expression is valid as long as there are regions before and after the resonance where the mixing angle in Eq.~\eqref{eq:mix_angle} is small,~\ie where $|\Delta_{\rm pl}-\Delta_{\asc}| \gg \Delta_{\asc\gamma}$, or
\begin{equation}
    |\delta m^2| \gg 2\omega \ga |\mathbf{B}| \simeq (10^{-17}\, \rm{eV})^2 \left(\frac{\omega}{10^{-4}\, \rm{eV}}\right)\left(\frac{\ga}{10^{-12} \,{\rm GeV}^{-1}}\right)\left(\frac{|\mathbf{B}|}{0.1 \, \mu\rm{G}}\right).
\end{equation}
Assuming that the plasma mass,~\ie the electron number density, varies linearly across the resonance, the level crossing probability can be computed using the Landau-Zener expression~\cite{PhysRevLett.57.1275,Mirizzi:2009nq, Tashiro:2013yea}
\begin{equation}\label{eq:probres2}
    P(A_{\parallel}\rightarrow \asc)_{\res} \simeq \frac{\pi\omega \ga^2 |\mathbf{B}|^2}{m_{\asc}^2}\left|\frac{\dd \ln \ompl^2}{\dd t}\right|_{t_{\res}}^{-1} ,
\end{equation}
where we used the small coupling approximation and the fact that $P(A_{\parallel}\rightarrow \asc)_{\res} \ll 1 $ and $ p\simeq 1$. For the low energy CMB photons, axion masses, and small couplings that we are interested in, the above assumptions are always satisfied. The resonance is very narrow and extends over a time scale $\Delta t_{\res}$, defined as the time over which $\delta m^2/(2\omega)$ becomes of order $\ga |\mathbf{B}|$, 
\begin{align}\label{eq:deltat_res}
    \Delta t_{\res} & \simeq \frac{2\omega \ga |\mathbf{B}|}{m_{\asc}^2} \left|\frac{\dd \ln \ompl^2}{\dd t}\right|_{t_{\res}}^{-1} \sim \frac{\omega \ga |\mathbf{B}|}{m_{\asc}^2} r_{\res} \nonumber \\
    & \simeq 10^{-10} r_{\res} \left(\frac{\omega}{10^{-4} {\rm eV} }\right)\left(\frac{10^{-12}\, {\rm eV}}{ m_{\asc} }\right)^2\left(\frac{\ga}{10^{-12} \,{\rm GeV}^{-1}}\right) \left(\frac{|\mathbf{B}|}{0.1 \, \mu\rm{G}}\right),
\end{align}
where the second expression assumes a power law scaling of $\ompl^2$ and that the photon trajectory crosses the resonant region in the direction of the number density gradient ($\dd r/\dd t|_{t_{\res}} = 1$). In a realistic scenario, the plasma mass does not change monotonically and the photon trajectory will cross multiple resonances. In particular, we are interested in photons that convert within a halo, that will typically cross a resonance twice. The contributions from two resonances at $r_1$ and $r_2$ can be simply added incoherently, as long as~\cite{Brahma:2023zcw}
\begin{equation}
    \left| \int_{r_{2}}^{r_{1}} \dd r^{\prime} \frac{\delta m^2(r)}{2\omega} \right| \gg 2\pi.
\end{equation}
The above condition is usually easily satisfied, since the integral is typically larger than
\begin{equation}
    \frac{m_{\asc}^2}{2\omega} \Delta r \simeq 10^6 \, \left( \frac{ m_{\asc} }{10^{-12}\, {\rm eV}}\right)^2\left(\frac{10^{-4}\, {\rm eV}}{\omega}\right)\left(\frac{\Delta r}{{\rm kpc}}\right).
\end{equation}
There are two exceptions where interference (or {\it phase effects}) between the two resonances might be relevant. First, when the photon trajectory crosses the halo close to the edge of the resonance radius, so that $\Delta r$ is small; however, since the resonance is very narrow, the contribution from these regions to the integrated probability of conversion inside the halo are negligible (see~\citetalias{Pirvu:2023lch}). Second, when the resonance radius is in the innermost part of the halo, where the number density profile is almost flat, so that $\delta m^2(r)$ is small. Also in this case, however, the contribution to the signal is negligible, as it corresponds to small angular scales that are inaccessible to observations. 

So far we have considered the unrealistic scenario of a constant magnetic field along the photon trajectory. In the next section we consider a spatially varying magnetic field.

\subsection{Inhomogeneous magnetic field}

Photon-axion mixing for an inhomogeneous plasma and magnetic field can be computed using time-dependent perturbation theory in the limit of small coupling, which is always valid given current constraints on the axion-photon coupling. In this case the conversion probability for a photon traveling over a distance $r$ is given by~\cite{Raffelt:1987im, Marsh:2021ajy, Brahma:2023zcw}
\begin{align}\label{eq:prob_int}
    P(A_{\parallel}\rightarrow \asc) = \left| \int_0^r \dd r^{\prime} \Delta_{\asc\gamma}(r^{\prime}) e^{i \Phi(r^{\prime})} \right|^2, \quad
    \Phi(r) = \int_0^{r} \dd r^{\prime}\frac{\delta m^2(r^{\prime})}{2 \, \omega}.
\end{align}
The expression above can be evaluated using the stationary phase approximation to obtain the same result given in Eq.~\eqref{eq:probres2}~\cite{Brahma:2023zcw}. The stationary phase approximation is valid as long as the magnetic domain size $r_{\dom}$ is larger than $4\pi \omega/\ma^2$. However, as also noted in Ref.~\cite{Marsh:2021ajy}, the result in Eq.~\eqref{eq:probres2} is valid only if the magnetic varies slowly over a coherence length 
\begin{equation} \label{eq:b_coherence}
    r_{\dom} > \sqrt{\frac{4\pi \omega}{\ma^2} \left|\frac{\dd \ln \ompl^2}{\dd r}\right|_{r_{\res}}^{-1}}, 
\end{equation}
which is at most hundred parsec over the region of masses and photon frequencies considered here. For higher frequencies, when this approximation breaks down, a resonant conversion condition can still be met, but the conversation probability will no longer depend on frequency, as it will be dictated by the smallest length scale at play, $r_{\dom}$, instead of the frequency-dependent resonance length scale. We note that the resonance condition could be met even in the limit of $\ompl \ll m_{\asc}$, if the magnetic field oscillates with a period of $m_{\asc}^2/(2\omega)$, as pointed out in Ref.~\cite{Raffelt:1987im}.

Sub-kpc length scales are too small to be resolved by hydrodynamical cosmological simulations, but given the physics driving the magnetic fields it is reasonable to expect coherent magnetic fields over distances larger than sub-kpc. It is worth emphasizing that although the discussions in this paper are mostly in position space, the resonant conversion problem can actually be treated more elegantly in Fourier space~\cite{Baryakhtar:2018doz,Marsh:2021ajy}, where magnetic field consists of Fourier modes with amplitudes described by the power spectrum (see~\cite{2020MNRAS.498.3125P,2023arXiv230913104P} for more details) and uncorrelated phases. In Fourier space, most of the conversion comes from Fourier modes corresponding to the peak of the power spectrum, and small scale fluctuations in the magnetic field do not affect the correlation functions of photon-axion conversion, as long as the superposition principle remains valid and magnetic field power spectrum has a red Kolmogorov scaling~\cite{1941DoSSR..32...16K}. An additional requirement is that the magnetic field must be approximately constant over the length scale of resonance crossing~\eqref{eq:deltat_res}. However, in the limit of small coupling that we are working in, this requirement is always weaker compared to Eq.~\eqref{eq:b_coherence}.

\section{Polarization auto-correlations}\label{app:pol_cell}

To derive the polarization auto-power-spectra given in Eq.~\eqref{eq:cellEscEsc}, we start from the axion signal contribution to the Stokes parameters along the line of sight $\nhat$, introduced in Eq.~\eqref{eq:pfluc}:
\begin{align} 
    (Q\pm i U)^{\asc}(\hat n) &= -\frac{1-e^{-x}}{x} \bar{T} \int_{z_{\rm min}}^{z_{\rm max}} \dd z \, \frac{\dd  \tau^{\asc}(\hat n, \chi)}{\dd z} \gamma^{\pm}(\hat{n}, \chi),
\end{align}
where all the quantities appearing here are defined in Sec.~\ref{sec:cmbsign}. The left-hand side of the above equation is a spin-2 function and can be expanded in spin-2 spherical harmonics. The right-hand side contains the product of a scalar $\dd \tau^\asc/\dd z(\hat{n}, \chi)$ and a spin-2 function $\gamma^{\pm}(\hat{n}, \chi)$. Expanding using the appropriate spherical harmonics for each directional-dependent function on both sides of the equation results in
\begin{align}\label{eq:pol_expansion}
    \sum_{ \ell m} a^{\pm 2}_{\ell m} \, \leftindex_{\pm 2}Y_{\ell m}(\hat{n}) = & -\frac{1-e^{-x}}{x} \bar{T}  \int_{z_{\rm min}}^{z_{\rm max}} \dd z \, \frac{\chi^2}{H} \int \dd m \sum_i \frac{\delta(\chi - \chi_i)}{\chi^2}\delta^2(\nhat^{\prime}-\nhat_i)\delta(m-m_i) \times \nonumber \\ 
    & \sum_{\ell^\prime 0} \sum_{\ell^{\prime\prime} m^{\prime\prime}} \tau^{\asc}_{\ell^{\prime} 0}(\chi, m) \, \gamma^{\pm 2}_{\ell^{\prime\prime} m^{\prime\prime}}(\chi) \, Y_{\ell^{\prime} m^\prime}(\hat{n}) \, \leftindex_{\pm 2}Y_{\ell^{\prime\prime} m^{\prime\prime}}(\hat{n}) \int \dd^2\nhat'  Y_{\ell^{\prime} m^\prime}(\nhat^\prime),
\end{align}
where $\tau^\asc_{\ell 0}$ was defined in Eq.~\eqref{eq:tau_multipoles}. The coefficients $a^{\pm 2}_{\ell m}$ on the left-hand side are conventionally replaced by $E$ and $B$-modes defined in harmonic space
\begin{equation}
    a^{\pm 2}_{\ell m} = E^{\asc}_{\ell m} \pm i B^{\asc}_{\ell m}, \quad
    E^{\asc}_{\ell m} = \frac{a^{+2}_{\ell m}+a^{-2}_{\ell m}}{2}, \quad
    B^{\asc}_{\ell m} = i\frac{a^{+2}_{\ell m}-a^{-2}_{\ell m}}{2}.
\end{equation}
The product of the two spherical harmonics on the right hand side of Eq.~\eqref{eq:pol_expansion} can be rewritten using the relation
\begin{equation}\label{eq:y_prod}
   Y_{\ell^{\prime} m^{\prime}}(\nhat) \leftindex_{\pm 2}Y_{\ell^{\prime\prime} m^{\prime\prime}}(\nhat) = \sum_{\ell m} (-1)^{m} \sqrt{\frac{(2 \ell + 1)(2 \ell^{\prime} + 1)(2 \ell^{\prime\prime} + 1)}{4\pi}} 
    \begin{pmatrix}
			\ell & \ell^{\prime} & \ell^{\prime\prime} \\
			-m & m^{\prime} & m^{\prime\prime}
	\end{pmatrix}
    \begin{pmatrix}
			\ell & \ell^{\prime} & \ell^{\prime\prime} \\
			\mp 2 & 0 & \pm 2 
	\end{pmatrix}\, \leftindex_{\mp 2}Y_{\ell m}(\nhat).
\end{equation}
Therefore, we can read off the expansion coefficients $a^{\pm 2}_{\ell m}$ directly from Eq.~\eqref{eq:pol_expansion} and~\eqref{eq:y_prod},
\begin{align}
   a^{\pm 2}_{\ell m} = & -\frac{1-e^{-x}}{x} \bar{T}  \int_{z_{\rm min}}^{z_{\rm max}} \dd z \, \frac{\chi^2}{H} \int \dd m \sum_i \frac{\delta(\chi - \chi_i)}{\chi^2}\delta^2(\nhat^{\prime}-\nhat_i)\delta(m-m_i) \times \nonumber \\ 
    & \sum_{\ell^\prime m^\prime} \sum_{\ell^{\prime\prime} m^{\prime\prime}} (-1)^{m}\sqrt{2\ell+1}   W_{\ell \ell^{\prime} \ell^{\prime\prime}}^{2 0 2} 
    \begin{pmatrix}
			\ell & \ell^{\prime} & \ell^{\prime\prime} \\
			-m & m^{\prime} & m^{\prime\prime}
	\end{pmatrix}  \tau^{\asc}_{\ell^{\prime} 0}(\chi, m) \, \gamma^{\pm 2}_{\ell^{\prime\prime} m^{\prime\prime}}(\chi) \int \dd^2\nhat'  Y_{\ell^{\prime} m^\prime}(\nhat^\prime).
\end{align}
where we introduced 
\begin{equation} \label{eq:w_def}
    W_{\ell \ell^{\prime} \ell^{\prime\prime}}^{m m^{\prime} m^{\prime\prime}} = \sqrt{\frac{(2 \ell^\prime + 1) (2 \ell^{\prime\prime} +1)}{4 \pi}}  
	\begin{pmatrix}
        \ell & \ell^{\prime} & \ell^{\prime\prime} \\
        -m & m^{\prime} & m^{\prime\prime}
    \end{pmatrix}.
\end{equation}
To compute the correlators, note that, from the definition of the functions $\gamma^{\pm}(\nhat, \chi)$ in Eq.~\eqref{eq:gammas}, the only non-vanishing correlators (see Eq.~\eqref{eq:gamma_corr_pm}) are, in harmonic space,
\begin{align}\label{eq:polwindow}
    \ev{\gamma^{\pm 2 *}_{\ell^{\prime \prime} m^{\prime \prime}}(\chi) \gamma^{\pm 2}_{L^{\prime \prime} M^{\prime \prime}} (\chi)} = \frac{9}{N_{\res}(\chi)} \frac{2}{15} 2\pi \theta_\dom(\chi)^2 \, e^{- \ell^{\prime \prime}(\ell^{\prime \prime}+1) \theta_\dom^2 / 2} \delta_{\ell^{\prime \prime} L^{\prime \prime}} \delta_{m^{\prime \prime} M^{\prime \prime}} \equiv \mathcal{C}_{\ell^{\prime\prime}}^{\pm}(\chi)\delta_{\ell^{\prime \prime} L^{\prime \prime}} \delta_{m^{\prime \prime} M^{\prime \prime}},
\end{align}
while $ \ev{\gamma^{\pm 2 *}_{\ell^{\prime \prime} m^{\prime \prime}}(\chi) \gamma^{\mp 2}_{L^{\prime \prime} M^{\prime \prime}} (\chi)} = 0$, and correlations between different halos also vanish. The $\mathcal{C}_{\ell}^{\pm}$ above have been obtained using the flat-sky approximation, since the magnetic field domains extend over a small angular scale. Therefore, the only contribution to the polarization power spectra comes from the 1-halo term and reads
\begin{align}
  \ev{a^{\pm 2*}_{\ell m} a^{\pm 2}_{L M}} & = \left(\frac{1-e^{-x}}{x} \bar{T}\right)^2   \int_{z_{\rm min}}^{z_{\rm max}} \dd z \, \frac{\chi^2}{H} \int \dd m \, n(\chi, m) \left\lbrace \sum_{\ell^\prime \ell^{\prime \prime}}  (W_{\ell \ell^{\prime} \ell^{\prime\prime}}^{2 2 0 } )^2
     [\tau^\asc_{\ell^{\prime\prime}}(\chi, m)]^2  \mathcal{C}_{\ell^{\prime}}^{\mp}(\chi) \right\rbrace \delta_{\ell L} \delta_{m M},
\end{align}
where the orthonormality of the spherical harmonics $\int \dd^2 \nhat^{\prime} Y_{\ell^{\prime} m^\prime}(\nhat^\prime) Y_{L^{\prime} M^\prime}(\nhat^\prime) = \delta_{\ell^\prime L^\prime}\delta_{m^\prime M^\prime}$ has been used, together with the following properties of the 3j-symbols
\begin{equation}
    \sum_{m^{\prime} m^{\prime\prime}} 
	\begin{pmatrix}
        \ell & \ell^{\prime} & \ell^{\prime\prime} \\
        -m & m^{\prime} & m^{\prime\prime}
    \end{pmatrix}
    \begin{pmatrix}
        L & \ell^{\prime} & \ell^{\prime\prime} \\
        -M & m^{\prime} & m^{\prime\prime}
    \end{pmatrix} = \frac{1}{2\ell+1} \delta_{\ell L} \delta_{m M}, \,\, \quad (W_{\ell \ell^{\prime} \ell^{\prime\prime}}^{m m^{\prime} m^{\prime \prime}} )^2 = (W_{\ell \ell^{\prime \prime} \ell^{\prime}}^{m m^{\prime \prime} m^{\prime}} )^2.
\end{equation}
The $E$ and $B$ mode correlations can be written in terms of the correlations of the $a^{\pm 2}_{\ell m}$ coefficients from their definition,
\begin{equation}
	\ev{E^{\asc *}_{\ell m} E^{\asc}_{\ell^{\prime} m^{\prime}}} = \ev{B^{\asc *}_{\ell m} B^{\asc}_{\ell^{\prime} m^{\prime}}} = \frac{1}{4} \left[ \ev{a^{+2 *}_{\ell m}a^{+2}_{\ell^{\prime} m^{\prime}}} + \ev{a^{-2 *}_{\ell m}a^{-2}_{\ell^{\prime} m^{\prime}}} \right],
\end{equation}
noting that $\ev{a^{\pm2 *}_{\ell m}a^{\mp2}_{\ell^{\prime} m^{\prime}}} = 0$. The resulting $\mathcal{C}_\ell^{E^\asc E^\asc}$ and $\mathcal{C}_\ell^{B^\asc B^\asc}$ correspond to the expressions given in Eq.~\eqref{eq:cellEscEsc}. The cross-correlation $\ev{E_{\ell m}^{\asc \ast}B^{\asc}_{\ell m}}$ vanishes because it is proportional to $C_{\ell}^{+} - C_{\ell}^{-} = 0$. 

\section{Halo occupation distribution and galaxy power spectra}\label{app:gal_powerspectra}

In this appendix we summarize the HOD and the associated power spectra of Ref.~\cite{unwiseHOD} which we use to model the unWISE blue sample in our forecasts. The galaxy-galaxy power spectrum is
\begin{equation}
\begin{gathered}
    C_{\ell}^{g g} = C_{\ell}^{g g, \, 1-{\rm halo}} + C_{\ell}^{g g, \, 2-{\rm halo}} + A_{\rm SN}, \\
	C_{\ell}^{g g, \, 1-{\rm halo}} = \int \dd z \frac{\chi(z)^2}{H(z)}  \int  \dd m \, n(z,m) \ev{\left|u_{\ell}^g(z,m)\right|^2}, \\
	C_{\ell}^{g g, \, 2-{\rm halo}} = \int \dd z \frac{\chi(z)^2}{H(z)} \left[ \int \dd m \, n(z,m) b(z,m) u_{\ell}^g(z, m) \right]^2 P^{\rm lin}\left(\frac{\ell+\frac{1}{2}}{\chi(z)}, z\right), 
\end{gathered}
\end{equation}
The galaxy multipole space kernel $u_{\ell}^g (z,m)$ is defined as: 
\begin{equation}\label{eq:ug_full}
\begin{aligned}
    u_{\ell}^g(z, m) &= W(z) \bar{n}_g^{-1} \left[ N_c(m) + N_s(m) u_{\ell}^{\rm NFW}(z,m) \right], \\
    \bar{n}_g(z) &= \int \dd  m \, n(z,m) \left[ N_c(m) + N_s(m) \right], \\
    W(z) &=  \frac{H(z)}{\chi(z)^2} \frac{\dd N_g}{\dd z},
\end{aligned}
\end{equation}
and its second moment is
\begin{equation}
    \ev{\left|u_{\ell}^g(z,m)\right|^2} = W_g(z)^2 \bar{n}_g^{-2} \left[ N_s(m)^2 u_{\ell}^{\rm NFW}(z,m)^2 + 2 N_s(m) u_{\ell}^{\rm NFW}(z,m) \right].
\end{equation}
$\dd N_g / \dd z$ is the redshift distribution of the unWISE blue galaxies normalized to 1. This distribution has a median redshift of $z=0.6$, and is relatively flat in the redshift range between $0.2 \lesssim z \lesssim 0.8$. The functions $N_c(m)$ and $N_s(m)$ represent the expectation values for the number of central and satellite galaxies respectively in a halo of mass $m$. These are parametrized as:
\begin{equation}
    N_c(m) = \frac{1}{2} + \frac{1}{2}\operatorname{erf}\left(\frac{\log m-\log m_{\rm min }}{\sigma_{\log m}}\right), \quad N_s(m) = N_c(m) \left( \frac{m}{m_*}\right)^{\alpha_s}.
\end{equation}
Central galaxies lie exactly at the center of the halo profile and their number is modelled as a smoothed step function. Meanwhile, satellites are distributed inside halos according to an NFW profile. The function $u_{\ell}^{\rm NFW} (z,m)$ is the normalized harmonic transform of the truncated NFW density profile~\cite{NFWprofile}. Given a truncation radius $r = \lambda r_{\Delta}$, the Fourier transform has an exact analytical form given by~\cite{10.1093/mnras/stz3351, Scoccimarro_2001}:
\begin{eqnarray}
    u^{\rm NFW}(k |z,m) &=& \left[\ln \left(1+\lambda c_{\Delta}\right)-\frac{\lambda c_{\Delta}}{\left(1+\lambda c_{\Delta}\right)}\right]^{-1} \nonumber \\
    &\times& \left[\cos (q)\left[{\rm Ci}\left(\tilde{q} \right)-{\rm Ci}(q)\right] +
    \sin (q)\left[{\rm Si}\left( \tilde{q} \right)-{\rm Si}(q)\right] - \frac{\sin \left(\tilde{q} - 1\right) }{ \tilde{q} } \right],
\end{eqnarray}
where Si, Ci are the cosine and sine integrals with arguments $q \equiv k r_{\Delta} / c_{\Delta}$ and $\tilde{q} \equiv 1+\lambda c_{\Delta} q$. The scales are chosen with respect to a halo boundary defined at $\Delta = 200$ times the critical density. Under this convention, $r_{\Delta}$ is the halo radius that encloses mass $m_{\Delta}$ and $c_{\Delta}$ is the corresponding concentration. Making the substitution $k \to (\ell+1 / 2) / \chi$ gives the multipole projection function $u_{\ell}^{\rm NFW}(z,m)$. This HOD model is defined by $6$ free parameters. The values we use are taken from Table VI of~\cite{unwiseHOD_newer}:
\begin{equation}\label{eq:HODparams}
    {\rm HOD:} 
    \left\{\alpha_{\mathrm{s}} = 1.06, 
    \sigma_{\log m} = 0.02, 
    \log m_{\rm min } = 11.86 M_{\odot}, 
    \log m_* = 12.78 M_{\odot}, 
    \lambda = 1.80, 
    10^7 A_{\rm SN} = 0.87 \right\},
\end{equation}
where $A_{\rm SN}$ is the shot noise. It only appears in the definition for the galaxy auto-correlation function and for our purposes, it acts as a noise term.

To simplify the computation of the polarization-polarization-galaxy bispectrum, we assume that only central galaxies contribute to the signal. This is equivalent to setting $N_s(m) = 0$ everywhere in the auto- and cross-power spectra defined above. The centrals-only power spectrum is given by \begin{equation}\label{eq:celgg_centralsonly_unwise}
\begin{aligned}
    C_{\ell}^{g g, {\rm cen}} &= C_{\ell}^{g g, {\rm cen}, \, 2-{\rm halo}} + A^{\rm cen}_{\rm SN}, \\
	C_{\ell}^{g g, {\rm cen}, \, 2-{\rm halo}} &= \int \dd z \frac{\chi(z)^2}{H(z)} \left[ \int \dd m \, n(z,m) b(z,m) u_{\ell}^{g, {\rm cen}}(z, m) \right]^2 P^{\rm lin}\left(\frac{\ell+\frac{1}{2}}{\chi(z)}, z\right),
\end{aligned}
\end{equation}
where
\begin{equation}\label{eq:ug_centrals}
\begin{aligned}
    u^{g, {\rm cen}}(z, m) &= W(z) \bar{n}_{g, {\rm cen}}^{-1} N_c(m),\\
    \bar{n}_{g, {\rm cen}}(z) &= \int \dd  m \, n(z,m) N_c(m),
\end{aligned}
\end{equation}
and we define the shot noise from the total number of expected centrals in the unWISE blue sample:
\begin{equation}
    A^{\rm cen}_{\rm SN} = 4\pi \left( \int \dd z \frac{\chi(z)^2}{H(z)} \frac{\dd N_g}{\dd z} \bar{n}_{g, {\rm cen}}(z) \right)^{-1} \approx 2.87 \times 10^{-7}.
\end{equation}

\section{Bispectrum derivation}\label{app:pol_bispectra}

The axion-induced polarization signal is correlated with the location of LSS. The leading-order non-vanishing cross-correlation between CMB and LSS is the three-point function:
\begin{eqnarray}\label{eq:QQg}
    \ev{(Q \pm i U)^{\asc\ast}(\hat{n}_1)(Q \pm i U)(\hat{n}_2) g(\hat{n}_3)},
\end{eqnarray}
where $(Q \pm i U)^{\asc}$ is defined in Eq.~\eqref{eq:pfluc} and $g(\hat{n})$ represents the galaxy density field. We include only central galaxies, and model this map as
\begin{equation}
    g(\hat{n},\chi) = \sum_i u^{g, {\rm cen}}(\chi, m_i) \delta^{2} (\hat{n}-\hat{n}_i)
\end{equation}
at each redshift, where $u^{g, {\rm cen}}$ was defined in~\eqref{eq:ug_centrals}. Two terms contribute to the bispectrum in Eq.~\eqref{eq:QQg}: a 1-halo term for points $\lbrace \hat{n}_1, \hat{n}_2, \hat{n}_3\rbrace$ crossing the same halo, and a 2-halo term for $\lbrace \hat{n}_1, \hat{n}_2\rbrace$ crossing one halo and $\lbrace \hat{n}_3 \rbrace$ crossing a different halo. The first term contributes at small scales, while the second term includes the large-scale clustering of structure. Due to the hierarchy of scales between the magnetic field coherence length, the characteristic radius of photon-axion resonance conversion, and the distance between halos, both terms are dominated by squeezed triangle configurations. Note that there is no three-halo term, because the polarization signal from different halos is uncorrelated. 

\subsection{One-halo term}

We write explicitly the three-point function by summing the contributions from all halos $i$, such that the 1-halo contribution to Eq.~\eqref{eq:QQg} becomes
\begin{equation}
\begin{aligned}
    & \left(\frac{1-e^{-x}}{x} \bar{T}\right)^2 \int \dd \chi \chi^2\, \dd m \, \dd^2\hat{n} \, \frac{1}{9} P^2(\chi, m) N^2_{\rm res}(\chi, m) u(\nhat_1-\nhat) u(\nhat_2-\nhat) u^{g, {\rm cen}}(\chi, m) \delta^{2} (\nhat_3-\nhat)  \times \\ 
    &  \ev{\sum_{i} \delta(m-m_i) \frac{\delta(\chi -\chi_i)}{\chi^2} \delta^{2} (\nhat-\nhat_i)} \ev{\gamma^{\pm}(\hat{n}_1, \chi)\gamma^{\pm}(\hat{n}_2, \chi)} = \\
    & \left(\frac{1-e^{-x}}{x} \bar{T}\right)^2 \int \dd \chi \chi^2\, \dd m \, n(\chi, m)\frac{1}{9} P^2(\chi, m) N^2_{\rm res}(\chi, m) u^{g, {\rm cen}}(\chi, m) \times \\ 
    & u(\nhat_1-\nhat_3) u(\nhat_2-\nhat_3) \ev{\gamma^{\pm}(\hat{n}_1, \chi)\gamma^{\pm}(\hat{n}_2, \chi)}.
 \end{aligned}
\end{equation}
The last line of the equation above contains all the angular dependent functions. We can simplify the calculation by noting that the magnetic field domains are much smaller than the typical size of a halo, therefore, expect for the a small area around the center of the halo, most triangles will be such that $|\nhat_1-\nhat_2| \ll |\nhat_1-\nhat_3| \simeq |\nhat_2-\nhat_3|$. We therefore approximate $u(\nhat_1-\nhat_3) u(\nhat_2-\nhat_3) \simeq \left[ u(\nhat_1-\nhat_3)^2 + u(\nhat_2-\nhat_3)^2\right]/2$. We show the calculation for the first term with $\nhat_1$ below, since the one with $\nhat_2$ can be obtained in the same way. Expanding in spherical harmonics,
\begin{align}
u(\nhat_1-\nhat_3)^2 & = \sum_{L M}  \sum_{L^{\prime} M^\prime}\frac{4\pi}{\sqrt{(2L+1)(2L^\prime+1)}}  u_{L 0}(m, z) u_{L^\prime 0}(m, z) Y_{L M}(\nhat_1)Y_{L M}(\nhat_3)  Y_{L^\prime M^\prime}(\nhat_1)Y_{L^\prime M^\prime}(\nhat_3) \nonumber \\
	&  =  \sum_{L L^{\prime}}  \frac{4\pi}{\sqrt{(2L+1)(2L^\prime+1)}}  u_{L 0}(m, z) u_{L^\prime 0}(m, z) \sum_{\ell^{\prime \prime} m^{\prime \prime}} \left(W_{\ell^{\prime \prime} L L^{\prime}}^{000} \right)^2 Y_{\ell^{\prime \prime} m^{\prime \prime}}(\nhat_1)Y_{\ell^{\prime \prime} m^{\prime \prime}}(\nhat_3), 
\end{align}
where $u_{\ell 0}$ was defined in Eq.~\eqref{eq:tau_multipoles} and we used the result from Eq.~\eqref{eq:y_prod}, adapted to the case of spin-0 spherical harmonics, to contract their product. Now we expand also the correlator $\ev{\gamma^{\pm}(\hat{n}_1, \chi)\gamma^{\pm}(\hat{n}_2, \chi)} = \mathcal{C}^{\pm}_{\ell^{\prime}}(\chi)  \leftindex_{\pm 2}Y_{\ell^{\prime}m^{\prime}}(\nhat_1) \leftindex_{\pm 2}Y_{\ell^{\prime} m^{\prime}}(\nhat_2)$, where $\mathcal{C}^{\pm}_{\ell^{\prime}}$ was introduced in Eq.~\eqref{eq:polwindow}. Using again Eq.~\eqref{eq:y_prod}, we further combine the two remaining spin-0 and spin-2 spherical harmonics evaluated at $\nhat_1$, to get the bispectrum
\begin{equation}
\begin{aligned}
    & \left(\frac{1-e^{-x}}{x} \bar{T}\right)^2 \int \dd \chi \chi^2\, \dd m \, n(\chi, m) u^{g, {\rm cen}}(\chi, m) \times \\
    & \sum_{\ell m}  \sum\limits_{\substack{\ell^{\prime} m^{\prime} }} \sum_{\ell^{\prime\prime} m^{\prime\prime}} 
 \sum_{L L^{\prime}} (-1)^{m}  \sqrt{\frac{(2 \ell + 1)(2 \ell^{\prime} + 1)(2 \ell^{\prime\prime} + 1)}{4\pi}} 
    \begin{pmatrix}
		\ell & \ell^{\prime} & \ell^{\prime\prime} \\
        -m & m^{\prime} & m^{\prime\prime} 
	\end{pmatrix}
    \begin{pmatrix}
		\ell & \ell^{\prime}  & \ell^{\prime\prime} \\
		\mp 2 & \pm 2 & 0 
	\end{pmatrix}\times \\ 
	& \frac{1}{2}  \left(W_{\ell^{\prime \prime} L L^{\prime}}^{000} \right)^2 \tau^\asc_{L 0}(m, z)\tau^\asc_{L^\prime 0}(m, z)  \mathcal{C}^{\pm}_{\ell^{\prime}}(\chi)  \leftindex_{\mp 2}Y_{\ell m}(\nhat_1) \leftindex_{\pm 2}Y_{\ell^{\prime} m^{\prime}}(\nhat_2)Y_{\ell^{\prime \prime} m^{\prime \prime}}(\nhat_3),
\end{aligned}
\end{equation}
where $\tau^\asc_{\ell 0}$ was defined in Eq.~\eqref{eq:tau_multipoles}. Now, adding the second piece coming from doing the same calculation but replacing $\nhat_1 \rightarrow \nhat_2$ in the $u$ screening function, we can read off the bispectra
\begin{equation}
\begin{aligned}    
   \label{eq:QQg_1halo_final} 
        \ev{a_{\ell m}^{\pm 2} a_{\ell^{\prime} m^{\prime}}^{\pm 2} g_{\ell^{\prime\prime} m^{\prime\prime}}} & = (-1)^{m}  \sqrt{\frac{(2 \ell + 1)(2 \ell^{\prime} + 1)(2 \ell^{\prime\prime} + 1)}{4\pi}} 
    \begin{pmatrix}
			\ell & \ell^{\prime} & \ell^{\prime\prime} \\
			-m & m^{\prime} & m^{\prime\prime} 
    \end{pmatrix}
    \begin{pmatrix}
			\ell & \ell^{\prime}  & \ell^{\prime\prime} \\
			\mp 2 & \pm 2 & 0 
    \end{pmatrix} \times \\
	& \left(\frac{1-e^{-x}}{x} \bar{T}\right)^2 \int \dd \chi \, \dd m \, n(\chi,m) u^{g, {\rm cen}}(\chi,m) \times \\
    & \sum_{L L^{\prime}}  \left(W_{\ell^{\prime\prime} L L^{\prime}}^{000}\right)^2  \tau^\asc_{L0}(\chi, m)\tau^\asc_{L^{\prime}0}(\chi, m) \frac{\mathcal{C}^{\pm}_{\ell^{\prime}}(\chi) + 
    (-1)^{\ell + \ell^\prime + \ell^{\prime \prime}} \mathcal{C}^{\pm}_{\ell}(\chi) }{2}.
\end{aligned}
\end{equation}
To write the three-point function in terms of the $E$ and $B$-modes, we note that $\ev{a^{\pm 2\ast}_{\ell m}a^{\mp 2}_{\ell^{\prime} m^{\prime}} g_{\ell^{\prime\prime} m^{\prime\prime}}} = 0 $, which from the definition of $E$ and $B$ means that $\ev{E^{*}_{\ell m} E^{*}_{\ell^{\prime} m^{\prime}} g_{\ell^{\prime\prime} m^{\prime\prime}}} = \ev{B^{*}_{\ell m} B^{*}_{\ell^{\prime} m^{\prime}} g_{\ell^{\prime\prime} m^{\prime\prime}}} $. Similarly to what obtained in App.~\ref{app:pol_cell} for the power spectrum, it can then be shown that 
\begin{equation}
\begin{aligned}
	\ev{E^{\asc *}_{\ell m} E^{\asc}_{\ell^{\prime} m^{\prime}} g_{\ell^{\prime\prime} m^{\prime\prime}}} &= \ev{B^{\asc *}_{\ell m} B^{\asc}_{\ell^{\prime} m^{\prime}} g_{\ell^{\prime\prime} m^{\prime\prime}}} = \frac{1}{4} \left( \ev{a^{+2 *}_{\ell m}a^{+2}_{\ell^{\prime} m^{\prime}} g_{\ell^{\prime\prime} m^{\prime\prime}}} + \ev{a^{-2 *}_{\ell m}a^{-2}_{\ell^{\prime} m^{\prime}} g_{\ell^{\prime\prime} m^{\prime\prime}}} \right) \\
	& = (-1)^{m}  \sqrt{\frac{(2 \ell + 1)(2 \ell^{\prime} + 1)(2 \ell^{\prime\prime} + 1)}{4\pi}} 
    \begin{pmatrix}
			\ell & \ell^{\prime} & \ell^{\prime\prime} \\
			-m & m^{\prime} & m^{\prime\prime} 
    \end{pmatrix}
    \begin{pmatrix}
			\ell & \ell^{\prime}  & \ell^{\prime\prime} \\
			+ 2 & - 2 & 0 
    \end{pmatrix} e_{\ell \ell^{\prime} \ell^{\prime\prime}} \times \\
	& \left(\frac{1-e^{-x}}{x} \bar{T}\right)^2 \int \dd \chi \, \dd m \, n(\chi,m) u^{g, {\rm cen}}(\chi,m) \times \\
    & \sum_{L L^{\prime}}  \left(W_{\ell^{\prime\prime} L L^{\prime}}^{000}\right)^2  \tau^\asc_{L0}(\chi, m)\tau^\asc_{L^{\prime}0}(\chi, m) \frac{\mathcal{C}^{\rm pol}_{\ell}(\chi) + \mathcal{C}^{\rm pol}_{\ell^{\prime}}(\chi)}{2},
\end{aligned}
\end{equation}
where $C_{\ell}^{\rm pol}$ is defined in Eq.~\eqref{eq:cell_pol_f_dom} and
\begin{equation} \label{eq:e_def}
    e_{\ell \ell^{\prime} \ell^{\prime\prime}} \equiv \frac{1}{2} \left[1+ (-1)^{\ell+\ell^{\prime}+\ell^{\prime\prime}} \right].
\end{equation}
From the equation above we can read off the angle-averaged bispectrum $\mathcal{B}$ as
\begin{align}
\label{eq:red_bispectra}
	\ev{X^{\asc}_{\ell m} X^{\asc}_{\ell^{\prime} m^{\prime}} g_{\ell^{\prime\prime} m^{\prime\prime}}} &  = (-1)^m
      \begin{pmatrix}
			\ell & \ell^{\prime} & \ell^{\prime\prime} \\
			-m & m^{\prime} & m^{\prime\prime} 
	\end{pmatrix}
    \mathcal{B}_{\ell \ell^{\prime} \ell^{\prime\prime}}^{X^\asc X^\asc g},
\end{align}
which results in the expression given in Sec.~\ref{sec:cmb_bispectrum}, in Eq.~\eqref{eq:red_bispectrum_B_1halo}. $\mathcal{B}_{\ell \ell^{\prime} \ell^{\prime\prime}}^{X^\asc X^\asc g}$ is symmetric under the exchange of $\ell \leftrightarrow \ell^{\prime}$, as expected from the approximations made in the calculation above when taking the limit of squeezed triangles. In the signal-to-noise ratio only the angle-averaged bispectrum enters, since the sum over $m$ simplifies as 
\begin{align}
	\sum_{m m^{\prime} m^{\prime \prime}} \left[ (-1)^m
      \begin{pmatrix}
			\ell & \ell^{\prime} & \ell^{\prime\prime} \\
			-m & m^{\prime} & m^{\prime\prime} 
	\end{pmatrix}\right]^2 = 1.
\end{align}

\subsection{Two-halo term}

We write explicitly the three-point function by summing the contributions from two different halos $i$ and $j$, such that the 2-halo contribution to Eq.~\eqref{eq:QQg} becomes
\begin{equation}\label{eq:QQg_derivation}
    \begin{aligned}
    & \left(\frac{1-e^{-x}}{x} \bar{T}\right)^2 \int \dd \chi_a \chi_a^2 \, \dd \chi_b \chi_b^2 \, \dd m_a \, \dd m_b \, \dd^2\hat{n}_a \, \dd^2\hat{n}_b \times \\
    & \frac{1}{9} P^2(\chi_a, m_a) N^2_{\rm res}(\chi_a, m_a) u(\nhat_1-\nhat_a) u(\nhat_2-\nhat_a) \ev{\gamma^{\pm}(\hat{n}_1, \chi_a)\gamma^{\pm}(\hat{n}_2, \chi_a)} \times \\ 
    & u^{g, {\rm cen}}(\chi_b,m_b) \delta^{2} (\nhat_3-\nhat_b)  \ev{\sum_{i\neq j} \delta(m_a-m_i)\delta(m_b-m_j) \frac{\delta(\chi_a -\chi_i)}{\chi_a^2} \frac{\delta(\chi_b - \chi_j)}{\chi_b^2} \delta^{2} (\nhat_a-\nhat_i)\delta^{2} (\nhat_b-\nhat_j)}.
    \end{aligned}
\end{equation}
The average in the last line of the equation above is related to the halo-halo auto-correlation function $\xi^{hh}$,
\begin{equation}
\begin{aligned}\label{eq:hh_xi}
    \ev{\sum_{i\neq j} \delta(m_a-m_i)\delta(m_b-m_j) \frac{\delta(\chi_a -\chi_i)}{\chi_a^2} \frac{\delta(\chi_b - \chi_j)}{\chi_b^2} \delta^{2} (\nhat_a-\nhat_i)\delta^{2} (\nhat_b-\nhat_j)} = \\
    = n(m_a, \chi_a) n(m_b, \chi_b) \xi^{hh}(\nhat_a-\nhat_b  | m_a, \chi_a, m_b, \chi_b).
    \end{aligned}
\end{equation}
Eq.~\eqref{eq:QQg_derivation} then becomes
\begin{equation}
\begin{aligned}
    & \left(\frac{1-e^{-x}}{x} \bar{T}\right)^2 \int \dd \chi_a \chi_a^2 \, \dd \chi_b \chi_b^2 \, \dd m_a \, \dd m_b \, n(\chi_a,m_a)n( \chi_b, m_b) u^{g, {\rm cen}}(\chi_b,m_b) \times \\
    & \frac{1}{9} P^2(\chi_a, m_a) N^2_{\rm res}(\chi_a, m_a) \ev{\gamma^{\pm}(\hat{n}_1, \chi_a)\gamma^{\pm}(\hat{n}_2, \chi_a)}  \int \dd^2\hat{n}_a \,  u(\nhat_1-\nhat_a) u(\nhat_2-\nhat_a) \xi^{hh}(\nhat_a-\nhat_3).
\end{aligned}
\end{equation}
The integral over $\nhat_a$ can be simplified by noting that the magnetic field domains are much smaller than the typical size of a halo, such that $|\nhat_1-\nhat_2| \ll |\nhat_a-\nhat_1| \simeq |\nhat_a-\nhat_2|$. On the other hand, the halo-halo auto-correlation is dominated by larger scales, which means $|\nhat_a-\nhat_1| \simeq |\nhat_a-\nhat_2| \ll|\nhat_a-\nhat_3|$. Therefore, the bispectrum is dominated by the squeezed triangles, with $|\nhat_1-\nhat_2| \ll |\nhat_3-\nhat_1|\simeq |\nhat_3-\nhat_2|$. In the equation above we can then replace $\nhat_a \rightarrow \nhat_{1,2}$ inside the halo-halo auto-correlation $\xi^{hh}$ and simply perform the integral over $\nhat_a$. We therefore approximate $\xi^{hh}(\nhat_a-\nhat_3) \simeq \left[\xi^{hh}(\nhat_1-\nhat_3)+\xi^{hh}(\nhat_2-\nhat_3) \right]/2$. We show the calculation with $\nhat_1$ below, since the one with $\nhat_2$ can be obtained in the same way. Using the result from App.~B.1 of~\citetalias{Pirvu:2023lch}, 
\begin{align}\label{eq:uu_int}
    \int \dd^2\hat{n}_a\, u(\nhat_1 - \nhat_a) u(\nhat_2- \nhat_a) = \sum_{\ell m} \frac{4\pi}{2\ell+1} u_{\ell 0}^2(\chi_a, m_a)Y_{\ell m}(\nhat_1)Y_{\ell m}(\nhat_2),
\end{align}
where $u_{\ell 0}$ was defined in Eq.~\eqref{eq:tau_multipoles}. Expanding all the angular-dependent functions into spherical harmonics, the three-point function simplifies to
\begin{equation}
\begin{aligned}    
   \label{eq:QQg_derivation_3}
    & \left(\frac{1-e^{-x}}{x} \bar{T}\right)^2 \int \dd \chi_a \chi_a^2 \, \dd \chi_b \chi_b^2 \, \dd m_a \, \dd m_b \, n(\chi_a,m_a)n( \chi_b, m_b) u^{g, {\rm cen}}(\chi_b,m_b) \times \\
    & \sum_{\ell m}  \sum_{\ell^{\prime} m^{\prime}}  \sum_{\ell^{\prime\prime} m^{\prime\prime}} \frac{1}{2} [\tau^\asc_{\ell 0}(\chi_a, m_a)]^2 \mathcal{C}^{\pm}_{\ell^{\prime}}(\chi_a)  C^{hh}_{\ell^{\prime\prime}}(m_a, \chi_a, m_b, \chi_b) \times \\ 
    & Y_{\ell m}(\nhat_1)Y_{\ell m}(\nhat_2) \leftindex_{\pm 2}Y_{\ell^{\prime}m^{\prime}}(\nhat_1) \leftindex_{\pm 2}Y_{\ell^{\prime} m^{\prime}}(\nhat_2) Y_{\ell^{\prime\prime} m^{\prime\prime}}(\nhat_1)Y_{\ell^{\prime\prime} m^{\prime\prime}}(\nhat_3),
\end{aligned}
\end{equation}
where $\tau^\asc_{\ell 0}$ was defined in Eq.~\eqref{eq:tau_multipoles}, $\mathcal{C}^{\pm}_{\ell^{\prime}}$ was introduced in Eq.~\eqref{eq:polwindow}, and $C_{\ell}^{hh}$ is the power spectrum of the real-space halo-halo auto-correlation $\xi^{hh}$. This takes the following form:
\begin{equation}\label{eq:Cell_linear_hh}
    C_{\ell}^{hh} \left(m_a, \chi_a, m_b, \chi_b\right)=\frac{2}{\pi} b(m_a, \chi_a) b(m_b, \chi_b) \int \mathrm{d} k k^2 j_{\ell}\left(k \chi_a\right) j_{\ell}\left(k \chi_b \right) \sqrt{P^{\rm lin}(k, \chi_a)P^{\rm lin}(k, \chi_b)},
\end{equation}
where $P^{\rm lin}(k)$ is the linear matter power spectrum, and $b(m, \chi)$ is the linear halo bias. In the limit of small angle, $\ell \to \infty$, we can approximate the spherical Bessel function by $j_{\ell}(x) \rightarrow \sqrt{\pi/ \left(2 \ell+1\right)} \delta(\ell+1 / 2-x)$, where $x = \chi k$. Thanks to the delta function, the integral over comoving wavenumber $k$ simplifies to
\begin{equation}\label{eq:Cell_linear_hh_2}
    C_{\ell}^{hh} \left(m_a, \chi_a, m_b, \chi_b\right) \approx b(m_a, \chi_a) b(m_b, \chi_b) \frac{\delta(\chi_a - \chi_b)}{\chi^2_a} P^{\rm lin} \left( \frac{\ell+\frac{1}{2}}{\chi_a}, \chi_a \right).
\end{equation}
This simplification is equivalent to the Limber approximation~\cite{1953ApJ...117..134L, PhysRevD.88.063526}. The delta function in the expression above further simplifies the integral over $\chi_b$ in the bispectrum.

Now we want to reduce the product of spherical harmonics in Eq.~\eqref{eq:QQg_derivation_3} down to three, to read off the coefficients of the bispectrum. From the derivation of the polarization power spectra in App.~\ref{app:pol_cell}, we already know that
\begin{equation}\label{eq:y_prod_1}
    \sum_{m m^{\prime}} Y_{\ell m}(\nhat_1)Y_{\ell m}(\nhat_2)\leftindex_{\pm 2}Y_{\ell^{\prime} m^{\prime}}(\hat{n}_1) \leftindex_{\pm 2}Y_{\ell^{\prime} m^{\prime}}(\hat{n}_2)  = \sum\limits_{\substack{L M }}  \left(W_{L \ell^{\prime} \ell}^{220}\right)^2 \leftindex_{\mp 2}Y_{L M}(\nhat_1) \leftindex_{\mp 2}Y_{L M}(\nhat_2).
\end{equation}
Therefore, after relabeling, the bispectrum from Eq.~\eqref{eq:QQg_derivation_3} becomes
\begin{equation}
\begin{aligned}
    & \left(\frac{1-e^{-x}}{x} \bar{T}\right)^2 \int \dd \chi_a \chi_a^2 \, \dd m_a \, \dd m_b \, n(\chi_a,m_a)n( \chi_a, m_b) u^{g, {\rm cen}}(\chi_a,m_b) b(m_a, \chi_a) b(m_b, \chi_a) \times \\
    & \sum_{\ell^{\prime} m^{\prime} }  \sum_{\ell^{\prime\prime} m^{\prime\prime}} \sum\limits_{\substack{L L^{\prime} }} \frac{1}{2}\left(W_{\ell^{\prime} L^{\prime} L}^{220}\right)^2 [\tau^\asc_{L 0}(\chi_a, m_a)]^2 \mathcal{C}^{\pm}_{L^{\prime}}(\chi_a)  P^{\rm lin} \left( \frac{\ell^{\prime\prime}+\frac{1}{2}}{\chi_a}, \chi_a \right) \times \\ 
    &  \leftindex_{\mp 2}Y_{\ell^{\prime} m^{\prime}}(\nhat_1) \ Y_{\ell^{\prime\prime} m^{\prime\prime}}(\nhat_1) \leftindex_{\mp 2}Y_{\ell^{\prime} m^{\prime}}(\nhat_2) Y_{\ell^{\prime\prime} m^{\prime\prime}}(\nhat_3).
\end{aligned}
\end{equation}
Finally, replacing the product $\leftindex_{\mp 2}Y_{\ell^{\prime} m^{\prime}}(\nhat_1) Y_{\ell^{\prime\prime} m^{\prime\prime}}(\nhat_1)$ with one spherical harmonics, we get
\begin{equation}
\begin{aligned}
    &  \left(\frac{1-e^{-x}}{x} \bar{T}\right)^2 \int \dd \chi_a \chi_a^2 \, \dd m_a \, \dd m_b \, n(\chi_a,m_a) n( \chi_a, m_b) u^{g, {\rm cen}}(\chi_a,m_b) b(m_a, \chi_a) b(m_b, \chi_a)  \times \\
    &\sum_{\ell m}  \sum\limits_{\substack{\ell^{\prime} m^{\prime} }} \sum_{\ell^{\prime\prime} m^{\prime\prime}} 
 \sum_{L L^{\prime}} (-1)^{m}  \sqrt{\frac{(2 \ell + 1)(2 \ell^{\prime} + 1)(2 \ell^{\prime\prime} + 1)}{4\pi}} 
    \begin{pmatrix}
		\ell & \ell^{\prime} & \ell^{\prime\prime} \\
    -m & m^{\prime} & m^{\prime\prime} 
	\end{pmatrix}
    \begin{pmatrix}
		\ell & \ell^{\prime}  & \ell^{\prime\prime} \\
		\pm 2 & \mp 2 & 0 
	\end{pmatrix}\times \\ 
 &  \frac{1}{2}\left(W_{\ell^{\prime} L^{\prime} L}^{220}\right)^2 [\tau^\asc_{L0}(\chi_a, m_a)]^2 \mathcal{C}^{\pm}_{L^{\prime}}(\chi_a)  P^{\rm lin} \left( \frac{\ell^{\prime\prime}+\frac{1}{2}}{\chi_a}, \chi_a \right) \leftindex_{\pm 2}Y_{\ell m}(\nhat_1) \leftindex_{\mp 2}Y_{\ell^{\prime} m^{\prime}}(\nhat_2) Y_{\ell^{\prime\prime} m^{\prime\prime}}(\nhat_3). 
\end{aligned}
\end{equation}
Now, adding the second piece coming from doing the same calculation but replacing $\nhat_a \rightarrow \nhat_2$ in the halo-halo auto-correlation function, we can read off the bispectra
\begin{equation}
\begin{aligned}    
   \label{eq:QQg_final} 
        \ev{a_{\ell m}^{\pm 2} a_{\ell^{\prime} m^{\prime}}^{\pm 2} g_{\ell^{\prime\prime} m^{\prime\prime}}} & = (-1)^{m}  \sqrt{\frac{(2 \ell + 1)(2 \ell^{\prime} + 1)(2 \ell^{\prime\prime} + 1)}{4\pi}} 
      \begin{pmatrix}
			\ell & \ell^{\prime} & \ell^{\prime\prime} \\
			-m & m^{\prime} & m^{\prime\prime} 
	\end{pmatrix}
    \begin{pmatrix}
			\ell & \ell^{\prime}  & \ell^{\prime\prime} \\
			\pm 2 & \mp 2 & 0 
	\end{pmatrix} \\
	& \left(\frac{1-e^{-x}}{x} \bar{T}\right)^2 \int \dd \chi_a \chi_a^2 \, \dd m_a \, \dd m_b \, n(\chi_a,m_a) n( \chi_a, m_b) u^{g, {\rm cen}}(\chi_a,m_b) b(m_a, \chi_a) b(m_b, \chi_a) \times \\
    & \sum_{L L^{\prime}}  \frac{\left(W_{\ell L^{\prime} L}^{220}\right)^2+\left(W_{\ell^{\prime} L^{\prime} L}^{220}\right)^2}{2}  [\tau^\asc_{L0}(\chi_a, m_a)]^2 \mathcal{C}^{\pm}_{L^{\prime}}(\chi_a)  P^{\rm lin} \left( \frac{\ell^{\prime\prime}+\frac{1}{2}}{\chi_a}, \chi_a \right).
\end{aligned}
\end{equation}
To write the three-point function in terms of the $E$ and $B$-modes, we note that $\ev{a^{\pm 2\ast}_{\ell m}a^{\mp 2}_{\ell^{\prime} m^{\prime}} g_{\ell^{\prime\prime} m^{\prime\prime}}} = 0 $, which from the definition of $E$ and $B$ means that $\ev{E^{*}_{\ell m} E^{*}_{\ell^{\prime} m^{\prime}} g_{\ell^{\prime\prime} m^{\prime\prime}}} = \ev{B^{*}_{\ell m} B^{*}_{\ell^{\prime} m^{\prime}} g_{\ell^{\prime\prime} m^{\prime\prime}}} $. Similarly to what obtained in App.~\ref{app:pol_cell} for the power spectrum, it can then be shown that 
\begin{align}
	\ev{E^{\asc *}_{\ell m} E^{\asc}_{\ell^{\prime} m^{\prime}} g_{\ell^{\prime\prime} m^{\prime\prime}}} &= \ev{B^{\asc *}_{\ell m} B^{\asc}_{\ell^{\prime} m^{\prime}} g_{\ell^{\prime\prime} m^{\prime\prime}}} = \frac{1}{4} \left( \ev{a^{+2 *}_{\ell m}a^{+2}_{\ell^{\prime} m^{\prime}} g_{\ell^{\prime\prime} m^{\prime\prime}}} + \ev{a^{-2 *}_{\ell m}a^{-2}_{\ell^{\prime} m^{\prime}} g_{\ell^{\prime\prime} m^{\prime\prime}}} \right) \nonumber \\
	&  = (-1)^{m}\sqrt{\frac{(2 \ell + 1)(2 \ell^{\prime} + 1)(2 \ell^{\prime\prime} + 1)}{4\pi}} 
      \begin{pmatrix}
			\ell & \ell^{\prime} & \ell^{\prime\prime} \\
			-m & m^{\prime} & m^{\prime\prime} 
	\end{pmatrix}
    \begin{pmatrix}
			\ell & \ell^{\prime}  & \ell^{\prime\prime} \\
			+ 2 & - 2 & 0 
	\end{pmatrix} e_{\ell \ell^{\prime} \ell^{\prime\prime}} \times \nonumber  \\
	& \left(\frac{1-e^{-x}}{x} \bar{T}\right)^2 \int \dd \chi_a \chi_a^2 \, \dd m_a \, \dd m_b \, n(\chi_a,m_a) n( \chi_a, m_b) u^{g, {\rm cen}}(\chi_a,m_b) b(m_a, \chi_a) b(m_b, \chi_a) \times \nonumber \\
    & \sum_{L L^{\prime}}  \frac{\left(W_{\ell L^{\prime} L}^{220}\right)^2+\left(W_{\ell^{\prime} L^{\prime} L}^{220}\right)^2}{2}  [\tau^\asc_{L0}(\chi_a, m_a)]^2 \mathcal{C}^{\rm{pol}}_{L^{\prime}}(\chi_a)   P^{\rm lin} \left( \frac{\ell^{\prime\prime}+\frac{1}{2}}{\chi_a}, \chi_a \right),
\end{align}
where $C_{\ell}^{\rm pol}$ is defined in Eq.~\eqref{eq:cell_pol_f_dom} and $e_{\ell \ell^{\prime} \ell^{\prime\prime}}$ in Eq.~\eqref{eq:e_def}. From the equation above we can read off the angle-averaged bispectrum which results in the expression given in Sec.~\ref{sec:cmb_bispectrum}, in Eq.~\eqref{eq:red_bispectrum_B_2halo}. As for the 1-halo term, also in this case $\mathcal{B}_{\ell \ell^{\prime} \ell^{\prime\prime}}^{X^\asc X^\asc g}$ is symmetric under the exchange of $\ell \leftrightarrow \ell^{\prime}$. 

\section{Foregrounds and noise}\label{app:foregroundandnoise}

In this appendix, we outline our prescription for estimating the noise covariance matrix $\mathbf{N}_\ell$ used in the ILC described in Sec.~\ref{sec:ilc} for temperature and polarization. We consider temperature and polarization data from two CMB experiments: the combination of the Low Frequency Instrument (LFI)~\cite{planck2015LFI} and High Frequency Instrument (HFI)~\cite{planck2015HFI} on the Planck satellite and CMB Stage-$4$~\cite{abazajian_cmb-s4_2016}. For each experiment, we specify the observed frequency channels and angular resolution as defined by a Gaussian beam 
\begin{equation}\label{eq:gaussbeam}
  G_{\ell}(\omega) = \exp \left[- \ell(\ell+1) \, \frac{\theta_{\rm FWHM}^2}{8 \ln 2}\right],  
\end{equation}
where the full width at half maximum $\theta_{\rm FWHM} \left[ \rm rad\right]$ varies with frequency. The assumed values are recorded in Table~\ref{tab:noisemodel}.
\begin{table}[ht!]
    \begin{center}
        \begin{tabular}{|c||c|c|c|c|c|c|c|} 
            \multicolumn{1}{l}{Planck: } \\
            \hline
            $\omega/(2\pi)$ $\rm[GHz]$ & $30$ & $44$ & $70$ & $100$ & $143$ & $217$ & $353$\\
            $\theta_{\rm FWHM} \, [\rm arcmin]$ & $32.41$ & $27.1$ & $13.32$ & $9.69$ & $7.3$ & $5.02$ & $4.94$\\
            \hline
        \end{tabular}

        \bigskip

        \begin{tabular}{|c||c|c|c|c|c|c|c|} 
          \multicolumn{1}{l}{CMB-S$4$: } \\
            \hline
            $\omega/(2\pi)$ $\rm[GHz]$ & $20$ & $27$ & $39$ & $93$ & $145$ & $225$ & $278$ \\
            $\Delta_T \, [\rm \mu K \, arcmin]$ & $10.41$ & $5.14$ & $3.28$ & $0.50$ & $0.46$ & $1.45$ & $3.43$ \\
            $\theta_{\rm FWHM} \, [\rm arcmin]$ & $11.0$ & $8.4$ & $5.8$ & $2.5$ & $1.6$ & $1.1$ & $1.0$ \\
            \hline
        \end{tabular}
    \end{center}
    \caption{The top panel shows the frequency bins and beam parameters used for the Planck forecast. The bottom panel shows frequencies, sensitivity and resolution parameters for the CMB-S4 V3R0 configuration. $\theta_{\rm FWHM}$ is full width at half maximum of the assumed Gaussian beams, which characterizes the resolution of the instrument in each frequency channel, while $\Delta_T$ represents the amplitude of the white uncorrelated noise in CMB temperature units.}\label{tab:noisemodel}
\end{table}

 For Planck, since data is readily available, we take an empirical approach. Our analysis is based on publicly available individual frequency and component-separated CMB maps from the Planck Public Data Release 3 (PR3)~\cite{PLA}. We first subtract the SMICA CMB from individual frequency maps at 30-353 GHz in intensity as well as $Q$ and $U$ Stokes parameters. We mask the resulting maps with a galactic cut retaining $40\%$ of the sky, apodized to 2 degrees. We compute the auto- and cross-spectra between all masked maps. We do not correct for mode-coupling from the mask, approximating full-sky power spectra by the cut-sky pseudo-$C_\ell$ spectra divided by the effective unmasked sky-fraction $\sim 0.4$. Additionally, for polarization we use the full-sky expressions to produce $E$ and $B$-mode spectra from the Stokes parameter. We populate the matrices $\mathbf{N}_\ell$ used in the ILC using these auto- and cross-spectra. Our treatment provides an estimate of the level of foregrounds and instrumental noise on the cleanest region of the sky. A more careful treatment accounting for mode-coupling would improve the accuracy of our estimate primarily on large angular scales, and particularly for polarization spectra where the $E$-$B$ decomposition is particularly sensitive to masking.

For CMB-S4, we take a hybrid approach to estimating the noise covariance, considering three contributions: galactic foregrounds empirically measured from Planck, instrumental noise, and simulated extragalactic foregrounds. To estimate the contribution from galactic foregrounds, we first fit the low-$\ell$ (defined as $\ell < 100$) entries in $\mathbf{N}_\ell$ for Planck temperature and Stokes parameters to a power law $[\mathbf{N}_\ell]_{ij} = A_{ij} \ell^{-n_{ij}}$. We then use linear interpolation/extrapolation to obtain entries at the S$4$ frequencies, listed in the top row of the bottom panel in Table~\ref{tab:noisemodel}. The instrumental noise contribution is modeled as
\begin{equation}
I^{TT}_{\ell} = I^{E^{\asc} E^{\asc}}_{\ell} / \sqrt{2} = I^{BB}_{\ell} / \sqrt{2} = \Delta_T^2 \left[ 1 + \left(\ell/\ell_{\rm knee}\right)^{\alpha_{\rm knee}} \right],
\end{equation}
where $\Delta_T \left[ \rm \mu K \, rad\right]$ is the level of white noise representing the sensitivity in each frequency channel (values recorded in Table~\ref{tab:noisemodel}), and the parameters $\alpha_{\rm knee}=-3$ and $\ell_{\rm knee}=100$ parameterize atmospheric systematics on large angular scales. The dominant extragalactic foreground at high frequencies (where the axion-induced screening signal is most important) is the cosmic infrared background (CIB). We model this by computing the auto- and cross- power spectra of CIB maps from the Websky suite of simulations at $93, 145, 225, 278 {\rm \; GHz}$ respectively~\cite{Stein_2020}. For polarization, we assume that the CIB is $1\%$ polarized, and use the temperature maps to estimate the polarization signal from extragalactic sources. The full noise covariance is obtained by summing the three components described above for all auto- and cross-power spectra of the CMB-S4 frequency channels in temperature and polarization.

\section{Effectively massless axions}\label{app:massless_axion}

When the axion is effectively massless ($m_{\asc} \ll 10^{-14} \,{\rm eV}$), we no longer expect to find plasma densities that yield resonant conversion. In this scenario, we can apply the treatment in App.~\ref{app:conversion} for a small magnetic field with domain size $r_{\dom}$ to obtain a conversion probability given by
\begin{equation} 
    P(A_{\parallel}\rightarrow \asc) = (\Delta_{\asc\gamma} r_{\dom})^2 \frac{\sin^2 (\Delta_{\rm{osc}} r_{\dom}/2)}{(\Delta_{\rm{osc}} r_{\dom}/2)^2} \approx 4   \left(\frac{\Delta_{\asc\gamma}}{\Delta_{\rm pl}} \right)^2 \sin^2 (\Delta_{\rm pl} r_{\dom}/2).
\end{equation}
We expect photon-axion conversion to happen in different astrophysical and cosmological environments, including
\begin{eqnarray}
    &\text{ISM:} \quad & |\mathbf{B}|\simeq \mu G,\quad     \ompl \simeq 10^{-11} \,\rm{eV}, \quad r_{{\dom}} \simeq 1 \,\rm{kpc} \nonumber\\
    &\text{CGM:} \quad & |\mathbf{B}|\simeq 0.1 \mu G,\quad \ompl \simeq 10^{-12.5} \,\rm{eV}, \quad r_{{\dom}} \simeq 10 \,\rm{kpc}  \\
    &\text{IGM:} \quad & |\mathbf{B}|\simeq nG,\quad        \ompl \simeq 10^{-14} \,\rm{eV}, \quad r_{{\dom}} \simeq 1 \,\rm{Mpc},\nonumber
\end{eqnarray}
where ISM refers to the interstellar medium around the location of the Solar system, CGM refers to the circumgalactic medium in the vicinity of the Milky Way and other galaxies, and IGM refers to the intergalactic medium between galaxies. In all three cases, $\Delta_{\rm pl} \gg \Delta_{\asc\gamma}$ and  $\Delta_{\rm pl} r_{\dom} \gg 1$. In this limit, the conversion probability per domain is 
\begin{equation} 
    P(A_{\parallel}\rightarrow \asc) = 2   \left(\frac{\Delta_{\asc\gamma}}{\Delta_{\rm pl}} \right)^2 \approx \frac{2 \ga^2 |\mathbf{B}|^2 \omega^2}{\ompl^4}.
\end{equation}
and the conversion rate per unit length in the three different environments is approximately
\begin{equation}
  \frac{{\mathrm d} P}{{\mathrm d} r}= \frac{2 \ga^2 |\mathbf{B}|^2 \omega^2}{\ompl^4 \, r_{\dom} }  \approx \begin{cases}
    & 4 \times 10^{-15}/{\rm Mpc} \quad \text{(ISM)}, \\
    & 4 \times 10^{-12}/{\rm Mpc} \quad \text{(CGM)}, \\
    & 4 \times 10^{-12}/{\rm Mpc} \quad \text{(IGM)}, \\
    \end{cases}
\end{equation}
for $\omega = 10^{-4} \,\rm{eV}$ and $\ga = 10^{-10} \,{\rm GeV}^{-1}$. This suggests that if the axion is effectively massless, the conversion rate is small towards the center of the galaxy and peaks somewhere in between the circumgalactic medium and the intergalactic medium. The exact dependence of this conversion probability on the distance from the halo center depends on how the plasma density, the magnetic field strength and the magnetic field domain sizes change, in particular in the region several virial radii away from the galaxy center. This region is notoriously hard to model as a result of baryonic feedback. Recent and upcoming observations and simulations~\cite{2023arXiv230913104P,Madhavacheril:2019buy,Amodeo:2020mmu,Pillepich:2019bmb} that target the missing baryon problem might also shed more light on this question, and we leave a more detailed study to future work when more information is available. 

However, to motivate further studies, we estimate the sensitivity of axion-induced screening of the CMB to this signal with a simple heuristic model, first studied in~\cite{Mirizzi:2005ng}. We add to the toy model in~\cite{Mirizzi:2005ng} galaxies with sharp boundaries at fixed $r_{\rm b}$ from the halo center. Inside the sharp boundary, we assume the properties of the medium is similar to the ISM around the location of the Sun, while outside the sharp boundary, the medium is similar to the environment of the IGM. In this toy model, axion-photon conversion only happens outside the sharp boundary, and as a result, is anti-correlated with the location of halos. Similar to~\cite{Mirizzi:2005ng}, we treat $m_{\gamma, {\rm eff}} = \sqrt{\ompl^2 - m_{\asc}^2}$ as well as $\ga |\mathbf{B}|$ in the intergalactic medium as free parameters.

\begin{figure}[h!]
    \centering
    \includegraphics[width=1\textwidth]{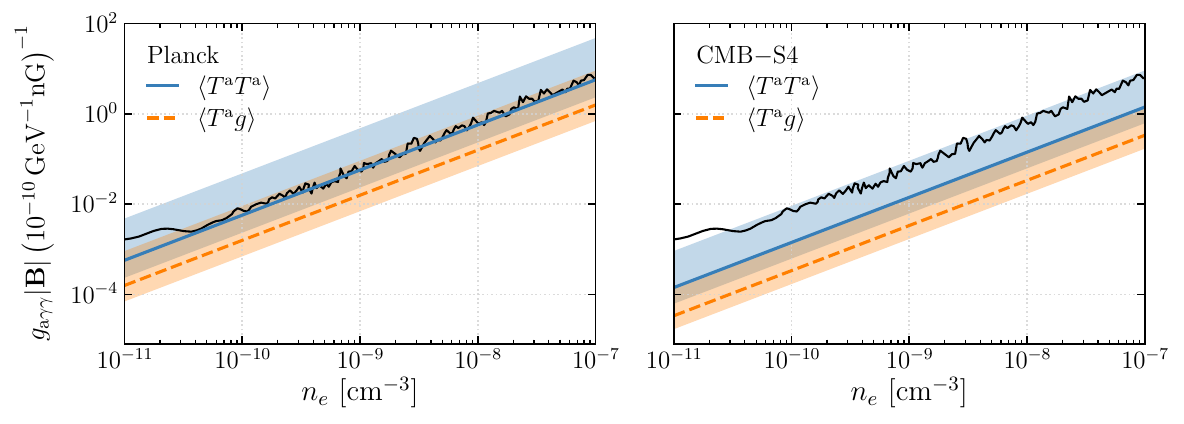}
    \caption{Projected sensitivity of axion-induced screening to the combination of axion-photon coupling $\ga/{10^{-10} {\rm GeV}^{-1}}$ and the extra-galactic magnetic field $|\mathbf{B}|/{\rm nG}$ in the simple sharp boundary model (see equation~\eqref{eq:masslessaxion}), assuming Planck (left) and CMB-S4 (right) sensitivity and the unWISE blue galaxy sample. The blue solid (orange-dashed) line is the projected sensitivity with temperature auto-correlation (cross-correlation with unWISE galaxies) assuming $r_b = 5 R_{\rm vir}$, while the shaded band is obtained for $r_b$ in the range $ (R_{\rm vir} ,  10 R_{\rm vir})$. The magnetic domain size is fixed at $r_{\dom} = 1\,{\rm Mpc}$ in the IGM and the effective electron density in the IGM $n_e = m_e m_{\gamma, \rm eff}^2/e^2$ is a free parameter, as in Ref.~\cite{Mirizzi:2005ng}. The region above the black solid line is excluded from the analysis of COBE/FIRAS data  in~\cite{Mirizzi:2005ng}.}\label{fig:massless_axion.pdf}
\end{figure}

In this toy model, we can compute the 1-halo and 2-halo contribution to the correlations of the conversion probability. In the limit where the 2-halo term completely dominates, the toy model qualitatively captures the anisotropies of this non-resonant conversion. The conversion probability along a line of sight direction $\hat{n}$ with a halo centered along the direction $\hat{n}_i$ can be computed to be
\begin{eqnarray}\label{eq:masslessaxion}
    P(\chi_i,m_i) = \bar{P} - \frac{4}{3}\frac{\ga^2 |\mathbf{B}|^2 \omega^2}{m_{\gamma, {\rm eff}}^4 \, r_{\dom}} r_{\rm b} \sqrt{1-\left(\frac{\chi_i\theta}{(1+z_i) r_{\rm b}}\right)^2},
\end{eqnarray}
where $\theta$ is again the angular separation between $\hat{n}$ and  $\hat{n}_i$, and $\bar{P}$ is the conversion probability along a line of sight with no halos. We neglect the latter contribution since it contributes only to the monopole. Note the additional {\it minus} sign, which suggests that this signal appears as an emission (lack of absorption) that is correlated with the location of halos. We generally expect a cross correlation between the halo location and the conversion probability, though the sign can depend on the exact shape of the density and magnetic field profile. The result of our estimate is presented in Fig.~\ref{fig:massless_axion.pdf}, where we forecast the sensitivity of Planck and CMB-S4 to the axion photon coupling in the massless axion limit. The result is presented in the units used in~\cite{Mirizzi:2005ng} to highlight the prospect of improvement with the methodologies presented in this paper. With CMB-S4, we expect an improvement of up to two orders of magnitude over the constraints presented in~\cite{Mirizzi:2005ng}. 

\section{Rough sensitivity estimate}\label{app:estimate}

In this appendix, we provide quick but clearly imprecise methods to estimate the sensitivity of dark screening to various beyond standard model scenarios of interest\footnote{We thank Diego Redigolo for his effort in trying to understand our results, which prompted this appendix.}. These estimates apply to cases where we inject or remove photons (in the CMB frequency band) in a frequency and position dependent manner. The estimates will be in a language that is familiar to researchers thinking about precision experiments to search for dark matter/dark sector, and will be increasing in rigor. 

The information carried by CMB photons is enormous, partly due to the fact the total number of observable CMB photons is very large. The total number of CMB photons we can collect with CMB-S4 will be $\mathcal{N}_{\gamma} \sim 10^{28}$. With zero noise and perfect distinguishability, we can in principle measure an optical depth, the probability of injecting or removing of CMB photons, as small as $1/\sqrt{\mathcal{N}_{\gamma}} \sim 10^{-14}$. This estimate is certainly very crude, since photon removal generally happens only in a small fraction of the universe, and more importantly, there are other contributions to the noise level that should be taken into account.

Generally, the temperature perturbation generated from axion-induced screening (similar for dark photon) is $\delta T^{\asc} \simeq \bar{T} \delta \tau^{\asc}$, which should be compared to the noise $\delta T^{\rm noise}$, consisting of the primary CMB perturbations, foregrounds, and instrumental noise. For signals that cannot be distinguished from background (including the primary CMB anisotropies), one can constrain an optical depth of $\delta \tau^{\asc} \sim 10^{-5}$. 

The dark screening signal can be distinguished from background in two major aspects. Firstly, the optical depth, from both photon to dark photon, or photon to axion conversion, is frequency dependent. This allows for significant reduction of contamination by using the ILC technique, which can reduce $\delta T^{\rm noise}$ by more than three orders of magnitude, for example, in the case of searches for dark photon screening with CMB-S4 (see Fig.~\ref{fig:ILC_improvement} and Fig.~8 of~\citetalias{Pirvu:2023lch}; note that this factor depends on the reference frequency used and can vary by about an order of magnitude across the whole frequency range). This reduction of noise depends on the instrument, as well as the frequency scaling of $\tau^{\asc}$, and has to be worked out explicitly. Secondly, the signal and noise generally have different spatial profile, or $\ell$-dependence. Qualitatively, this can be thought of as doing $\ell^2$-measurements at the same time, and the sensitivity increases when more $\ell$-modes get included in the experiment. For example, the sensitivity in $\delta\tau^{\asc}$ scales approximately as $1/\sqrt{\ell}$ for the auto-correlation observable (which contains two factors of $\delta\tau^{\asc}$), and as  $1/\ell$ for cross-correlation with large-scale structure observable (which has only one factor of $\delta\tau^{\asc}$). Combining the two improvements from the characteristic frequency and spatial dependence of the signal, we can estimate the sensitivity to $\delta \tau^{\asc} \sim 3\times 10^{-10}$ with auto-correlation, and $\delta \tau^{\asc} \sim 10^{-11}$ with cross-correlation. Keep in mind that optimizing the correlation functions only changes the scaling of the sensitivity with $\ell$, but the other parameters (such as $\bar{T}$, conversion radius, or magnetic field strength) appear in the same combination.

Depending on the BSM model, the optical depth can have different parametric dependence on the physical quantities of distant galaxies and halos, as well as BSM parameters. For example, the photon to axion conversion optical depth $\tau^{\asc}$ scales as
\begin{equation}
    \tau^{\asc} \sim \frac{\ga^2 \left|\mathbf{B}\right|^2 \omega \, r_{\rm res}}{m_{\asc}^2},
\end{equation}
and a rough sensitivity of $\ga$ can be estimated to be $ \ga \sim 10^{-13} {\rm GeV}^{-1}$ with cross correlation function $\ev{T^{\asc} g}$ and CMB-S4 sensitivity. Note that the parameters $r_{\rm res}$ and $\left|\mathbf{B}\right|$ both depend on the halo mass and the axion mass, which in turn gives a scaling of the sensitivity to $\ga$ with $m_{\asc}$ that is more complicated than the one appearing in the equation above. Similar estimates can be obtained for searches of polarization signals once the residual noise-level post-ILC is estimated.

\section{Likelihood and sensitivity forecast} \label{app:likelihood}

For a fixed axion mass, all the observables used in this work have a simple scaling with the axion-photon coupling as $\propto \ga^n$, with $n=2$ or $4$. Given some angular correlation function $d_\ell$ from the ILC subtracted maps and an expected axion-induced signal $\ga^n s_{\ell}^{\rm a}$, the likelihood can be written as
\begin{equation}\label{eq:likelihood}
    -2 \ln \mathcal{L}(\ga) = \sum_{\ell} \frac{\left(d_{\ell} - \ga^n s_{\ell}^{\rm a}\right)^2}{\sigma_\ell^2} + {\rm const.},
\end{equation}
where $\sigma_\ell$ represents the noise covariance at each $\ell$ whose form depends on the specific observable considered; note that in this notation $\sigma_\ell$ also includes the appropriate factor to account for the number of samples available at each scale, including the effect of fractional sky coverage. Assuming that the data has no signal, we have $\ev{d_{\ell} s_{\ell}^{\rm a}} = 0$ and $\ev{d_{\ell}^2} = \sigma_\ell^2$, and the likelihood is maximized at $\ga=0$. The expectation value of the likelihood is therefore 
\begin{equation}
    \ev{-2 \ln \mathcal{L}(\ga)} = \ga^{2n} \sum_{\ell} \frac{\left(s_{\ell}^{\rm a}\right)^2}{\sigma_\ell^2} + {\rm const.}
\end{equation}
Following a Bayesian approach, the posterior distribution of the parameter $\ga$ is then
\begin{equation}
    f(\ga) = \frac{e^{- \frac{\ga^{2n}}{2 \sigma_{n}^2}}}{(2\sigma_n^2)^{\frac{1}{2n}}\Gamma\left(1+\frac{1}{2n}\right)},
\end{equation}
where we have defined $\sigma_n^2 \equiv \sum_{\ell} \left[\left(s_{\ell}^{\rm a}\right)^2/\sigma_\ell^2\right]$ and obtained the denominator from normalizing the distribution to 1 for $\ga \geq 0$,~\ie assuming a flat prior over positive couplings. To estimate the $1-\sigma$ sensitivity on the parameter, $\sigma_{\ga}$, we compute the largest value of the parameter that is compatible with the observation at $68\%$ CL. In general this is given by $\sigma_{\ga} = x (\sigma_n)^{1/n}$, where the numerical coefficient $x$ can be obtained by solving the equation
\begin{equation}
    \int_{0}^{x(\sigma_n)^{1/n}} f(\ga) = 0.68.
\end{equation}
For the case of $n=1$, one simply recovers a Gaussian posterior distribution and $x=1$; in that case the parameter $\sigma_1^2$ is just the inverse of the Fisher matrix, where now the likelihood is simply quadratic in the parameter and therefore the usual 2nd order Taylor expansion defining the Fisher matrix is exact. For our purposes, we are interested in the cases of $n=2,4$ that give
\begin{equation}
    x \simeq 0.76 \quad {\rm for} \quad n=2, \quad \quad \quad x \simeq 0.7 \quad {\rm for} \quad n=4.
\end{equation}
As expected if the leading order term in likelihood is $\ga^4$ or $\ga^8$, it will change more rapidly away from the maximum when $\ga$ deviates from 0, leading to a smaller uncertainty on the parameter. The numerical factors obtained here are used to estimate the sensitivity on the photon-axion coupling from the two- and three-point functions in Sec.~\ref{sec:results} -- see Eqs.~\eqref{eq:sigma_XX},~\eqref{eq:sigma_Tg} and~\eqref{eq:sigma_BBg}. A similar procedure can be followed if one wants to combine all the observables that have different scaling with $\ga$, by adding up their contributions in Eq.~\eqref{eq:likelihood} (neglecting cross-correlations).

\bibliography{bib}
\bibliographystyle{minor-changes-JHEP}

\end{document}